\documentstyle[aps,graphicx,epsf]{revtex}\tighten
\draft

\def \D {\mbox{D}}
\def\curl {\mbox{curl}\,}
\def \ep {\varepsilon}

\def \D {\mbox{D}}
\def \curl {\mbox{curl}\,}
\def \ep {\varepsilon}
\def\be{\begin{equation}}
\def\ee{\end{equation}}
\def\bea{\begin{eqnarray}}
\def\eea{\end{eqnarray}}

\def\be{\begin{equation}}
\def\ee{\end{equation}}
\def\bea{\begin{eqnarray}}
\def\eea{\end{eqnarray}}
\def\lb{\label}

\def\mathbb{\mbox{\boldmath $R$}}  
\def\gam{\gamma}

\def\eps{\epsilon}

\def\sig{\sigma}

\def\sigm{\sigma_{-}}
\def\sigc{\sigma_{\times}}
\def\np{n_{+}}
\def\nm{n_{-}}
\def\nc{n_{\times}}

\def\Sigp{\Sigma_{+}}
\def\Sigm{\Sigma_{-}}
\def\Sigc{\Sigma_{\times}}
\def\Np{N_{+}}
\def\Nm{N_{-}}
\def\Nc{N_{\times}}

\def\Om{\Omega}

\def\udot{\dot{u}}
\def\Udot{\dot{U}}
\def\ca{{\cal A}}

\def\cn{{\cal N}}

\def\D{\mbox{D}}
\def\p{{\bf e}}
\def\ptl{\partial}

\def\e{{\rm e}}
\def\la{\langle}
\def\ra{\rangle}
\def\ti{\tilde}
\def\hsp5{\hspace{5mm}}
\def\ol{\overline}
\def\case#1/#2{\textstyle\frac{#1}{#2}}

\def\EEE{E_1{}^1}

\def \D {\mbox{D}}
\def \D {\mbox{D}}

\def \curl {\mbox{curl}\,}
\def\eps{\epsilon}
\def\udot{\dot{u}}
\def\3nab{\tilde{\nabla}}

\def\la {\langle}
\def\ra {\rangle}
\def\hs {\,-\,}

\def\be {\begin{equation}}
\def\ee {\end{equation}}
\def\bea {\begin{eqnarray}}
\def\eea {\end{eqnarray}}

\newcommand{\sfrac}[2]{{\textstyle{#1\over#2}}}
\def\case#1/#2{\textstyle\frac{#1}{#2}}

\begin{document}

\title{The Dynamics of Brane-World Cosmological Models}

\author{A. A. Coley$^{1,*}$} \address{$^1$Department of Mathematics and
Statistics,
Dalhousie University, Halifax, Nova Scotia, Canada}
\maketitle

\begin{abstract}

Brane-world cosmology is 
motivated by recent developments in string/M-theory and 
offers a new perspective on the hierarchy problem.
In the brane-world scenario,  our Universe is a four-dimensional subspace
or {\em brane}
embedded in a higher-dimensional {\em bulk} spacetime. 
Ordinary matter fields are
confined to the brane while the gravitational field can also
propagate in the
bulk, and it is not
necessary for the extra dimensions to be small, or even
compact, leading to modifications of Einstein's theory of general relativity at high
energies.
In particular, the Randall-Sundrum-type models  are
relatively simple
phenomenological models which
capture some of the essential features of the
dimensional reduction of eleven-dimensional supergravity
introduced by Ho$\check{\mbox{r}}$ava and Witten.  These
curved (or warped) models are self-consistent and simple and allow for an
investigation of the essential non-linear gravitational dynamics.
The governing field equations induced on the brane differ from
the general relativistic equations  in that there are nonlocal
effects from the free gravitational field in the bulk, transmitted
via the projection of the bulk Weyl tensor,
and  the local quadratic energy-momentum corrections, which are
significant in the high-energy regime close to the initial
singularity. In this review we investigate the dynamics of the five-dimensional 
warped Randall-Sundrum  brane-worlds  and
their generalizations, with particular emphasis on whether
the currently observed high degree of homogeneity and isotropy
can be explained. In particular, we discuss the
asymptotic dynamical evolution of spatially homogeneous
brane-world cosmological models containing both a perfect fluid and a scalar field 
close to the initial singularity.
Using dynamical systems techniques
it is found that, for
models with a physically relevant equation of state, an
isotropic singularity is a past-attractor in all
orthogonal 
spatially homogeneous models (including Bianchi type IX models).
In addition, we describe
the dynamics in a
class of inhomogeneous brane-world 
models, and show that these models also
have an isotropic initial singularity. These results 
 provide support for the conjecture that 
typically the initial  cosmological singularity is isotropic in brane-world
cosmology.
Consequently we argue that, unlike the situation in general relativity,
brane-world cosmological models may offer a plausible solution to the initial
conditions problem in cosmology.

\end{abstract}

\newpage

\section{Introduction}

Cosmological models in which our Universe is a four-dimensional brane
embedded in a higher-dimensional spacetime are of current
interest. In the brane-world scenario, ordinary matter fields are
confined to the brane while the gravitational field can also
propagate in the extra dimensions (i.e., in the
`bulk')~\cite{rubakov5,rubakov4,rubakov3,rubakov2,rubakov1}.
In this paradigm it is not
necessary for the extra dimensions to be small or  
compact, which differs from the standard Kaluza-Klein approach,
and Einstein's theory of general relativity (GR) must be modified at high
energies (i.e., at early times). At low energies gravity is
also localized at the brane (even when the extra dimensions are not
small) \cite{HDDK1,HDDK2,randall1,randall2}.

The five 10-dimensional (10D) superstring
theories and the 11D supergravity theory
are believed to arise as
different limits of a single theory, known as
M-theory \cite{mtheory1,mtheory2,mtheory3}. The 11th dimension in 
M-theory is related to the string
coupling strength, and at low energies M-theory is 
approximated by 11D supergravity. In Horava-Witten 
theory~\cite{Horava}, gauge
fields of the standard model are confined to two 10-branes
located at the end points of an $S^1/Z_2$ orbifold and the 
6 extra dimensions on the
branes are compactified on a very small scale effectly resulting in a number of
5D ``moduli" fields. A 5D realization
of the Horava-Witten theory and the corresponding brane-world
cosmology is given in~\cite{low2,low3,low1},
with an extra dimension that can be large relative to the
fundamental scale, providing a basis for the 5D Randall-Sundrum (RS)
models~\cite{randall1,randall2,Giddings}.
An important consequence of extra dimensions is that the
fundamental scale is $M_{4+d}$, where $d$ is the number of
extra dimensions, rather than the
4D Planck scale $M_p \equiv  M_{4}$.  If the
extra-dimensional volume is significantly above the Planck-scale, then
the true fundamental scale $M_{4+d}$ can be much less than the
effective scale $M_p \sim 10^{19}~{\rm GeV}$ and the weakness of gravity in
4D can be understood in terms of the fact that it
leaks into the extra curved or ``warped" dimensions.

As in the
Horava-Witten theory the RS branes are $Z_2$-symmetric
and have a self-gravitating  tension, which counteracts the
influence of the negative bulk cosmological constant on the brane.
The RS brane-worlds and their generalizations provide
phenomenological models which
capture  some of the features
of M-theory, 
they provide a self-consistent and simple 5D  
realization of the
Horava-Witten supergravity solutions~\cite{Horava}, and they allow for an
investigation of the essential non-linear gravitational dynamics
in the high-energy regime close to the initial
singularity when moduli effects from the 
extra dimensions can be neglected.
RS brane-world models have in common a five dimensional
space-time (bulk), governed by the Einstein equations with a cosmological
constant, in which gravity is localized at the brane due to 
the curvature of the bulk (i.e., ``warped
compactification").  A negative bulk cosmological
constant prevents gravity from leaking into the
extra dimension at low energies. In the  
RS2 model there are two $Z_2$-symmetric
branes, which have equal
and opposite tensions. Standard model fields are confined
on the negative tension brane, while the positive-tension brane has fundamental
scale $M$ and is ``hidden".
In the  RS1
model there is only one, positive tension, brane which can be obtained
in the limit as the negative tension brane goes
to infinity.  We concentrate here mainly on 
RS1 branes
in which the negative cosmological constant is offset by
the positive brane tension; the 
RS2 brane models introduce the added problem of radion
stabilization and complications arising from
negative tension.

It has recently become important to test the astrophysical and
cosmological implications of these higher dimensional theories derived
from string theory.  In particular, can these
cosmological models explain the high degree of
currently observed homogeneity and isotropy?  
Cosmological observations indicate that we live in a Universe which is
remarkably uniform on very large scales. However, the spatial homogeneity
and
isotropy of the Universe is difficult to explain within the
standard general relativistic framework since, in the presence of matter,
the class of solutions to the Einstein equations which evolve
towards a Friedmann universe is essentially a set of measure
zero~\cite{ch73}. In the
inflationary scenario, we live in
 an isotropic region of a potentially highly
irregular universe as the result of an accelerated expansion phase in the early
universe
thereby solving many of the problems of cosmology. Thus this
scenario can successfully generate a homogeneous and
isotropic Friedmann-like universe from initial conditions which, in the
absence of inflation, would have resulted in a universe far
removed from the one we live in today. However, still only a restricted
set
of initial data will lead to smooth enough conditions for the
onset of inflation \cite{js86,rellis86}, so the
issue of homogenization and isotropization is still not satisfactorily
solved.
Indeed, the initial conditions problem (that is, to explain why  the
universe is so isotropic and spatially homogeneous from generic initial
conditions),  is perhaps one of the central problems of modern theoretical
cosmology.
These issues were recently revisited in the context of brane cosmology
\cite{Coley:2002a,Coley:2002b}.

 A geometric formulation of the class of
RS brane-world models was given in~\cite{sms2,sms1}.
The dynamical equations on the 3-brane differ from the equations in GR. 
There are 
local (quadratic) energy-momentum corrections that
are significant only at high energies and the dynamical 
equations reduce to the regular Einstein
field equations of GR for late times.  However, 
for very high energies (i.e, at early
times), these additional energy momentum correction terms will 
play a critical role in the
evolutionary dynamics of the brane-world models.  In addition to 
the matter field corrections, there
are nonlocal effects from the free
gravitational field in the bulk, transmitted via the projection 
${\cal E}_{\mu\nu}$ of the bulk Weyl
tensor, that contribute further corrections to the Einstein equations on
the brane.
A particularly useful formulation of the governing equations has been derived
\cite{Maartens1,Maartens2}  adopting a covariant approach \cite{mac73,WE}.

Recently, much effort has been devoted to
understand the 
cosmological dynamics of RS brane-world
models.
There are many generalizations of the
original RS scenario which allow for matter with
cosmological symmetry on the brane (Friedmann branes) 
\cite{BinDefLan:2000a,BinDefLan:2000b,BinDefLan:2000c}
(in this case the bulk is Schwarzschild-Anti de Sitter space-time
\cite{msm2,cline14}), and non-empty bulks have also been considered
including models with a dilatonic scalar field
in the bulk \cite{mw3,mw1,mw2}.
The Friedmann brane models have been
extensively investigated
\cite{cline15,BinDefLan:2000a,BinDefLan:2000b,cline9,cline1,cline3,BinDefLan:2000c,cline6,cline16,cline4,cline7,cline8,cline10,cline5,cline17,cline2,cline11,cline14,cline13,cline12}
and  inflationary scalar perturbations \cite{mwbh}
and bulk perturbations 
\cite{new3,new6,new8,new2,new4,new7,new1,new5} in these models have also
been
considered.
The most important feature that distinguishes brane cosmology from the
standard
scenario is the fact that at high energies the Friedmann equation is
modified
by an extra term quadratic in the energy density.
The dynamics of a brane-world universe filled with a
perfect fluid have been intensively investigated
\cite{CamposSopuerta:2001a,CamposSopuerta:2001b,Coley:2002a,Coley:2002b}.
It has been found there exist new regimes that are not inherent in the
standard
cosmology, such as stable oscillation \cite{CamposSopuerta:2001b} and the
collapse
of a flat universe \cite{Santos:2001}. Some features of brane-world
inflation have
been studied in \cite{steep1,low2,low3,low1,mss,mwbh} and the cosmological
dynamics
for scalar field models with an exponential 
potential have been described in
\cite{Dunsby:2002,naureen2,inf5}.

The asymptotic dynamical evolution of spatially homogeneous
brane-world cosmological models 
containing both a perfect fluid and a scalar field
close to the initial singularity,
where the energy density of the matter is larger than the brane
tension and  the behaviour deviates significantly from the
classical GR  case, has been studied in detail using dynamical systems techniques. 
The physical case of a
perfect fluid with $p=(\gamma-1)\rho$ satisfying $\gamma \ge
4/3$, where $\gamma = 4/3$ corresponds to radiation and $\gamma
=2$ corresponds to a massless scalar field close to the initial
singularity, is of particular interest.
We shall show that an
isotropic singularity is a past-attractor in all
orthogonal Bianchi models (including Bianchi type IX models),
and is also a local past-attractor in a
class of inhomogeneous brane-world models.
The dynamics of
these inhomogeneous brane-world models, together
with the BKL conjectures 
\cite{bkl1,bkl2,bkl3}, provide support for the conjecture that the
initial cosmological singularity is isotropic in spatially
inhomogeneous brane-world models.
It is consequently argued that, unlike the situation in GR, it is plausible that
the initial singularity is isotropic and that
brane cosmology may naturally give rise to a set of initial data
that provide the conditions for inflation to subsequently take
place.

\newpage
\section{Brane dynamics}

The field equations induced on the brane lead to new
terms
that carry bulk effects onto the brane \cite{sms2,sms1}:
\begin{equation}
G_{\mu\nu}=-\Lambda g_{\mu\nu}+\kappa^2
T_{\mu\nu}+\widetilde{\kappa}^4S_{\mu\nu} - {\cal E}_{\mu\nu}\,,
\label{2}
\end{equation}
where $\kappa^2=8\pi/M_{\rm p}^2$,
 $\widetilde{\kappa}^2=
8\pi/\widetilde{M}_{\rm p}^3$, and $\widetilde{M}_{\rm p}$ is
the fundamental 5D Planck mass, which is typically much
less than the effective Planck mass on the brane,
the bulk cosmological constant $\widetilde{\Lambda}$ is negative
and is the only 5D source of the gravitational field,
and
\begin{equation}
\lambda=6{\kappa^2\over\widetilde\kappa^4} \,, ~~ \Lambda =
{\textstyle{1\over2}}\widetilde\kappa^2\left(\widetilde{\Lambda}+
{\textstyle{1\over6}}\widetilde\kappa^2\lambda^2\right)\,.
\label{3}
\end{equation}
We assume that $\widetilde{\Lambda}$ is chosen so that
$\Lambda=0$.
The bulk corrections to the Einstein equations on the brane are of
two forms: first, the matter fields contribute local quadratic
energy-momentum corrections via the tensor $S_{\mu\nu}$, and
second, there are nonlocal effects from the free gravitational
field in the bulk, transmitted via the projection ${\cal
E}_{\mu\nu}$ of the bulk Weyl tensor. The matter corrections are
given by
\begin{equation}
S_{\mu\nu}={\textstyle{1\over12}}T_\alpha{}^\alpha T_{\mu\nu}
-{\textstyle{1\over4}}T_{\mu\alpha}T^\alpha{}_\nu+
{\textstyle{1\over24}}g_{\mu\nu} \left[3 T_{\alpha\beta}
T^{\alpha\beta}-\left(T_\alpha{}^\alpha\right)^2 \right]\, ,
\label{3'}
\end{equation}
which reduces to
\[
S_{\mu\nu}={\textstyle{1\over12}}\rho^2
u_\mu u_\nu
+{\textstyle{1\over12}}\rho\left(\rho+2 p\right)h_{\mu\nu}\,,
\]
for a perfect fluid or minimally-coupled scalar field (or combination
thereof)
and $u^\mu$ is the 4-velocity comoving with matter 
and $h_{\mu\nu} \equiv g_{\mu\nu}+u_\mu u_\nu$.
The quadratic energy-momentum
corrections to standard GR will be significant for
$\widetilde{\kappa}^4\rho^2 \gtrsim \kappa^2\rho$ (i.e., in the
high-energy regime).

Due to its symmetry properties,
the projection of the bulk Weyl tensor can be irreducibly decomposed with
respect to
$u^\mu$ \cite{Maartens1,Maartens2} :
\[
{\cal E}_{\mu\nu}={-6\over\kappa^2\lambda}\left[{\cal
U}\left(u_\mu u_\nu+{\textstyle {1\over3}} h_{\mu\nu}\right)+{\cal
P}_{\mu\nu}+{\cal Q}_{\mu}u_{\nu}+{\cal Q}_{\nu}u_{\mu}\right]\,,
\]
where 
\[
{\cal U}=-{\textstyle{1\over6}}\kappa^2 \lambda\, {\cal
E}_{\mu\nu}u^\mu u^\nu
\]
is an effective nonlocal energy density on the brane (which need
not be positive), arising from the free gravitational field in the
bulk. There is an effective nonlocal anisotropic stress
\[
{\cal P}_{\mu\nu}=-{\textstyle{1\over6}}\kappa^2 \lambda\left[
h_\mu{}^\alpha h_\nu{}^\beta-{\textstyle{1\over3}}h^{\alpha\beta}
h_{\mu\nu}\right] {\cal E}_{\alpha\beta}
\]
on the brane, which carries 
gravitational wave effects of the free gravitational field in the
bulk. The effective nonlocal energy flux on the brane is given by
\[
{\cal Q}_\mu ={\textstyle{1\over6}}\kappa^2 \lambda\,
h_\mu{}^\alpha {\cal E}_{\alpha\beta}u^\beta\, .
\]

The local and nonlocal bulk modifications may be consolidated into
an effective total energy-momentum tensor:
\begin{equation}
G_{\mu\nu}=-\Lambda g_{\mu\nu}+\kappa^2 T^{\rm tot}_{\mu\nu}\,,
\label{6'}
\end{equation}
where
\begin{equation}
T^{\rm tot}_{\mu\nu}= T_{\mu\nu}+{6\over \lambda}S_{\mu\nu}-
{1\over\kappa^2}{\cal E}_{\mu\nu}\,.
\end{equation}
The effective total energy density, pressure, anisotropic stress
and energy flux are~\cite{Maartens1,Maartens2} (for a comoving fluid --
the tilting case is treated later):
\begin{eqnarray}
\rho^{\rm tot} &=& \rho\left(1+{\rho\over2\lambda}\right)+{6 {\cal
U}\over\kappa^4\lambda}\,, \label{a}\\ p^{\rm tot} &=& p+
{\rho\over2\lambda}(\rho+2p) +{2{\cal U}\over\kappa^4\lambda}\,,
\label{b}\\ \pi^{\rm tot}_{\mu\nu} &=&{6\over
\kappa^4\lambda}{\cal P}_{\mu\nu}\,, \label{c}\\ q^{\rm tot}_\mu
&=& {6\over \kappa^4\lambda}{\cal Q}_\mu \,.\label{d}
\end{eqnarray}

For an empty bulk,
the brane energy-momentum tensor separately satisfies the
conservation equations, $\nabla^\nu T_{\mu\nu}=0 $. The Bianchi
identities on the brane then imply that the projected Weyl tensor obeys
the constraint
\begin{equation}
\nabla^\mu{\cal E}_{\mu\nu}={6\kappa^2\over\lambda}\nabla^\mu
S_{\mu\nu}\,. \label{5}
\end{equation}
Consequently nonlocal bulk effects can be sourced by local bulk
effects.
The brane energy-momentum tensor and the consolidated
effective energy-momentum tensor are {\em both} conserved
separately. These conservation equations, as well as the brane
field equations and Bianchi identities, are given in covariant
form in~\cite{Maartens1,Maartens2}.

\subsection{Projected Weyl Tensor}

In general, the 4
independent conservation equations determine 4 of the 9 independent
components
of ${\cal E}_{\mu\nu}$ on the brane. There is no
evolution equation for the  nonlocal anisotropic stress ${\cal
P}_{\mu\nu}$. Thus, in general, the projection of the
5D field equations onto the brane does not lead to a
closed system, since
there are bulk degrees of freedom whose impact on the brane cannot
be predicted by brane observers. These degrees of freedom could
arise from propagating gravity waves in the bulk which are governed by
off-brane
5D bulk dynamical equations.

If the nonlocal anisotropic stress contribution from the bulk
field vanishes, i.e., if
${\cal P}_{\mu\nu}=0$,
then the evolution of ${\cal E}_{\mu\nu}$ is fully determined.
A special case of this arises when
the brane is Robertson-Walker (RW)
and in some special cases in which
the source is  perfect fluid
matter and branes
with isotropic 3-Ricci curvature
(such as, for example,
Bianchi~I branes).
Although these special cases can give consistent closure on
the brane, there is no guarantee that the brane is embeddable in a
regular bulk, unlike the case of a Friedmann brane whose
symmetries imply (together with $Z_2$ matching) that the bulk is
Schwarzschild-AdS$_5$~\cite{msm2,cline14}.

The nonlocal anisotropic stress terms enter into crucial dynamical
equations,
such as the Raychaudhuri equation and the shear propagation
equation, and can lead to important changes from the
GR case.
The correction terms must be consistently derived from the
higher-dimensional equations.
In the analysis we shall primarily assume that the effective nonlocal
anisotropic stress
is zero (in the fluid comoving frame).
This is supported by a
dynamical analysis \cite{CHL} and the fact
that since ${\cal P}_{\mu\nu}$
corresponds to gravitational waves in higher-dimensions it is expected
that the dynamics
will not be affected significantly at early times close to the singularity  
\cite{lmw1,lmw2}.
This is the only assumption we shall make, and it is expected that
inclusion
of  ${\cal P}_{\mu\nu}$ will {\em not} affect the
qualitative dynamical features of the models close to the initial
singularity.
Indeed, recent
analysis provides further support for  ${\cal P}_{\mu\nu}=0$
\cite{ColHer1,ColHer2}.
We expect that ${\cal P}_{\mu\nu} \sim  {\cal U}  g_{\mu\nu}$ on
dimensional
grounds, and so for a Friedmann brane close to the initial singularity
we expect that ${\cal P}_{\mu\nu} \sim a^2$ ${\cal U}C_{\mu\nu}$ (where
$C_{\mu\nu}$
is slowly varying), which is consistent with the linear (gravitational)
perturbation
analysis (in a pure AdS bulk background) \cite{new1}. Hence, ${\cal
P}_{\mu\nu}$ is
dynamically negligible  close to the initial singularity.
In addition, from an analysis of the evolution 
equation for ${\cal Q}_{\mu}$  it can be shown that a small ${\cal Q}_{\mu}$ does
not affect 
the dynamical evolution of ${\cal U}$ close to
the 
initial singularity 
\cite{Coley:2002a,gm,Maartens1,Maartens2}.
Let us discuss this in more detail.

\newpage

\subsubsection{Type N spacetimes}

In brane-world cosmology gravitational waves can 
propagate in the higher dimensions (i.e., in the `bulk').
In some appropriate regimes the
bulk gravitational waves may be approximated by plane waves
\cite{ColHer1}.
For example, there might occur thermal radiation of bulk gravitons
\cite{Lang}.
In particular, at sufficiently
high energies particle interactions
can produce 5D gravitons which are emitted into the bulk. Conversely,
in models with a bulk black hole, there may be gravitational waves
hitting the brane. At sufficiently large distances from the black hole
these gravitational waves may be approximated as of type N
\cite{ColHer1}. Alternatively, 
if the brane has low energy initially, 
energy can be transferred onto the brane by bulk particles 
such as gravitions; an equilibrium is expected to set in once the 
brane energy density reaches  a limiting value.  
We can study 5D gravitational waves that are
algebra\-ically special and of type N.

Let us assume that the 5D bulk
is algebraically special and of type N.  This puts a constraint on the 
5D Weyl tensor which
makes it possible to deduce the form of the non-local stresses from a 
brane point of view.
For a 5D type N spacetime there exists a frame $\ell_a$, $\tilde{\ell}_a$, 
$m^i_a$, $i=1,2,3$  
such that \cite{classification2,classification1}:
\begin{equation}
C_{abcd}=4C_{1i1j}\ell_{\{a}m^i_b\ell_cm^j_{d\}}.
\end{equation}
The ``electric part'' of the Weyl tensor, ${\cal E}_{ab}=C_{acbd}n^cn^d$, 
where $n^c$ is the normal vector on the brane, can for the type N bulk be
written as
\begin{equation}
{\cal E}_{ab}=C_{1i1j}\left[\ell_a(m_c^in^c)-m^i_a(\ell_cn^c)\right]
\left[\ell_b(m_c^jn^c)-m^j_b(\ell_cn^c)\right], 
\end{equation}
where ${\cal E}_{ab}n^b={\cal E}^a_{~a}=0$. 
Note  that for a type N bulk we also have  
${\cal E}_{ab}\ell^b=0$,
which can be rewritten using the projection operator on the brane,
as
${\cal E}_{ab}\hat{\ell}^b=0, \quad 
\hat{\ell}^b\equiv g^b_{~c}\ell^c$.
Hence, the vector $\hat{\ell}_b$ is the projection of the null vector
$\ell_b$ 
onto the brane, and
\begin{equation}
\hat{\ell}^b\hat{\ell}_b=-(\ell^an_a)^2. 
\end{equation}

In the case that 
$\hat{\ell}^a$ is time-like $(\ell^an_a\neq 0)$ 
we can set $u^{\mu}$ parallel to 
$\hat{\ell}^{\mu}$. In this frame, the requirement 
$\hat{\ell}^{\mu}{\cal E}_{\mu\nu}=0$ 
implies that we have ${\cal U}=0={\cal Q}_{\mu}$, and hence we can write 
${\cal E}_{\mu\nu}=-(\tilde{\kappa}/\kappa)^4{\cal P}_{\mu\nu}$.
This case was discussed in  
\cite{ColHer1}.
More importantly, in the case $\ell^an_a=0$,
in which $\hat{\ell}_{\mu}$ is a null vector, 
${\cal E}_{\mu\nu}$ 
can be written
\begin{equation}
{\cal E}_{\mu\nu}=-\left(\frac{\tilde{\kappa}}{\kappa}
\right)^4\epsilon\hat{\ell}_{\mu}\hat{\ell}_{\nu}.
\end{equation}
Hence, this case is formally equivalent to the energy-momentum tensor of a 
\emph{null} fluid or an extreme 
tilted perfect fluid. Using a covariant decomposition of ${\cal
E}_{\mu\nu}$, 
the non-local energy terms are given by:
\begin{equation}
{\cal U}=\epsilon(\hat{\ell}_{\nu}u^{\nu})^2, \quad
{\cal Q}_{\mu}=\epsilon(\hat{\ell}_{\nu}u^{\nu})\hat{\ell}_{\mu}, \quad
{\cal P}_{\mu\nu}=\epsilon\hat{\ell}_{\langle\mu}\hat{\ell}_{\nu\rangle}.
\end{equation}
The equations on the brane now close and the dynamical behaviour can be
analysed
\cite{ColHer2}. Note that
\begin{equation}
{\cal U}{\cal P}_{\mu\nu}={\cal Q}_{\mu}{\cal Q}_{\nu} -
\frac{1}{3}{g}_{\mu\nu}
{\cal Q}_{\lambda}{\cal Q}^{\lambda},
\end{equation}
so that in this case ${\cal E}_{\mu\nu}$ is determined completely by
${\cal U}$ and ${\cal Q}_{\mu}$.



We can investigate the effect this type N bulk may have on the
cosmological 
evolution of the brane. 
If we assume that we are in the regime of 
an isotropic past singularity and the cosmological evolution 
is dominated by an isotropic perfect 
fluid with equation of state $p=(\gamma-1)\rho$,
the equations for ${\cal U}$ and ${\cal Q}_{\mu}$ are
\begin{equation}
\dot{{\cal U}}+\frac 43\Theta{\cal U}=0, \quad \quad 
\dot{{\cal Q}}_{\mu}+\frac 43\Theta{\cal Q}_{\mu}=0
\end{equation}
(see also \cite{Lang}).
For the isotropic singularity, the expansion factor is given by 
$\Theta=1/(\gamma t)$. Defining the expansion-normalised non-local
density, 
$U\equiv{\cal U}/\Theta^2$ and non-local energy flux, 
$Q_{\mu}\equiv{\cal Q}_{\mu}/\Theta^2$, we obtain
\begin{equation}
\dot{U} = \frac{2}{3\gamma t}(3\gamma-2)U, \quad \quad
\dot{Q}_{\mu} = \frac{2}{3\gamma t}(3\gamma-2)Q_{\mu}.
\end{equation} 
Hence, the isotropic singularity 
is stable to the past with regards to these stresses if $\gamma>2/3$. 
Similarly, we have a Friedmann universe to the future with 
$\Theta=2/(\gamma t)$, 
and thus 
\begin{equation}
\dot{U} = \frac{2}{3\gamma t}(3\gamma-4)U, \quad \quad
\dot{Q}_{\mu} = \frac{2}{3\gamma t}(3\gamma-4)Q_{\mu}.
\end{equation} 
This implies that if the isotropic fluid on the brane is less stiff 
than radiation at late times 
($\gamma<4/3$), then the Friedmann universe is stable to the future with respect
to the non-local 
stresses.

These qualitative results are further supported by a more detailed
 dynamical systems analysis  \cite{COLEY2,WE}  of the asymptotic
behaviour of two tilting $\gamma$-law fluids in a class of Bianchi type
VI$_0$ models 
\cite{ColHer2}.
In particular, the physically relevant case of interest here, namely that
the second fluid is a null
fluid or fluid with extreme tilt, was investigated. All 
of the equilibrium points were found and their stability
determined, so that the local attractors were established. 
The effect a type N bulk may
have on the cosmological evolution of the brane
was then investigated. 
It was found that if $\gamma > 2/3$,
then the null fluid is not dynamically important at early times;
that is, the effects of the projected Weyl tensor will not affect
the dynamical behaviour close to the initial singularity. This
supports the result that an isotropic singularity is a local
stable past attractor.
Also, if $\gamma
< 4/3$, the extreme tilting fluid is dynamically negligible to
the future.
Therefore, the null brane fluid is not dynamically important
asymptotically at early times for all values of $\gamma$
of physical import, supporting the qualitative analysis
above \cite{ColHer2}.
In particular, this implies that the effects of ${\mathcal
E}_{\mu\nu}$ are not dynamically important (at least in this class
of models) in the asymptotic regime close to the singularity.

\subsubsection{Bulk spacetimes}

Most work on
anisotropic brane-world universes use the {\it effective}
4D Einstein equations on the brane. Since the 
brane equations are not closed, 
various additional conditions on ${\cal P}_{\mu \nu}$ have been
proposed in a rather ad hoc way. There has been some work on
determining the  effects of a non-zero ${\cal P}_{\mu\nu}$ on the
behaviour of brane-world cosmological 
models~\cite{Ruth,Barrow:2002b,mss,Savchenko:2002,Toporensky:2001}.
For example, in~\cite{frolov} a bulk metric with a Kasner brane
was presented. Since the Kasner metric is a solution of
the 4D Einstein vacuum equations, the bulk metric is a
simple warped extension; for the choice of bulk metric made 
in~\cite{frolov}
it was shown that the brane can only be anisotropic if it contains
a constant tension with $\rho = P$! Other examples include the
Schwarzschild black string solution~\cite{chr,CHSS}, the spatially
homogeneous 
Einstein
space with a black 
hole~\cite{Cadeau:2000t}  and the brane 
wave spacetimes~\cite{ColHer1}, but 
all of these result in Bianchi models
with an anisotropic fluid source.

In order to have a fully consistent picture, we cannot avoid
specifying  the bulk geometry; i.e., we need to construct explicit
anisotropic brane cosmologies by solving the full 5D
vacuum Einstein equations with negative cosmological constant.
This has non trivial implications for the brane itself because, in
brane-world models, the Israel junction conditions (relating
the extrinsic curvature to the matter on the brane) must be
satisfied.  For instance, if the brane contains perfect
fluid matter, then the junction conditions will impose
constraints on the components of the extrinsic curvature. 

In~\cite{CMMS}
a moving brane admitting a spatially homogeneous but anisotropic Bianchi I
3D slicing in a static anisotropic bulk
was considered.
Having
solved the bulk Einstein equations, it was shown that the
junction conditions induce anisotropic stresses in the matter on
the brane (i.e., the brane cannot contain a perfect fluid).
Indeed, from the bulk Einstein equations and the junction
conditions, it follows that this anisotropic stress can only
vanish if the bulk is isotropic and hence the brane is isotropic.
It was also noted that the
anisotropic stresses obtained on the brane in \cite{CMMS} are 
fixed by the  bulk metric.  
Some explicit Bianchi~I
brane-world cosmological
models using the freedom still available in embedding the 3-brane 
were constructed.
In one example the cosmological models do not isotropize as
the initial singularity is approached, but in this example $w
\rightarrow -1$ as $t \rightarrow 0$!  
(In other examples,
the models do isotropize in
the past, but the matter content never behaves as a
perfect fluid.)

This work was generalized in  \cite{FLSZ} to a class of non-static  bulks.
The bulk was assumed to be empty but endowed with
a negative cosmological constant
 and explicit
analytic bulk solutions
 with anisotropic 3D
spatial slices were found. The $Z_2$-symmetric
branes were embedded in the anisotropic spacetimes and the constraints
on the brane energy-momentum tensor due to the 5D
anisotropic geometry were discussed.  
The question of whether it is
possible to find a bulk geometry and a brane 
with perfect fluid matter consistent with the Israel junction conditions
was studied. 
Einstein's equations were integrated explicitly in the cases in which the
shear does not depend on the extra-dimension and in which the
metric is separable.
  Remarkably, in the separable bulk solution
the analysis implies that the brane {\it can support perfect fluid
type matter} for certain parameter values.
In a particular example, by solving for the cosmological
evolution on the brane it was found that as $\tau \rightarrow
-\infty$, $\rho$ converges towards the RS tension
and the effective
pressure is {\it negative} (but converges towards zero more rapidly than
the effective energy). 
Therefore, unlike \cite{CMMS}, it was found \cite{FLSZ}
that it is possible to find a perfect fluid anisotropic brane in some 
such bulk geometries, but again there is no flexibility in
the choice of the perfect fluid in contrast with the isotropic
case (i.e., 
only a very particular type of perfect fluid is compatible with
the given bulk geometry).

Therefore, it does seem difficult to
construct examples of anisotropic branes with perfect fluid matter.
However, these solutions \cite{CMMS,FLSZ} 
are very special examples and say nothing about
the general case. 
An exact solution in the bulk gives rise to a precise 
relationship between the density and
the pressure on the brane. But exact closed form solutions are few 
sparse; indeed, even in 4D GR there are
only a few very special exact Bianchi solutions 
(models are usually only specified up to a set of ODE),  so
this approach will not shed light on the general case. Moreover, 
the isotropic singularity conjecture is not contradicted  
by these results.
First, it is not necessary
for ${\cal P}_{\mu\nu}$ to be zero but rather that it be dynamically
negligible (i.e., some appropriate normalized quantity is negligible) as
the singularity is approached. Second, the values for the
parameter $\gamma$ is in the wrong range
in the examples of \cite{CMMS} and \cite{FLSZ}
(the isotropic singularity theorems only apply for $\gamma \ge 4/3$).

In order to illustrate this point further, let us consider the
following Bianchi I metric:
\begin{eqnarray}
ds^2 & = & -t^{2(\ol{\gamma}-1)} [1 + \alpha t^{\ol{\gamma}-2}]
^{-\frac{2(\ol{\gamma}-1)}{(\ol{\gamma}-2)}} \quad dt^2 
+ t^{4/3} [1+\alpha t^{\ol{\gamma}-2}]
^{-\frac{2(1+2 \cos \psi)}{3(\ol{\gamma}-2)}} \quad
dx^2  \nonumber \\ 
&& + t^{4/3} [1+\alpha t^{\ol{\gamma}-2}]
^{-\frac{2(1- \cos \psi - \sqrt{3} \sin \psi)}
{3(\ol{\gamma}-2)}} \quad
dy^2  + t^{4/3} [1+\alpha t^{\ol{\gamma}-2}]
^{-\frac{2(1- \cos \psi + \sqrt{3} \sin \psi)}
{3(\ol{\gamma}-2)}} \quad
dz^2 \label{themetric}
\end{eqnarray}
where $\alpha$ and $\psi$ are constants and $\ol{\gamma} >2$.
For $\alpha = 0$, the metric is the Bin\'etruy, Deffayet
and Langlois or BRW (see later) solution 
with $a(T) = T^{\frac{1}{3\ol{\gamma}}}$
\cite{BinDefLan:2000a,BinDefLan:2000b,BinDefLan:2000c}
(where $T \equiv t^{\frac{1}{\ol{\gamma}}}$).  We note that 
for the metric (\ref{themetric}) 
$G^{}_\mu \,\! ^\nu=$ diag $(-A, B, B, B)$, where 
$B = (\ol{\gamma} -1)A$ and 
$$ A \equiv A_0 t^{-2\ol{\gamma}}[1+
\alpha t ^{\ol{\gamma} -2}]^{\frac{\ol{\gamma}}{\ol{\gamma}-2}}.  $$
We shall show that a (anisotropic) Bianchi I  
perfect fluid RS brane can be embedded in a 5D Einstein
space using the Campbell-Magaard theorem.  Following 
Dahlia and Romero \cite{DaRo} we consider
the 5D metric in Gaussian normal form 
(with 5th coordinate $\ell$).  If we wish to embed a RS
brane into a $Z_2$ symmetric bulk (with cosmologial
constant $\Lambda$)
we need to consider the equations (on the brane 
$\ell = \ell_0$, a constant).
\begin{equation}
T_\mu \,\!^\nu _{\,\!;\nu} = \dot{\rho} + (\rho+p) \Theta =0, \label{A1}   
\end{equation}

\begin{equation}
\frac{\kappa^2}{4} (T_{\mu \nu} T^{\mu \nu} - \frac{1}{3} T^2)
 =  \frac{\kappa^2}{6} [\rho^2 + 3p\rho + \rho \lambda - 
3 \lambda p - 2 \lambda^2] = 2 \Lambda + \tilde{R}, \label{A2} 
\end{equation}
where $T_{\mu\nu}$ is the energy-momentum tensor for a perfect
fluid and $\tilde{R}$ is the scalar curvature of the 
hypersurface $\ell = \ell_0$.  If a solution of Eqs. 
(21) and (22) can be found, then an 
analytic solution of the 5D bulk equations 
$G_{\mu \nu}= -\Lambda g_{\mu \nu}$ (satisfying the Eqs. 
(\ref{A1}) and (\ref{A2}) on the brane) is guaranteed by the 
Cauchy-Kowalewski theorem \cite{DaRo}.

By a direct calculation, for metric (\ref{themetric}) we obtain
\begin{eqnarray}
\tilde{R} &=& R_0 t^{- 2 \ol{\gamma}} 
[1+ \alpha
t^{\ol{\gamma}-2}]^{\frac{\ol{\gamma}}{\ol{\gamma}-2}}\label{A3} \\
\Theta & = & \theta_0 t^{-  \ol{\gamma}} 
[1+ \alpha t^{\ol{\gamma}-2}]^{\frac{1}{\ol{\gamma}-2}} 
[1+\frac{1}{2} \alpha t^{\ol{\gamma}-2}]. \label{A4} 
\end{eqnarray}
With these expressions, Eqs. (\ref{A1}) and (\ref{A2}) 
can always be solved to find $\rho=\rho(t), p=p(t)$.
To lowest order in $t$, we have that 
$\tilde{R}_0 = R_0t^{-2\ol{\gamma}} (1+ {\cal O}(t))$,
$\Theta = \theta_0 t^{-\ol{\gamma}}(1+ {\cal O}(t))$,
and we can solve Eqs. (21) and (22) to obtain 
$\mu = \rho = \rho_0 t{-\ol{\gamma}}(1+{\cal O}(t))$ and 
$p = (\ol{\gamma}-1) \rho(1+{\cal O}(t))$.  That is,
to lowest order we obtain the BRW solution.

Therefore, using 
extended Campbell-Magaard theory \cite{DaRo}, we have demonstrated
that there are exact anisotropic (of Bianchi type I)
perfect fluid brane cosmological models that can be 
embedded into a 5D bulk Einstein space (see also \cite{Katz}). This construction does not,
of course, gaurentee that the bulk is regular. Note that for the
example (\ref{themetric}) constructed, as $t \to 0$ 
the model
isotropizes (towards the BRW solution) and the singularity 
is isotropic.

\subsubsection{Dynamical systems approach}

In order to illustrate the possible effects of ${\cal P}_{\mu\nu}$
on the dynamics of a brane-world cosmological model, an alternative
approach was taken in \cite{HH}. Again it was assumed that the effective
cosmological constant on a Bianchi I brane is zero, 
and the matter on the brane is of the form of a
non-tilted perfect fluid. 
Utilizing new dependent dynamical variables, 
the effective Einstein field equations
yield a dynamical system describing the evolution of the Bianchi I
brane-world.  In this analysis there is
no evolution equation for the quantity  
\begin{equation}
{\cal P}\equiv \frac{\sqrt{3}}{\kappa^2\lambda}
\frac{\sigma_{\mu\nu}{{\cal P}}^{\mu\nu}}{\sigma H^2}, \label{modelling}
\end{equation}
modelling the
non-local bulk corrections. 
In order to obtain a closed system of equations, it was
 then assumed that ${\cal P}$ is a differentiable function
of the remaining dynamical variables; that is, ${\cal P} = 
{\cal P}(\Sigma,\Omega,\Omega_{\lambda},\Omega_{{\cal U}})$. With this
assumption, the resulting brane-world spacetimes represented by
the equilibrium points of the system are geometrically self-similar
\cite{CarrColey:1999}.

A number of models were investigated.
In general, the
dynamics in the case  ${\cal P}={{\cal P}}_0$ (a constant) is
quite different from that found currently in the literature.
For ${\cal P} \not=1$ (${\cal P}_0\not = 0$), 
none of the equilibrium points are isotropic, and all have some
form of dark matter with ${\cal U}
\not = 0$.  It appears that the non-local bulk corrections have
a significant impact on the dynamical behaviour
and  the initial singularity is
not isotropic.
In the case 
${\cal P} = {\cal P}_{\Sigma}\Sigma^\alpha$  ($\alpha>0$),
the BRW brane-world solution ${\cal F}_b$ may
not represent the past asymptotic state 
(there are
parameter values of $P_\Sigma$ such that the initial singularity
is represented by a Kasner-like
solution).  In
the case
${\cal P} = {\cal P}_{\Omega_{{\cal U}}}\Omega_{{\cal U}}^\beta$
($\beta>0$)
\cite{Barrow:2002b,Savchenko:2002}, 
${\cal F}_b$ is always a source, although there
will exist cosmological models that will asymptote to the past
towards anisotropic bulk-dominated Kasner-like models (i.e.,
 the initial singularity is not necessarily isotropic).
Thus in each of these three cases, the effects of the
non-local ${\cal P}_{\mu\nu}$ could be significant. Unlike
the standard cosmological brane-world situation in which ${\cal F}_b$
 represents the 'generic' initial behaviour
in brane-world cosmologies containing ordinary matter, it was found
 that Kasner-like bulk-dominated models can also act as
sources and that ${{\cal F}}_b$  does not even
exist as an equilibrium state in one of the cases studied.
The
future behaviour of the models is determined primarily by the parameter
$\gamma$, and was studied in \cite{HH}.

We note that, from (\ref{modelling}), ${\cal P}$ is not defined in
the (isotropic limit) as $\sigma \rightarrow 0$. This is
problematic when attempting to analyse isotropic equilibrium 
points; indeed,  the isotropic limit cannot be studied and the analysis 
in \cite{HH} is
essentially
concerned with the local stability of non-isotropic equilibrium points.
However, the DEs may be well-defined and have an isotropic limit
depending on  the chosen 
ansatz for ${\cal P}$. In the second and third cases, 
 a well-defined isotropic limit exists and in these cases
${\cal F}_b$ is a local attractor.  In the first case
(${\cal P}$ constant), there is no
isotropic limit possible. Indeed, the models do not even allow
for an isotropic solution at any time (the BRW models do not
exist), and any
 conclusions concerning an isotopic singularity are
inappropriate in this case.
Consequently the analysis in \cite{HH} supports the existence of a local
isotropic singularity.


\newpage
\section{Dynamical Equations}

The general form of the brane energy-momentum tensor for any
matter fields of perfect fluid form (including scalar fields) can
be covariantly decomposed. We adopt the notation of
\cite{Maartens1,Maartens2}.
Angled brackets
denote the projected, symmetric and tracefree part:
 \be
V_{\langle\mu\rangle}=h_\mu{}^\nu V_\nu\,,~~
W_{\langle\mu\nu\rangle}=\left[h_{(\mu}{}^\alpha h_{\nu)}{}^\beta-
{\textstyle{1\over3}}h^{\alpha\beta}h_{\mu\nu}\right]W_{\alpha\beta}\,,
 \ee
with round brackets denoting symmetrization.
An overdot denotes $u^\nu\nabla_\nu$
(i.e., timelike directional derivative),
$\Theta=\nabla^\mu u_\mu$ is the volume expansion rate of the
$u^\mu$ congruence, $A_\mu=\dot{u}_\mu =A_{\langle\mu\rangle}$ is
its 4-acceleration,
$\sigma_{\mu\nu}=\D_{\langle\mu}u_{\nu\rangle}$ is its shear rate,
and $\omega_\mu=-{1\over2}\curl u_\mu=\omega_{\langle\mu\rangle}$
is its vorticity rate.
The covariant spatial curl is given by \cite{Maartens1,Maartens2} 
 \be
\curl V_\mu=\ep_{\mu\alpha\beta}\D^\alpha V^\beta\,,~~ \curl W
_{\mu\nu}=\ep_{\alpha\beta(\mu}\D^\alpha W^\beta{}_{\nu)}\,,
 \ee
where $\ep_{\mu\nu\sigma}$ is the projection orthogonal to $u^\mu$
of the brane alternating tensor, and $\D_\mu$ is the projected
part of the brane covariant derivative (i.e., spatial derivative), defined
by
 \be
\D_\mu F^{\alpha\cdots}{}{}_{\cdots\beta}=\left(\nabla_\mu
F^{\alpha\cdots}{}{}_{\cdots\beta}\right)_\perp= h_\mu{}^\nu
h^\alpha{}_\gamma \cdots h_\beta{}^\delta \nabla_\nu
F^{\gamma\cdots}{}{}_{\cdots\delta}\,.
 \ee
The remaining
covariant equations on the brane are the propagation and
constraint equations for the kinematic quantities $\Theta$,
$A_\mu$, $\omega_\mu$, $\sigma_{\mu\nu}$, and for the nonlocal
gravitational field on the brane. The
nonlocal gravitational field on the brane is given by the 
brane Weyl tensor $C_{\mu\nu\alpha\beta}$. This splits into the
gravito-electric and gravito-magnetic fields on the brane:
 \be
E_{\mu\nu}=C_{\mu\alpha\nu\beta}u^\alpha u^\beta
=E_{\langle\mu\nu\rangle }\,,~~
H_{\mu\nu}={\textstyle{1\over2}}\ep_{\mu\alpha \beta}
C^{\alpha\beta}{}{}_{\nu\gamma}u^\gamma=H_{\langle\mu\nu\rangle}
\,.
 \ee
The Ricci identity for $u^\mu$ and the Bianchi identities
$\nabla^\beta C_{\mu\nu\alpha\beta} =
\nabla_{[\mu}(-R_{\nu]\alpha} + {1\over6}Rg_{\nu]\alpha})$ 
give rise to the evolution and constraint equations governing the
above covariant quantities 
once the Ricci tensor
$R_{\mu\nu}$ is replaced  
by the effective total energy-momentum tensor
from the field equations \cite{ee3,cov2,cov1}. 
In the
following
$R^\perp_{\mu\nu}$ is the Ricci tensor for 3-surfaces
orthogonal to $u^\mu$ on the brane and
$R^\perp=h^{\mu\nu}R^\perp_{\mu\nu}$. 
Note that in the 
GR limit, $\lambda^{-1}\to0$, we have 
that ${\cal E}_{\mu\nu}\to0$.

\subsubsection{Vorticity-free models with a linear barotropic equation of
state}

In the case of a perfect fluid with an equation of state
$p = (\gamma-1)\rho$, we obtain the following equations
when the vorticity is zero 
($\Lambda \ne 0$) \cite{Maartens1,Maartens2}:

\begin{eqnarray}
&&A_\mu = -{(\gamma-1)\over\gamma} {1\over\rho} \D_\mu\rho
\,,\label{BB}\\
&&\dot{\rho}=-\gamma\Theta\rho\,,\label{B1}\\
&& \dot{{\cal U}}+{\textstyle{4\over3}}\Theta{{\cal U}}+
\D^\mu{{\cal Q}}_\mu+2A^\mu{{\cal Q}}_\mu+\sigma^{\mu\nu}{{\cal
P}}_{\mu\nu}=0\,,
\label{B2}
\\&& \dot{{\cal Q}}_{\langle\mu\rangle}+{\textstyle{4\over3}}\Theta{{\cal
Q}}_\mu
+{\textstyle{1\over3}}\D_\mu{{\cal U}}+{\textstyle{4\over3}}{\cal
U}A_\mu +\D^\nu{{\cal P}}_{\mu\nu}+A^\nu
{{\cal P}}_{\mu\nu}+\sigma_{\mu\nu}{{\cal Q}}^\nu
=-{\textstyle{1\over6}} \kappa^4 \gamma \rho\D_\mu
\rho\,.\label{B3}
\end{eqnarray}

 \begin{eqnarray}
 &&\dot{\Theta}+{\textstyle{1\over3}}\Theta^2+\sigma_{\mu\nu}
 \sigma^{\mu\nu}-{\rm D}^\mu A_\mu+A_\mu
 A^\mu+{\textstyle{1\over2}}\kappa^2(3\gamma-2)\rho -\Lambda=
 -{\textstyle{\kappa^2\over2\lambda}}(3\gamma-1)\rho^2-
 {6\over\kappa^2\lambda}{{\cal U}}\,. \label{B4}
 \end{eqnarray}

 \begin{equation}
 \dot{\sigma}_{\langle \mu\nu \rangle }
 +{\textstyle{2\over3}}\Theta\sigma_{\mu\nu}
 +E_{\mu\nu}-\D_{\langle \mu}A_{\nu\rangle } +\sigma_{\alpha\langle
 \mu}\sigma_{\nu\rangle }{}^\alpha- A_{\langle \mu}A_{\nu\rangle}
 ={3\over\kappa^2\lambda}{{\cal P}}_{\mu\nu}\,. \label{B5}
 \end{equation}

 \begin{eqnarray}
  && \dot{E}_{\langle \mu\nu \rangle }
  +\Theta E_{\mu\nu} -\curl H_{\mu\nu}
  +{\textstyle{\gamma\kappa^2\over2}}\rho\sigma_{\mu\nu}
  -2A^\alpha\ep_{\alpha\beta(\mu}H_{\nu)}{}^\beta
  -3\sigma_{\alpha\langle \mu}E_{\nu\rangle }{}^\alpha
  \nonumber\\&&~~{}= -{\textstyle{\gamma\kappa^2\over2\lambda}}
  \rho^2\sigma_{\mu\nu}
  -{1\over\kappa^2\lambda}\left[4{{\cal U}}\sigma_{\mu\nu}+3\dot
  {{\cal P}}_{\langle \mu\nu \rangle} +\Theta{{\cal P}}_{\mu\nu}
  +3\D_{\langle\mu}{{\cal Q}}_{\nu\rangle}
  +6A_{\langle\mu}{{\cal Q}}_{\nu\rangle}+ 3\sigma^\alpha{}_{\langle\mu} 
  {{\cal P}}_{\nu\rangle\alpha}\right] \,. \label{B6}
  \end{eqnarray}

  \begin{eqnarray}
   &&\dot{H}_{\langle
   \mu\nu \rangle } +\Theta H_{\mu\nu} +\curl E_{\mu\nu}-
   3\sigma_{\alpha\langle \mu}H_{\nu\rangle }{}^\alpha
   +2A^\alpha\ep_{\alpha\beta(\mu}E_{\nu)}{}^\beta =
   {3\over\kappa^2\lambda}\left[ \curl{{\cal P}}_{\mu\nu}
   +\sigma^\alpha{}_{(\mu}\ep_{\nu)\alpha\beta} {{\cal Q}}^\beta\right]
   \,. \label{B7}
   \end{eqnarray}

   \begin{equation}
   \D^\nu\sigma_{\mu\nu}
   -{\textstyle{2\over3}}\D_\mu\Theta  =
   -{6\over\kappa^2\lambda} {{\cal Q}}_\mu
    \,.\label{B9}
    \end{equation}

    \begin{equation}
     \curl\sigma_{\mu\nu}
      -H_{\mu\nu}=0 \,.\label{B10}
      \end{equation}

      \begin{eqnarray}
       && \D^\nu E_{\mu\nu}
    -{\textstyle{1\over3}}\kappa^2\D_\mu\rho
     -[\sigma,H]_\mu= {\kappa^2 \rho\over
     3\lambda} \D_\mu\rho +{1\over\kappa^2\lambda}\left(2\D_\mu{{\cal
U}}-2
     \Theta{{\cal Q}}_\mu-3\D^\nu{{\cal P}}_{\mu\nu}
     +3\sigma_{\mu\nu}{{\cal Q}}^\nu\right)\!,
     \label{B12}
     \end{eqnarray}

     \begin{eqnarray}
      &&\D^\nu H_{\mu\nu}
       +[\sigma,E]_\mu
       =-{1\over\kappa^2\lambda}\left(3\curl{{\cal Q}}_\mu+3[\sigma,
       {{\cal P}}]_\mu\right)
       \,,\label{B14}
       \end{eqnarray}
       
\noindent
and the Gauss-Codazzi equations on the brane:
        \bea
        &&R^\perp_{\langle
        \mu\nu\rangle}+{\textstyle{1\over3}}\Theta\sigma_{\mu\nu}
        -E_{\mu\nu} - \sigma_{\alpha\langle
        \mu}\sigma_{\nu\rangle }{}^\alpha
        ={3\over\kappa^2\lambda}{\cal
        P}_{\mu\nu}\,, \label{B15}\\ &&R^\perp+
        {\textstyle{2\over3}}\Theta^2-\sigma_{\mu\nu} \sigma^{\mu\nu}
        -2\kappa^2\rho -2\Lambda = {\kappa^2\over\lambda}\rho^2+
        {12\over\kappa^2\lambda}{\cal U}\,. \label{B16}
         \eea

\subsubsection{Integrability conditions for ${\cal P}_{\mu\nu} =0, {\cal
Q}_\mu
= 0$}

When ${\cal P}_{\mu\nu} =0$ and ${\cal Q}_\mu = 0$, these equations
simplify.
Using Eq.(\ref{BB}) , Eq. (\ref{B3}) becomes for $\gamma \neq 1$:
\begin{equation}
\label{calU}
\D_\mu {\cal U} = \left[ \frac{4(\gamma -1)}{\gamma}
\frac{\cal U}{\rho} - \frac{\kappa^4 \gamma}{2} \rho  \right]
\D_\mu \rho.
\end{equation}
Taking the directional time derivative of Eq. (\ref{calU}), and
using the relations for interchanging space and time
derivatives, e.g.,
\begin{equation}
h_\mu{}^\nu[\D_\nu f]^\cdot
= h_\mu{}^\nu [\nabla_\nu + A_\nu]
\dot{f}
- \left[\frac{1}{3} \Theta h_\mu{}^\nu +
\sigma_\mu{}^\nu  \right] \D_\nu f,
\end{equation}
we obtain the integrability condition
\begin{eqnarray}
&&\left[\frac{4}{3}(3 \gamma -4)
{\cal U} - \frac{\kappa^4 \gamma^2}{2} \rho^2  \right] \rho \D_\mu \Theta
=\nonumber\\
&&\qquad\qquad
\left[\frac{4}{3\gamma}(3\gamma -4) (\gamma-1)
{\cal U}+ \frac{\kappa^4\gamma}{6}(3\gamma-1)\rho^2\right]
\Theta \D_\mu \rho.
\label{condition}
\end{eqnarray}
Taking the directional time derivative of (\ref{B9}) and interchanging
space and time derivatives,  we obtain
\begin{equation}
\label{equalszero}
\frac{(\gamma-1)}{\gamma} [12 {\cal U} + \kappa^4
\rho^2] \frac{1}{\rho} \D_\mu\rho =0.
\end{equation}
Hence for $\gamma \neq 1$ and $[12 {\cal U} + \kappa^4
\rho^2] \neq 0$, we have
$$ \D_\mu \rho = 0,$$
so that
$$ A_\mu = \D_\mu {\cal U} = \D_\mu \Theta = \D^\nu \sigma_{\mu \nu}
= \D_\mu \sigma^2 =0$$
and hence, in general, {\it the brane is spatially homogeneous}.
The special case
$\gamma = \frac{2}{3}$,  ${\cal U} + \kappa^4
\rho^2 = 0$,
in which the integrability conditions yield no constraints, corresponds
to the case in which there are no corrections to the general relativistic
equations.
Finally, in the special case $\gamma =1$, in which
$p =0$ and $A_\mu =0$, we obtain analogues of Eqs (\ref{calU})
and (\ref{condition}) 
from which it follows that
$\dot{\rho} =0$, and hence $\Theta=0$.

\subsection{Spatially Homogeneous Branes}

In the case of spatial homogeneity we have that
$$ \omega_\mu = 0, \quad A_\mu = 0, \quad \D_\mu \Theta =
\D_\mu \rho = \D_\mu p = \D_\mu {\cal U} =0. $$
With a general equation of state, we then obtain
\begin{equation}
\dot{\rho}+\Theta(\rho+p)=0,
\end{equation}

\begin{eqnarray}
&& \dot{\cal U}+{\textstyle{4\over3}}\Theta{\cal U}+\D^\mu{\cal
Q}_\mu  +\sigma^{\mu\nu}{\cal P}_{\mu\nu}=0\, ,
 \\
 && \dot{\cal
Q}_{\langle\mu\rangle}+{\textstyle{4\over3}}\Theta{\cal Q}_\mu
+ \D^\nu{\cal P}_{\mu\nu} +\sigma_{\mu\nu}{\cal Q}^\nu =0.
\end{eqnarray}

\begin{eqnarray}
&&\dot{\Theta}+{\textstyle{1\over3}}\Theta^2+\sigma_{\mu\nu}
 +{\textstyle{1\over2}}\kappa^2(\rho + 3p) -\Lambda=
-{\textstyle{\kappa^2\over2\lambda}}(2\rho+3p)\rho-
{6\over\kappa^2\lambda}{\cal U}\,  \label{gray}.
\end{eqnarray}

\begin{equation}
\dot{\sigma}_{\langle \mu\nu \rangle }
+{\textstyle{2\over3}}\Theta\sigma_{\mu\nu}
+E_{\mu\nu}  +\sigma_{\alpha\langle
\mu}\sigma_{\nu\rangle }{}^\alpha
={3\over\kappa^2\lambda}{\cal P}_{\mu\nu}\,.
\end{equation}
\begin{eqnarray}
 && \dot{E}_{\langle \mu\nu \rangle }
+\Theta E_{\mu\nu} -\curl H_{\mu\nu}
+{\textstyle{1\over2}}\kappa^2(\rho+p)\sigma_{\mu\nu}
-3\sigma_{\alpha\langle \mu}E_{\nu\rangle }{}^\alpha
\nonumber\\
&&~~{}= -{\textstyle{1\over2}}\kappa^2
(\rho+p){\rho\over\lambda}\sigma_{\mu\nu}
-{1\over\kappa^2\lambda}\left[4{\cal U}\sigma_{\mu\nu}+3\dot{\cal
P}_{\langle \mu\nu \rangle} +\Theta{\cal P}_{\mu\nu}
+3\D_{\langle\mu}{\cal Q}_{\nu\rangle}
+ 3\sigma^\alpha{}_{\langle\mu} {\cal
P}_{\nu\rangle\alpha}\right] \,.
\end{eqnarray}
\begin{eqnarray}
 &&\dot{H}_{\langle
\mu\nu \rangle } +\Theta H_{\mu\nu} +\curl E_{\mu\nu}-
3\sigma_{\alpha\langle \mu}H_{\nu\rangle }{}^\alpha =
{3\over\kappa^2\lambda}\left[ \curl{\cal
P}_{\mu\nu}
+\sigma^\alpha{}_{(\mu}\ep_{\nu)\alpha\beta} {\cal Q}^\beta\right]
\,.
\end{eqnarray}
\begin{equation}
\D^\nu\sigma_{\mu\nu} =
-{6\over\kappa^2\lambda} {\cal Q}_\mu
 \,.
\end{equation}
\begin{equation}
 \curl\sigma_{\mu\nu}
 -H_{\mu\nu}=0 \,.\label{pcc3}
\end{equation}
\begin{eqnarray}
 && \D^\nu E_{\mu\nu}
 -[\sigma,H]_\mu =  - {1\over\kappa^2\lambda}\left(2\Theta{\cal
Q}_\mu+3\D^\nu{\cal
P}_{\mu\nu}
-3\sigma_{\mu\nu}{\cal Q}^\nu\right)\!,
\end{eqnarray}
\begin{eqnarray}
 &&\D^\nu H_{\mu\nu}
+[\sigma,E]_\mu =
-{1\over\kappa^2\lambda}\left(3\curl{\cal
Q}_\mu+3[\sigma,{\cal P}]_\mu\right)
\,.
\end{eqnarray}
The Gauss-Codazzi equations on the brane are:
 \bea
&&R^\perp_{\langle \mu\nu\rangle}+\dot{\sigma}_{\langle \mu\nu
\rangle }+\Theta\sigma_{\mu\nu} ={6\over\kappa^2\lambda}{\cal
P}_{\mu\nu}\,,  \\
&&R^\perp+
{\textstyle{2\over3}}\Theta^2-\sigma_{\mu\nu} \sigma^{\mu\nu}
-2\kappa^2\rho -2\Lambda = {\kappa^2\over\lambda}\rho^2+
{12\over\kappa^2\lambda}{\cal U}\,  \label{gfr}.
 \eea

\subsubsection{RW brane}

It follows from the symmetries that for a RW brane
${\cal Q}_\mu=0={\cal P}_{\mu\nu}$. The generalized Friedmann
equation (\ref{gfr}) on a spatially homogeneous and isotropic brane
is~\cite{BinDefLan:2000a,BinDefLan:2000b,BinDefLan:2000c}
\begin{equation}\label{f}
H^2={\textstyle{1\over3}}\kappa^2\rho\left(1+{\rho\over
2\lambda}\right)+{\textstyle{1\over3}}\Lambda -{k\over a^2} +
{2{\cal U}_o\over\kappa^2\lambda} \left({a_o\over a}\right)^4\,,
\end{equation}
where $k=0,\pm1$. The nonlocal term that arises from bulk Coulomb
effects (also called the dark radiation term) is strongly limited
by conventional nucleosynthesis~\cite{lmw1,lmw2}:
$({\cal U} / \rho)_{\rm nucl} / \kappa^2 \lambda < 0.005$,
where ${\cal U}={\cal U}_o(a_o/a)^4$.
A more stringent constraint comes from small scale gravity experiments
\cite{Lang}.  This implies
that the unconventional evolution in the high energy 
regime ends, resulting in a transition to the conventional 
regime, and
the nonlocal term is sub-dominant during
the radiation era and rapidly becomes negligible thereafter.
The generalized Raychaudhuri equation (\ref{gray}) (with
$\Lambda=0={\cal U}$) reduces to
 \be
\dot H+H^2= - {\textstyle{1\over6}}
\kappa^2\left[\rho+3p+(2\rho+3p) {\rho\over \lambda}\right]\,.
 \ee

The bulk metric for a flat Friedmann brane is given
explicitly in~\cite{BinDefLan:2000a,BinDefLan:2000b,BinDefLan:2000c}.
In natural static and spherically symmetric coordinates, 
the bulk metric is Schwarzchild-anti de Sitter 
(with the
$Z_2$-symmetric RW brane at the boundary)
and is given
in~\cite{msm2,cline14}. The RW brane moves radially
along the 5th dimension, with $R=a(T)$, where $a$ is the RW scale
factor, and the junction conditions determine the velocity via the
Friedmann equation for $a$. Thus one can interpret the expansion
of the universe as motion of the brane through the static bulk.
In Gaussian normal coordinates, the
brane is fixed but the bulk metric is not manifestly static.
The bulk black hole gives rise to dark radiation
on the brane via its Coulomb effect.
 The mass parameter of the black
hole in the bulk is proportional to ${\cal U}_o$.
In order to avoid a naked singularity, we assume
that the black hole mass is non-negative.

In general, ${\cal U}\neq0$ in the Friedmann
background~\cite{BinDefLan:2000a,BinDefLan:2000b}.
Thus for a perturbed Friedmann model, the nonlocal bulk effects
are covariantly and gauge-invariantly described by the first-order
quantities $\D_\mu{\cal U}$, ${\cal Q}_{\mu}$, ${\cal
P}_{\mu\nu}$.

\subsection{Static Branes}

Assuming spatial homogeneity  (so that $A_\mu=0$) with ${\cal
Q}_\mu={\cal P}_{\mu\nu}=0$, in the
isotropic curvature subcase
with $R^\perp_{\langle \mu\nu\rangle}=0$, so that
${R^\perp}=ka^{-2}$ and
 \be \dot{R^\perp}= -{\textstyle{2\over3}}\Theta {R^\perp},
\ee
we have that
\begin{eqnarray}
&&\dot{\rho}=-\gamma\Theta\rho\,,\label{BBB1}\\
&& \dot{\cal {U}}=-{\textstyle{4\over3}}\Theta\cal{U}\,,\label{BBB2}\\
&&\dot{\Theta}+{\textstyle{1\over3}}\Theta^2+2\sigma^2+{\textstyle{1\over2}}\kappa^2(3\gamma-2)\rho
-\Lambda=
 -{\textstyle{\kappa^2\over2\lambda}}(3\gamma-1)\rho^2-
 {6\over\kappa^2\lambda}{\cal U}\,. \label{BBB4}\\
&&R^\perp+
 {\textstyle{2\over3}}\Theta^2-2\sigma^2
 -2\kappa^2\rho -2\Lambda = {\kappa^2\over\lambda}\rho^2+
 {12\over\kappa^2\lambda}{\cal U}\,, \label{BBB16}\\
&& \dot{\sigma^2} = -2\Theta \sigma^2,
\eea 
and the system of equations is
closed.

In the static models we have that $\Theta=0$ (and we shall 
assume that $\sigma=0$), and

\begin{eqnarray}
&&{\textstyle{1\over2}}\kappa^2(3\gamma-2)\rho_0 -\Lambda=
 -{\textstyle{\kappa^2\over2\lambda}}(3\gamma-1)\rho_0^2-
 {6\over\kappa^2\lambda}{\cal U}_0\,. \label{BBBB4}\\
&&R^\perp_0+
 -2\kappa^2\rho_0 -2\Lambda = {\kappa^2\over\lambda}\rho_0^2+
 {12\over\kappa^2\lambda}{\cal U}_0\, , \label{BBB16}
\eea 
where
$\rho_0$, ${\cal{U}}_0$ and ${R^\perp}_0$ are constants.
These constitute two algebraic equations for four constants (and $\gamma$,
which need not be equal to $2/3$). Since ${\cal{U}}_0$ need not be
positive, ${R^\perp}_0$ need not be positive definite.
There are static RW branes that can be embedded in a bulk that is not
Schwarzschild-AdS \cite{Gerg}.

\subsection{The EGS Theorem in Brane-World Models}

The isotropy of the cosmic microwave background (CMB) radiation
has crucial implications for the spatial homogeneity of the universe. If
all fundamental observers after last scattering observe an
isotropic CMB, then it follows from a theorem of Ehlers, Geren and
Sachs (hereafter EGS)  \cite{EGS} that in GR the universe
must have a
RW geometry. More precisely, the EGS
theorem states that if all
observers in a dust universe see an `isotropic radiation field', which is
implicitly
identified with the CMB, then that
spacetime is spatially homogeneous and isotropic. This can trivially be
generalised to the case of a geodesic and barotropic perfect
fluid~\cite{egsother1,egsother2}. However, as has been emphasised recently
\cite{cla-bar99b,cla-bar99a}, the resulting spacetime 
will be RW only if
the matter content is that of a perfect fluid form and the observers are
geodesic and
irrotational.  The EGS theorem has recently been investigated in
inhomogeneous universes models with non-geodesic observers, and
inhomogeneous spacetimes have been found which also allow every observer
to see
an isotropic CMB \cite{cla-bar99b,cla-bar99a}.

 The EGS
theorem is based on the collisionless Boltzmann equation and on
the dynamical field equations.
Bulk effects in brane-world models do not change the Boltzmann equation,
but they do
change the dynamical field equations 
\cite{coley,Maartens1,cline14,sms2,sms1},
so that  in principle the resulting  geometry need not be
RW on the brane. 
The  theorem  is concerned with universes in which all
observers see an isotropic radiation field. 
It follows from the multipole expansions of the
 Einstein-Boltzman equations for photons in a curved spacetime that even 
 in the brane-world scenario a spacetime with an
 isotropic radiation field must have a velocity field of the photons
which is shearfree
 and obeys~\cite{egsother1,egsother2}
 \be
 A_\mu=\D_{\mu} q,~~~\Theta=3\dot{q},\label{irf}
 \ee
 where $q$ is a function of the energy density of the radiation field. Any
 observers traveling on this congruence will then observe the isotropic
radiation.
 This velocity field is also a conformal Killing vector of the spacetime,
and hence
a spacetime admitting an isotropic radiation field must be conformally
 stationary, in which case the metric can be given in local coordinates by
 \be
 ds^2=e^{2q(t,x^\alpha)}\left\{-dt^2+h_{\alpha\beta}dx^\alpha
 dx^\beta\right\}\label{metric}
 \ee
 where $h_{\alpha\beta}(x^\gamma)$ can be diagonalised. If
 $q=q(t)$, then the acceleration is zero.

 Therefore, assuming~(\ref{irf}) and $\sigma =0$
 (as well as $\omega =0$),
 we immediately have that

 \begin{equation}
 {\textstyle{2\over3}}\D_\mu\Theta  =
 {6\over\kappa^2\lambda} {\cal Q}_\mu
  \,,\label{D9}
  \end{equation}

  \begin{equation}
   H_{\mu\nu}=0 \,.\label{D10}
   \end{equation}
From  Eqs.~(\ref{BB})
 and (\ref{B1})  we then obtain (assuming $p = (\gamma - 1)\rho$)
   \begin{equation}
   \D_\mu\Theta  = 3\D_\mu(\dot{q}) = 3h_\mu{}^\nu[\D_\nu q]^\cdot =
3(\gamma-1)\D_\mu\Theta
    \,,\label{D111}
    \end{equation}
    which implies that  (for $\gamma \ne {4\over3}$)
    \begin{equation}
    \D_\mu\Theta  = 0\,.\label{DD9}
    \end{equation}
    (When $\gamma = {4\over3}$ we obtain  $\rho = \rho_0 q^{-4}$).
    Consequently, from Eq.~(\ref{D9}) we have that

    \begin{equation}
    {\cal Q}_\mu = 0\,.\label{Dd9}
    \end{equation}
From (\ref{D111}) we have that
$\D_\mu(\dot{q}) = 0$ and $[\D_\nu q]^\cdot = 0$, which then implies that
$A_\mu = 0$.
This can best be seen by using local coordinates (\ref{metric}), in which
these conditions
yield $q(t,x^\alpha) = f(t) + g_1(x^\alpha)$ and $e^{-q} = e^{-f(t)} +
g_2(x^\alpha)$,
so that for consistency either $f(t)=0$ (whence $\Theta = 0$) or $q=q(t)$,
and hence $A_\mu = 0$.
Therefore,  we find that
\begin{equation}
A_\mu = 0\,,\label{DDd9}
\end{equation}
and hence $\D_\mu\rho = 0$.

Taking $H_{\mu\nu}=0, {\cal Q}_\mu = 0,\D_\mu\rho= 0, \D_\mu\Theta= 0,
A_\mu =0$, which implies that
\begin{equation}
E_{\mu\nu}={3\over\kappa^2\lambda}{\cal P}_{\mu\nu}\,, \label{D5}
\end{equation}
Eqs. (\ref{B3}) and (\ref{B12}) then yield
\begin{equation}
\D_\mu{\cal U}=0, ~  \D^\nu{\cal P}_{\mu\nu} =0, ~\D^\nu E_{\mu\nu}=
0\,,\label{DDD}
\end{equation}
and Eqs. (\ref{B6}) and (\ref{B15}) become
\begin{equation}
 \dot{E}_{\langle \mu\nu \rangle }
 +{2\over3}\Theta E_{\mu\nu}
 = 0 \, , \quad R^\perp_{\langle
 \mu\nu\rangle} = 2E_{\mu\nu}\, .\label{D6}
 \end{equation}
For the metric (\ref{metric}) with $q=q(t)$, it then follows that
 \begin{equation}
 R^\perp_{\langle
 \mu\nu\rangle} =  R^\perp_{\langle
 \mu\nu\rangle}(x^\gamma),   \label{perp}
 \end{equation}
 where $R^\perp_{\mu\nu}$ is the 
 Ricci tensor of the 3-metric $h_{\alpha\beta}(x^\gamma)$
 \cite{coleymcmanus}. Consequently
   \begin{equation}
    \partial_t [R^\perp_{\langle\mu\nu\rangle}] = 0,
   \end{equation}
 and hence
 $\partial_t [E_{\mu\nu}] = 0$ and  $\partial_t [{\cal P}_{\mu\nu} ] = 0$.
The remaining non-trivial equations are the conservation law
 for ${\cal U}$, the generalized Raychaudhuri equation, and the
generalized Friedmann equation.

\subsubsection{Discussion}

The 5D metric in the bulk is given in Gaussian normal
coordinates by
  \begin{equation}
  ds^2 =  \widetilde{g}_{AB} dx^A dx^B = {g}_{ab}(t,x^\gamma,y) dx^a dx^b
+ dy^2,
  \end{equation}
where $A,B= 0, 1-3, 4=y$ and $a,b=0, 1-3$, the brane is
located at $y =0$ and the unit normal to the brane is  $n^A =
\delta^A_y$. Assuming that the metric functions ${g}_{ab}$ are
smooth, they can be expanded in powers of $y$ close to the brane
up to $O(y^2)$.
The metric functions are determined up to  $O(y^0)$ by Eq. (\ref{metric}).
The extrinsic curvature on the brane,    $K_{AB}^{+}$, which is
discontinuous
at $y=0$, restricts the form of  ${g}_{ab}(t,x^\gamma,y)$ to  $O(y)$.
Assuming that there are no 5D fluxes the stress-energy of
the bulk,
$\widetilde{T}_{AB}$,  in the neighbourhood
of the brane satisfies   $\widetilde{T}_{0\alpha}=0$, which is valid for
scalar
fields, perfect fluids, a cosmological constant and combinations thereof,
 and we have that
${g}_{0\alpha}=0$  (which then implies that    ${\cal Q}_\mu =0$). The
non-trivial
components of   ${g}_{ab}$  up to  $O(y^2)$ can then be written (with a
slight change of
notation)  in the  $y>0$ ($^{+}$-region)   as
  \begin{equation}
  {g}_{00}= - (1 + f(t)y + F(t,x^\gamma) y^2),
  \end{equation}
   \begin{equation}
   {g}_{\alpha\beta} = Q(t)(1 + q(t)y){h}_{\alpha\beta} +
   {G}_{\alpha\beta}(t,x^\gamma)y^2.
   \end{equation}

Noting that
${\cal E}_{ab} = \widetilde{C}_{aCbD}n^Cn^D = \widetilde{C}_{a5b5}$,
by a direct calculation we find that on the brane ($y=0$)
\begin{equation}
\widetilde{C}_{0505}  =  {1\over4Q(t)}{\cal F}(t) +
{1\over12Q(t)}(6Q(t)F(t,x^\gamma) - ^3R(x^\gamma) - 2G(t,x^\gamma)),
\end{equation}
\begin{equation}
 \widetilde{C}_{\alpha5\beta5}
  =  -{1\over3}^3R_{\alpha\beta}(x^\gamma) + {1\over12}(^3R(x^\gamma)
    + {\cal F}(t) +2Q(t)F(t,x^\gamma) +
2G(t,x^\gamma)){h}_{\alpha\beta}(x^\gamma)  -
{2\over3}{G}_{\alpha\beta}(t,x^\gamma),
 \end{equation}
 where
\begin{equation}
 {\cal F}(t) \equiv Q_{tt} - {1\over Q}(Q_{t})^2 - {1\over2}Qfq  + Qq^2 -
 {1\over2}Qf^2,
\end{equation}
\begin{equation}
  G(t,x^\gamma) \equiv    {h}^{\alpha\beta} {G}_{\alpha\beta},
\end{equation}
 and
 $\widetilde{C}_{\mu505} = 0$   (so that   ${\cal Q}_\mu = 0$).

From
${\cal E}_{\langle\mu\nu\rangle}= -
R^\perp_{\langle\mu\nu\rangle}(x^\gamma)$  and
${\cal E}_{00}= - {6\over\kappa^2\lambda} {\cal U}(t)$, we then find that
\begin{equation}
 {G}_{\alpha\beta}(t,x^\gamma) = R^\perp_{\alpha\beta}(x^\gamma) +  
 {1\over3}(G(t,x^\gamma)
 -R^\perp(x^\gamma))h_{\alpha\beta}(x^\gamma),
 \end{equation}
 and
\begin{equation}
 G(t,x^\gamma) = {3\over2} {\cal F}(t) +  {36Q(t)\over\kappa^2\lambda}
{\cal U}(t)
 +3Q(t)F(t,x^\gamma) -  {1\over2} R^\perp(x^\gamma).
 \end{equation}

 Finally, for  a bulk source consisting of a  combination
 of scalar fields, perfect fluids and a cosmological constant, we then
also have that
 \begin{equation}
 F(t,x^\gamma)= F(t), ~~ G(t,x^\gamma)= G(t),
 \end{equation}
 and it follows that $R^\perp(x^\gamma)$ is constant.

 We can solve the full equations in the bulk iteratively.
 In general it seems that
 the remaining bulk field equations consist
 of five ODEs for the five non-trivial functions (of $t$) in the metric
and additional
 matter fields and so there will be many possible solutions.
 It may be possible to integrate the bulk equations explicitly in the case
 of a 5D cosmological constant.

\newpage
\section{Essential Features}

The generalized Friedmann equation, which determines 
the volume expansion of the universe
or the Hubble function
$H = \frac{1}{3}\Theta$, in the case of spatially homogeneous
cosmological models is
\begin{equation}
H^2 = \frac{1}{3}\kappa^2\rho\left(1+\frac{\rho}{2\lambda}\right)
-\frac{1}{6}{} R^\perp +\frac{1}{3}\sigma^2+\frac{1}{3}\Lambda +
\frac{2{\cal U}}{\lambda\kappa^2}\,, \label{frie}
\end{equation}
where 
$2\sigma^2\equiv\sigma^{ab}\sigma_{ab}$ is the
shear scalar  and
$H = {\dot a}/a $ \cite{Maartens1,sms2,sms1}, and
the background induced metric 
satisfies
\begin{equation}\label{6c}
\D_\mu{\cal U}={\cal Q}_{\mu}={\cal P}_{\mu\nu}=0\,.
\end{equation}

For a flat isotropic brane (with $\Lambda=0={\cal U}$) 
we obtain
\begin{equation}\label{ff}
H^2={\kappa^2\over3} \rho\left(1 +{\rho\over2\lambda}\right).
\end{equation}
The conservation equation is given by
\begin{equation}\label{cons}
\dot{\rho}+3H(\rho+p)=0.
\end{equation}
For a minimally coupled scalar field the energy density and 
pressure are, respectively,
\begin{equation}
\rho={1\over2}\dot\phi^2+ V(\phi),~ p={1\over2}\dot\phi^2-V(\phi),
\end{equation}
so that the conservation law is equivalent to the Klein-Gordon equation
\begin{equation}\label{kg}
{\ddot \phi} + 3H {\dot \phi} + V'(\phi) = 0.
\end{equation}
At early times, assuming that the energy density is dominated by a scalar
field and that
$\rho$ is very large (and ${\cal U}=0$), 
the conservation equation~(\ref{cons})
becomes
\begin{equation}
\dot{\rho}= -3\dot\phi^2\rho,
\end{equation}
so that $\rho$ is monotonically decreasing and the models will
eventually
evolve to the low density regime and hence the usual general
relativitistic
scenario.

\subsection{Anisotropic and Curved Branes}

There are many reasons to consider cosmological models that
are more general than RW, both spatially homogeneous and anisotropic
and spatially inhomogeneous.
The 3-curvature in RW models is given by
$R^\perp = {6k \over a^2}$, where $k=0, \pm1$ is the curvature constant. 
The structure of the initial singularity
in RW brane-models was studied in~\cite{Ishihara}.
An equivalent 3-curvature occurs in orthogonal spatially homogeneous
and isotropic curvature models, and a similar term occurs 
in other models such as
 Bianchi V cosmological models.
For a
RW model on the brane, Eq. (\ref{6c}) follows, which implies that
${\cal U}={\cal U}(t)$.
In the case of Bianchi type I we get
${\cal Q}_\mu=0$ but we do not get any restriction on
${\cal P}^{\mu \nu}$.  Since there is no way of fixing the dynamics
of this tensor we can study the particular case in which it is zero.
Thus, in RW and Bianchi I models the evolution equation for ${\cal U}$ 
is
$\dot{\cal U} = - 4H{\cal U}$ ~\cite{Maartens1,Maartens2}, 
which integrates to ${\cal U}=\frac{{\cal
U}_0}{a^4}$,
which has the structure of a radiation fluid (sometimes referred to as
the "dark radiation term",
but where
${\cal U}_0$ can be negative).

A Bianchi~I brane is covariantly characterized in ~\cite{mss}.
The conservation equations reduce to the conservation 
equation~(\ref{cons}), an
evolution equation for the effective nonlocal energy density on the brane
${\cal U}$, and a differential constraint 
on the effective nonlocal anisotropic stress
${\cal P}_{\mu\nu}$.  
In the Bianchi I case,
the presence of the nonlocal bulk tensor ${\cal P}_{\mu\nu}$ in
the Gauss-Codazzi equations on the brane 
means that we cannot simply integrate
to find the shear as in GR.
However, when
the nonlocal energy density vanishes
or is negligible, i.e., ${\cal U}=0$, then the
conservation equations imply $\sigma^{\mu\nu}{\cal
P}_{\mu\nu}=0$.
This consistency condition implies 
$\sigma^{\mu\nu}\dot{\cal
P}_{\mu\nu}=6{\cal P}^{\mu\nu} {\cal P}_{\mu\nu}/\kappa^2\lambda$.
Since there is no evolution
equation for ${\cal P}_{\mu\nu}$ on the 
brane~\cite{Maartens1,Maartens2}, this is
consistent on the brane.
This assumption is often made in the case of RW branes, and in
that case it leads to a conformally flat bulk
geometry~\cite{BinDefLan:2000a,BinDefLan:2000b,BinDefLan:2000c}.
The shear evolution equation may be integrated after contracting it with
the shear, to give
\begin{equation}\label{s}
\sigma^{\mu\nu}\sigma_{\mu\nu}  =
{6\Sigma_0^2\over a^6}\, ,
\end{equation}
where $\Sigma_0$ is constant.
Bianchi I models on the brane 
have been studied with a massive scalar field \cite{mss},
with a perfect fluid with linear equation of 
state
\cite{CamposSopuerta:2001a,CamposSopuerta:2001b,Toporensky:2001} 
and a perfect fluid and scalar field~\cite{paul}.
A similar shear term occurs in Bianchi cosmological models (such as
Bianchi type V models)
in which the hypersurfaces of homogeneity
are orthogonal to the fluid velocity.

The conservation equation is given by equation
(\ref{cons}).
If we assume that the matter content is equivalent
to that of a non-tilting perfect fluid with a linear barotropic
equation of state (i.e.,
$p = (\gamma-1)\rho$, where  
$\rho\geq 0$, and $\gamma\in[0,2]$)  
the conservation equation then yields
$\rho=\rho_0 a^{-3\gamma}$, where $\rho_0 > 0$.
A dynamical analysis of scalar
field models indicates that at early times the scalar field is effectively
massless.
A massless scalar field is equivalent to such a perfect fluid
with a stiff equation of state parameter  $\gamma=2$.

We can therefore write down a phenomenological (spatially homogeneous)
generalized
Friedmann equation:
\begin{equation}\label{f}
H^2={\kappa^2\rho_0 \over 3a^{3\gamma}}  +  {\kappa^2\rho_0^2\over
6\lambda a^{6\gamma}} - {k \over a^2}
 + \frac{1}{3}\Lambda + {\Sigma_0^2\over a^6} +
{{\cal U}_0 \over a^4}.
\end{equation}
In many applications the 4D
cosmological constant is assumed to be zero~\cite{randall1,randall2}; here
we shall assume that if
it is non-zero it is positive;
i.e., $\Lambda\geq 0$.

In particular, the cosmological
evolution of the RS brane-world scenarios in RW
and the Bianchi I and V
perfect fluid models with a linear barotropic
equation of state with ${\cal U} = 0$ were studied 
in~\cite{CamposSopuerta:2001a,CamposSopuerta:2001b}
using dynamical systems techniques. This work was
generalized in~\cite{CamposSopuerta:2001b},
in which the 5D Weyl tensor has
a non-vanishing projection onto the three-brane where matter fields are
confined, i.e., ${\cal U} \ne 0$; in the case
of RW models the study was completely general whereas in the
Bianchi type I case the Weyl tensor components 
were neglected (the
Bianchi
V models were not considered).

The particular models considered so far, in which the curvature and shear
are given by
the expressions in the above phenomenological equation, are very special.
In particular,
the Bianchi I and V models are not generic, and so the study of the
dynamics of these models
does not shed light on the typical behaviour of spatially homogeneous
brane models. Clearly, more
general scenarios must be considered to obtain physical insights, and we
shall
study Bianchi type IX models in detail later.

\subsubsection{Late times:}

Recent observations of distant supernovae and galaxy clusters seem to
suggest
that our universe is presently undergoing a phase of accelerated 
expansion~\cite{perlmutter3,perlmutter1,perlmutter2,perlmutter4},
indicating the dominance of dark energy with negative pressure in our
present
universe. The idea that a slowly rolling scalar field provides the
dominant
contribution to the present energy density has gained prominence in recent
times~\cite{ratra1,ratra3,ratra2}.
In~\cite{Mazumbar}  a potential consisting of a general combination of two
exponential terms
was considered
within the context of brane cosmology to determine whether it is possible
to obtain both early time inflation and  accelerated
expansion during the present epoch through the dynamics
of the same scalar field.
Such a potential,
motivated by phenomenological considerations
from scalar tensor theories of gravity 
and higher
dimensional quantum effects or a cosmological constant together with a
4D potential, has been recently claimed to
conform to all
the current observational constraints  on quintessence 
(for certain values
of the parameters)~\cite{barreiro}.
The brane-world dynamics was assumed to be governed 
by the modified Friedmann Eq.
(\ref{ff})
where, for reasons of simplicity, the contributions from bulk
gravitons and a higher dimensional cosmological constant were set to
zero~\cite{steep1,steep6,mwbh}. 
It was found that the brane-world inflationary scenario is feasible with
steep potentials (in contrast
to the situation in standard cosmology). During inflation
the necessity of sufficient inflation and the COBE normalised amplitude
of density perturbations can be used to fix the values of the brane
tension
and the scale of the potential at this stage.
The emergence of a second phase of accelerated expansion
necessitates the introduction of a second exponential term in the
potential.
A successful brane quintessence model  with a scalar field (and a
power-law potential)
and a dark radiation component (${\cal U}_0 \ne 0$) 
was discussed in~\cite{mw3,mw1,mw2}, with particular emphasis
on how the quadratic matter term affects the evolution of the scalar
field.

When a positive cosmological constant is present 
($\Lambda\geq 0$),  the
de Sitter model is always the global attractor for ${\cal U}\geq 0$.
For ${\cal U} < 0$,  models
can (re)collapse (even without a positive curvature); i.e., there exist
(re)collapsing models both for RW and Bianchi type I models for any values
of
$\gamma$.
This means that for
${\cal U} < 0$ the  de Sitter model is only a local attractor
\cite{CamposSopuerta:2001a,CamposSopuerta:2001b}
(and the cosmic no-hair theorem is consequently
violated).
In~\cite{Santos:2001} a set of sufficient conditions which must be
satisfied by
the brane matter and bulk metric so that a homogeneous and anisotropic
brane
asymptotically evolves to a de Sitter spacetime in the presence of a positive
cosmological constant on the brane  was derived.
It was shown that from violations of these
sufficient conditions a negative nonlocal energy density or
the presence of strong anisotropic stress (i.e., a magnetic field) may lead
the
brane to collapse.
For  ${\cal U}\leq \ 0$ and  positive curvature (as in the $k=1$ 
RW models), not
only can models recollapse, but
there exist oscillating universes in which the
physical variables oscillate periodically between
a minimum and a maximum value without reaching any spacelike
singularity~\cite{CamposSopuerta:2001b}.
Indeed, when there are bulk effects, such as for example in case of a
negative contribution
of the dark-energy density term ${\cal U}$, the singularity
theorems are violated and a singularity can be avoided.

\subsubsection{Intermediate times: inflation}

The qualitative properties of
Bianchi I  brane models  with a perfect fluid and a linear equation
of state (${\cal U} = 0$ and ${\cal U} \ne 0$)
were studied in~\cite{CamposSopuerta:2001a,CamposSopuerta:2001b}.
A variety of intermediate behaviours can occur, and the various
bifurcations in these models were discussed in some detail. In addition,
the maximum value of the
shear in Bianchi I models was studied in~\cite{Toporensky:2001}. In
particular,
models with a positive curvature can recollapse,
although the condition for recollapse in the brane-world can be different
to that
in GR~\cite{CamposSopuerta:2001a,CamposSopuerta:2001b}.

Recent measurements of the power spectrum of  the CMB 
anisotropy~\cite{bernandis2,bernandis1,bernandis3}
provide strong justification for
inflation~\cite{lidsey3,lidsey2,lidsey1}. 
Since the Friedmann equation is modified
by an extra term quadratic in energy density $\rho$ at high 
energies,
it is of interest to study
the implications of such a modification on the inflationary
paradigm.
The issue of inflation
on the brane was first investigated in~\cite{mwbh}, where it was shown
that on a RW brane in 5D anti de Sitter space,
extra-dimensional effects are conducive to the advent of inflation.
It has also been
realized that the brane-world scenario is more suitable for inflation with
steep potentials because the quadratic term in $\rho$
increases friction in the inflaton field equation~\cite{mwbh}. This
feature has been
exploited to construct inflationary models using both large inverse power
law~\cite{steep6} as well as steep exponential~\cite{steep1} potentials
for
the scalar field.

In particular, for a RW brane inflation at high energies
($\rho>\lambda$) proceeds at a higher rate than the corresponding
rate in GR. As noted earlier, this introduces important changes to
the dynamics of the early 
universe, and
accounts for an increase in the amplitude of scalar~\cite{mwbh}
and tensor~\cite{lmw1,lmw2} fluctuations at Hubble-crossing, and for a
change to the evolution of large-scale density perturbations
during inflation~\cite{gm}.
The condition that radiation domination sets
in before nucleosynthesis can be used to impose constraints on the
parameters
of brane inflation models~\cite{steep1,steep6}. A definite prediction of
these
models is the parameter independence of the spectral index of scalar
density perturbations~\cite{steep1,steep6,mwbh}.
The behaviour of an anisotropic brane-world in the
presence of inflationary scalar fields was examined in~\cite{mss}.
The evolution equations on the brane (with $\Sigma_0 \ne 0$) 
 for a minimally coupled massive scalar field 
 with  $V={1\over2}m^2\phi^2$ was studied
and it was shown that, contrary to
expectations, a large anisotropy does not adversely affect
inflation~\cite{mss}. In fact, a large initial anisotropy introduces
more damping into the scalar field equation of motion, resulting
in greater inflation.
After the kinetic term and the anisotropy have dropped to a very small
value, a transient anisotropy dominated regime begins and
the anisotropy
then rises, gradually
approaching its asymptotic `slow-roll' value
for the simple `chaotic' potential. The kinetic energy does not remain
constant
but gradually increases as the field amplitude decreases during
slow-roll. Since  the decay of anisotropy is found to be generically
accompanied by a corresponding decrease in the kinetic energy of
the scalar field~\cite{mss},  this effect leads to greater inflation.
Subsequently, the anisotropy dissipates.

In RS brane-worlds, where the bulk has only a vacuum
energy, inflation on the brane must be driven by a 4D scalar field
trapped on the brane \cite{lidrev}. In more general brane-worlds, where
the bulk
contains a 5D scalar field, it is possible that the 5D field
induces inflation on the brane via its effective
projection~\cite{hs11,Brandenberger1,hs3,hs10,cline16,hs4,hs1,hs5,hs6,hs9,hs8}.
More exotic possibilities arise from the interaction between two
branes, including possible collisions, which is mediated by a 5D
scalar field and which can induce either inflation~\cite{kss1,kss2} or a
hot big-bang radiation era, as in the ``ekpyrotic" or cyclic
scenario~\cite{ek2,ek1,ek3}, or colliding bubble 
scenarios~\cite{bub1,bub2,bub3}
(for colliding branes in an M~theory approach see~\cite{ek5a,ek5b}).
In general, 
high-energy brane-world modifications to the dynamics of inflation
on the brane have been 
investigated~\cite{inf3,inf4,inf6,inf2,cline17,inf5,inf7}. Essentially,
the
high-energy corrections provide increased Hubble damping
making slow-roll inflation possible
even for potentials that would be too steep in standard
cosmology~\cite{steep7,steep1,steep6,steep5,steep2,steep4,steep3}.

\subsubsection{Early times: initial singularity}

An important question is how the higher-dimensional bulk
effects modify the picture of gravitational collapse/
singularities. 
The generalized Raychaudhuri equation governs
gravitational collapse and the initial singularity on the
brane. The local energy density and pressure corrections,
${1\over12}\widetilde{\kappa}^4\rho(2\rho+3p)$,
further enhance the tendency to collapse~\cite{mwbh} if $2\rho+3p>0$
(which
is satisfied in thermal collapse).
The nonlocal term,
which is proportional to ${\cal U}$,
can act either way depending on its sign;
the effect of a negative
${\cal U}$ is to counteract gravitational collapse, and hence the 
singularity can be avoided in this case.

A unique feature of brane cosmology is that the effective equation of
state
at high densities can become ultra stiff. Consequently
matter can overwhelm shear
for equations of state which are
stiffer than dust, leading to quasi-isotropic early expansion of
the universe.
For example, in
Bianchi I models \cite{CamposSopuerta:2001a,CamposSopuerta:2001b,mss}
the approach to the initial singularity is matter
dominated and not shear dominated, due to the predominance of the matter
term
$\rho^2/2\lambda^2$ relative to the shear term
$\Sigma_0^2/a^6$ (and 
$\sigma^2/H^2 \to 0$ as  $t \to 0$).
The fact that the density effectively grows faster than $1/a^6$ for
$\gamma > 1$ is a uniquely brane effect (i.e., it is not possible in GR).

Indeed, a spatially homogeneous and isotropic non-general-relativistic brane-world
(without brane tension; ${\cal U} = 0$)
solution, first discussed by Bin\'etruy, Deffayet and
Langlois \cite{BinDefLan:2000a,BinDefLan:2000b,BinDefLan:2000c}, 
is always a source/repeller for $\gamma \ge 1$
 in the presence of
non-zero shear. This solution is
self-similar, and is  sometimes referred to as the
Brane-Robertson-Walker model (BRW)~\cite{Coley:2002b}.
The equilibrium point corresponding to the BRW model is often denoted
by ${\cal F}_b$.
The BRW model, in which 
\begin{equation}
a(t)\sim t^{\frac{1}{3\gamma}},
\end{equation}
is valid at very high
energies ($\rho \gg \lambda$) as the initial singularity is approached ($t
\rightarrow 0$).
In the brane-world scenario
anisotropy dominates only for $\gamma<1$ (whereas in GR
it dominates for $\gamma<2$), and therefore
for all physically relevant values of $\gamma$
the singularity is
isotropic.

If $\rho^{tot} > 0$ and $\rho^{tot} + 3 p^{tot}> 0$ 
for all $t < t_0$, then from the generalized  Raychaudhuri equation
(\ref{gray})
 and using the generalized Friedmann equation (\ref{gfr})
and the conservation equations, it follows that
for $\dot {a_0} > 0$ (where $a_0 \equiv a(t_0)$) there exists a time $t_b$
with
$t_b<t_0$ such that $a(t_b)=0$, and there exists a singularity at $t_b$,
where
we can rescale time so that $t_b=0$ and the singularity occurs at the
origin.
We can find the precise constraints on $\Lambda$ and ${\cal U}_0$
in terms of $a_0$ in order for these conditions to be satisfied at $t =
t_0$. It follows that if these conditions are satisfied at
$t = t_0$
they are satisfied for all $0 < t < t_0$, and a singularity necessarily
results.
These conditions are indeed satisfied for regular matter undergoing
thermal collapse in which the
the local energy density and pressure satisfy
$\rho(2\rho+3p)>0$ (and is certainly satisfied for perfect fluid matter
satisfying
the weak energy condition $\rho \geq 0$ and a
linear barotropic equation of state with $\gamma \geq 1$).
On the other hand, it is known that a large positive cosmological constant
$\Lambda$ or
a significant negative  nonlocal term ${\cal U}$
counteracts gravitational collapse and can lead to the singularity being
avoided
in exceptional circumstances.

In more detail, we shall see later that since the 
dimensionless variables are 
bounded close to the singularity
it follows from Eq. (\ref{q}) that $0< q < 2$,
where $q$ is the deceleration parameter.
Hence, from Eq. (\ref{raych}), $H$ diverges as the initial singularity is
approached.
At an equilibrium point $q = q^*$, where $q^*$ is a constant with $0< q^*
< 2$,
so that from Eq. (\ref{raych}) we have that $H  \rightarrow (1 + q^*)^{-1}
t^{-1} $
as $ t  \rightarrow 0^{+}$ ($\tau  \rightarrow -\infty$).
From Eqs. (\ref{q}), (\ref{FRIEM}) and the conservation laws
it then follows that $\rho \rightarrow \infty$
as $ t  \rightarrow 0^{+}$. It then follows directly from the
conservation laws (\ref{CE}) that $\Omega_b= 
\kappa^2 \rho^2/ 6 \lambda H^2$ dominates as $ t  \rightarrow
0^{+}$
(and that all of the other $\Omega_i$ are negligable dynamically as the
singularity
is approached).  The fact that the effective equation of state
at high densities become ultra stiff, so that the
matter can dominate the shear dynamically, is a unique feature of brane
cosmology.
Hence, close to the singularity the matter
contribution is effectively given by
$\ti{\rho}^{\rm tot} = \ti{\rho}^2/2\lambda \equiv
\ti{\mu}_b , \quad
\ti{p}^{\rm tot} = (\ti{\rho}^2 +2\ti{\rho}
\ti{p})/2\lambda = (2\gamma -1)\ti{\mu}_b$.
Therefore, as the initial singularity is approached, the model
is approximated by an RW model in some
appropriately defined mathematical sense.
Goode et al. ~\cite{GW85a,GW85b} introduced the rigorous mathematical
concept of an isotropic singularity  into   cosmology in order to
define precisely the notion of a
``Friedmann-like'' singularity.  Although a number of
perfect fluid cosmologies are known to admit an
isotropic singularity, in GR 
a cosmological model will not 
admit an isotropic singularity in general.

\newpage
\section{Bianchi IX Brane-world Cosmologies}

The brane energy-momentum
tensor, including both  a perfect fluid and a minimally coupled scalar
field, is given by
\begin{equation}
T_{\mu\nu}=T_{\mu\nu}^{pf}+T_{\mu\nu}^{sf},\label{source}
\end{equation}
where
\begin{eqnarray}
T_{\mu\nu}^{pf}&=&\rho u_\mu u_\nu+ph_{\mu\nu}\,\\
T_{\mu\nu}^{sf} &=& \phi_{;\mu}\phi_{;\nu}
-g_{\mu\nu}\left(\frac{1}{2}\phi_{;\alpha}\phi^{;\alpha}+V(\phi)\right),
\end{eqnarray}
and $u^\mu$ is the fluid 4-velocity and
 $\phi$ is the minimally coupled scalar field having potential $V(\phi)$.
If $\phi_{;\mu}$ is timelike, then a
 scalar field with potential $V(\phi)$ is equivalent to a perfect fluid
 having an energy density and pressure
\begin{eqnarray}
\rho^{sf}&=&-\frac{1}{2}\phi_{;\mu}\phi^{;\mu}+V(\phi)\\
p^{sf} &=&-\frac{1}{2}\phi_{;\mu}\phi^{;\mu}-V(\phi).
\end{eqnarray}

The local matter corrections $S_{\mu\nu}$ to the Einstein equations 
on the brane  are
given by Eq. (3),
which is equivalent to
\begin{equation}
S_{\mu\nu}^{pf}=\frac{1}{12} \rho^2 u_\mu u_\nu
+\frac{1}{12}\rho\left(\rho+2 p\right)h_{\mu\nu}\,,
\end{equation}
for a perfect fluid
and
\begin{eqnarray}
S_{\mu\nu}^{sf}&=&\frac{1}{6}\left(-\frac{1}{2}\phi_{;\alpha}\phi^{;\alpha}
+V(\phi)\right)\phi_{;\mu}\phi_{;\nu}\nonumber\\
&&\qquad\qquad +\frac{1}{12}\left(-\frac{1}{2}\phi_{;\alpha}\phi^{;\alpha}
+V(\phi)\right)\left(-\frac{3}{2}\phi_{;\alpha}\phi^{;\alpha}
-V(\phi)\right)g_{\mu\nu}\,,
\end{eqnarray}
for a minimally coupled scalar field.  If we have both a perfect fluid and
a scalar field
and we assume that the gradient of the scalar field, $\phi^{;\,\mu}$,
is aligned with the fluid 4-velocity, $u^{\mu}$
(that is, $\phi^{;\,\mu}/\sqrt{-\phi_{;\,\alpha}\phi^{;\,\alpha}}=u^\mu$),
the local brane
effects due to a combination of a perfect fluid and a scalar field are
then
\begin{eqnarray}
S_{\mu\nu} &=&\frac{1}{12}
\left(\rho-\frac{1}{2}\phi_{;\,\alpha}\phi^{;\,\alpha}+V(\phi)\right)^2
u_\mu u_\nu\nonumber\\
&&+\frac{1}{12}\left(\rho-\frac{1}{2}\phi_{;\,\alpha}\phi^{;\,\alpha}+V\right)
\left(\rho+2 p-\frac{3}{2}\phi_{;\,\alpha}\phi^{;\,\alpha}-V(\phi)\right)
h_{\mu\nu} \, .
\end{eqnarray}

The non-local effects from the free gravitational field in the bulk
are given by ${\cal E}_{\mu\nu}$ \cite{Maartens1,Maartens2}.
We will assume here that
$\D_\mu{\cal U}={\cal Q}_{\mu}={\cal P}_{\mu\nu}=0$.
Since ${\cal P}_{\mu\nu}=0$,
in this case the evolution of ${\cal E}_{\mu\nu}$ is fully 
determined.
In general ${\cal U}={\cal U}(t)\neq0$ (and can be negative) 
\cite{BinDefLan:2000a,BinDefLan:2000b,BinDefLan:2000c}.
All of the bulk corrections mentioned above may be consolidated into an
effective total
energy density and pressure as follows. The modified Einstein equations
take the standard
Einstein form with a redefined energy-momentum tensor 
(assuming $\Lambda =0$) according to Eqs. (4) and (5), where
the redefined 
total energy density and pressure due to both a 
perfect fluid and a scalar field are given by
\begin{eqnarray}
\rho^{\rm total} \label{total1}
&=&\rho+\left(-\frac{1}{2}\phi_{;\,\alpha}\phi^{;\,\alpha}+V(\phi)\right)
 \nonumber\\&& \qquad+\frac{\widetilde{\kappa}^{4}}{\kappa^6}
\Biggl[\frac{\kappa^4}{12}\left(\rho-\frac{1}{2}\phi_{;\,\alpha}\phi^{;\,\alpha}
+V(\phi)\right)^2+{\cal U}\Biggr] \\
p^{\rm total} \label{total2} &=&
p+\left(-\frac{1}{2}\phi_{;\,\alpha}\phi^{;\,\alpha}-V(\phi)\right)\nonumber\\
&&+\frac{\widetilde{\kappa}^{4}}{\kappa^6}
\Biggl[
\frac{\kappa^4}{12}\left(\rho-\frac{1}{2}\phi_{;\,\alpha}\phi^{;\,\alpha}+V(\phi)\right)
\left(\rho+2p-\frac{3}{2}\phi_{;\,a}\phi^{;\,a}-V(\phi)\right)
+\frac{1}{3}{\cal U}\Biggr]
\end{eqnarray}

As a consequence of the form of the bulk energy-momentum tensor
and of $Z_2$ symmetry, it follows \cite{sms2,sms1} that
the brane energy-momentum tensor separately satisfies the
conservation equations (where we  
assume that the scalar field and the matter are non-interacting); i.e.,
\begin{eqnarray}
\dot\rho+3H(\rho+p)=0, \label{matter_conservation}\\
\ddot\phi+3H\dot\phi+\frac{\partial V}{\partial
\phi}=0\label{Klein-Gordon}\,,
\end{eqnarray}
whence  the Bianchi identities on the brane imply that the projected
Weyl tensor obeys the constraint
\begin{equation}
\dot{\cal U}+4H{\cal U}=0 . \label{Udot}
\end{equation}

\subsection{Bianchi IX Models: Setting up the Dynamical System}

The source term (restricted to the brane) is a 
non-interacting mixture of non-tilting perfect fluid 
ordinary matter with $p = (\gamma-1)\rho$,
and a minimally coupled scalar 
field with an exponential potential of the form
$V(\phi)=V_0e^{\kappa k\phi}$
 (where $\phi=\phi(t))$ \cite{La89d,La89a,La89c,La89b}.
The variables are the same as those introduced 
in \cite{WE}, with the addition of
\begin{eqnarray}\label{new_vars}
\tilde\Omega &=&\frac{\kappa^2 \rho}{3D^2},\qquad\qquad
\tilde\Omega_{\Lambda} = \frac{\Lambda}{3D^2},\qquad\qquad
\tilde\Phi=\frac{\kappa^2 V}{3D^2},\nonumber\\
&&\tilde\Psi=\sqrt{\frac{3}{2}}\frac{\kappa\dot\phi}{3D},\qquad\qquad
\tilde \Omega_{\cal U}= \frac{\cal U}{3D^2}
\end{eqnarray}
where $$ D\equiv \sqrt{H^2+\frac{1}{4}(n_1n_2n_3)^{2/3}}.$$
The total equivalent dimensionless energy density 
due to all sources and bulk corrections is
\begin{equation}
\tilde \Omega^{total}\equiv
\frac{\kappa^2\rho^{total}}{3D^2}
=\tilde\Omega
+\tilde\Omega_{\Lambda}
+\tilde\Phi+\tilde\Psi^2
+c^2\tilde\Omega_{\cal U}+\frac{c^2}{4}
D^2\left( \tilde\Omega + \tilde\Phi+\tilde\Psi^2\right)^2,
\end{equation}
where $c^2\equiv {\widetilde \kappa^4}/{\kappa^4}$.

The governing differential equations for ${\bf X}\equiv
[D,\tilde H,\tilde\Sigma_1,\tilde\Sigma_2,\tilde 
 N_1,\tilde N_2,\tilde N_3,
\tilde\Omega,\tilde\Omega_{\Lambda},\tilde\Psi,\tilde\Phi,\tilde
\Omega_{\cal U}] $
are as follows \cite{HCH}: 
\begin{eqnarray}
         D'        &=&-(1+\tilde q)\tilde H D \label{D_prime}\\
  \tilde H '       &=& \tilde q (\tilde H^2-1) \label{H_prime}\\
  \tilde \Sigma_+' &=& (\tilde q-2)\tilde H\tilde\Sigma_+-\tilde S_+
\label{Sigma+_prime}\\
  \tilde \Sigma_-' &=& (\tilde q-2)\tilde H\tilde\Sigma_--\tilde S_-
\label{Sigma-_prime}\\
  \tilde N_1 '     &=&  \tilde N_1(\tilde H\tilde q -4\tilde\Sigma_+)
\label{N1_prime}\\
  \tilde N_2 '     &=&  \tilde N_2(\tilde H\tilde q
                        +2\tilde\Sigma_++2\sqrt{3}\tilde\Sigma_-)
\label{N2_prime}\\
  \tilde N_3 '     &=&  \tilde N_3(\tilde H\tilde q
                        +2\tilde\Sigma_+-2\sqrt{3}\tilde\Sigma_-)
\label{N3_prime}\\
  \tilde\Omega'    &=& \tilde H\tilde\Omega\left(2(\tilde q
+1)-3\gamma\right)
                    \label{Omega_prime} \\
  \tilde\Omega_{\Lambda}'    &=& 2\tilde
H\tilde\Omega_{\Lambda}\left(\tilde q +1\right) \label{OmegaL_prime}\\
  \tilde\Psi'      &=&(\tilde q-2)\tilde
H\tilde\Psi-\frac{\sqrt{6}}{2}k\tilde\Phi
                    \label{Psi_prime}\\
  \tilde\Phi'      &=&2\tilde\Phi\left((1+\tilde q)\tilde H
                         +\frac{\sqrt{6}}{2}k\tilde\Psi\right)
\label{Phi_prime}  \\
  \tilde\Omega_{\cal U} ' &=& 2\tilde H\tilde\Omega_{\cal U}(\tilde q -1)
                    \label{Omega(U)_prime}
\end{eqnarray}
where $\Sigma_+,\Sigma_-$ are the shear variables,  $N_1,N_2,N_3$
are the curvature variables
(relative to a group-invariant orthonormal frame),
and a logarithmic (dimensionless) time variable, $\tau$, has been defined.
The quantity $\tilde q$ is the deceleration parameter and $\tilde S_+$
and $\tilde S_-$ are curvature terms that are defined by the following
expressions:
\begin{eqnarray}
\tilde q &\equiv&
2\tilde \Sigma_+^2
+2\tilde\Sigma_-^2
+ \frac{(3\gamma-2)}{2}\tilde\Omega
-\tilde\Omega_{\Lambda}
+2\tilde\Psi^2
-\tilde\Phi  \nonumber\\
&&+c^2\tilde\Omega_{\cal U}
 +\frac{c^2}{4}D^2\Biggl[\left( \tilde\Omega +
\tilde\Psi^2+\tilde\Phi\right)
                        \left((3\gamma-1)\tilde\Omega +
5\tilde\Psi^2-\tilde\Phi\right)
                 \Biggr] \label{qdef}\\
{\tilde{S}_{+}}&\equiv&{\displaystyle \frac {1}{6}}\, \left( \! \,
        {\tilde{N}_{2}} - {\tilde{N}_{3}}\, \!  \right) ^{2}
        - {\displaystyle \frac {1}{6}}\,{\tilde{N}_{1}}\,
        \left( \! \,2\,{\tilde {N}_{1}} - {\tilde{N}_{2}} -
        {\tilde{N}_{3}}\, \!  \right) \label{S+def}\\
{\tilde{S}_{-}}&\equiv& \frac{1}{6}\,\sqrt {3}\, \left( \!
\,{\tilde{N}_{2}} -
         {\tilde{N}_{3}}\, \!  \right) \, \left( \! \,
         - {\tilde{N}_{1}} + {\tilde{N}_{2}} + {\tilde{N}_{3}}\, \!
\right) \label{S-def}
\end{eqnarray}
In addition, there are two constraint equations that must also be
satisfied:
\begin{eqnarray}
G_1({\bf X})&\equiv& \tilde H^2 +\frac{1}{4}(\tilde N_1\tilde N_2 \tilde
N_3)^{2/3} -1=0 \label{constraint1}\\ G_2({\bf X})&\equiv&1- \tilde
\Sigma_+^2-\tilde
\Sigma_-^2-\tilde\Omega^{total} -\tilde V=0\label{constraint22}
\end{eqnarray}
where$$\tilde V =\frac{1}{12}\left[({\tilde N_1}^2+{\tilde N_2}^2+{\tilde
N_3}^2-2(\tilde
N_1\tilde N_2 + \tilde N_1\tilde N_3+\tilde N_2\tilde N_3)+3(\tilde
N_1\tilde N_2\tilde
N_3)^{2/3}\right].$$ 
Equation (\ref{constraint1}) follows from the definition of $D$, and
equation
(\ref{constraint22}) is the generalized Friedmann equation.
The resulting Bianchi type IX brane-world equations are now closed and
suitable for a qualitative analysis using techniques from
dynamical systems theory.  
The system of equations  ${\bf X}' =
{\bf F}({\bf X})$ are subject to the two constraint equations $G_{1}({\bf
X})=0$ and $G_{2}({\bf
X})=0$.  These constraint equations essentially restrict the dynamics of
the dynamical system
${\bf X}' = {\bf F}({\bf X})$ to lower dimensional surfaces 
in the state space.  In
principle, these constraint equations may be used to eliminate two of the
twelve variables
provided the constraints are not singular.

\subsubsection{Comments}

Since the dynamical system is invariant under the transformation $\tilde
\Phi
\to -\tilde\Phi$ we can restrict our state space to $D\geq 0$.
Note that the dynamical system is also invariant under the transformation
$(\tilde \Sigma_+,\tilde \Sigma_-,\tilde N_1, \tilde N_2, \tilde N_3)\to
(-\frac{1}{2}\tilde\Sigma_+-\frac{\sqrt{3}}{2}\tilde\Sigma_-,
\frac{\sqrt{3}}{2}\tilde \Sigma_+-\frac{1}{2}\tilde \Sigma_-,\tilde N_2,
\tilde N_3,
\tilde N_1)$.
This symmetry implies that any equilibrium point with a non-zero $\tilde
\Sigma_{\pm}$ term, will have two equivalent copies of that point located
at positions that are rotated through an angle of $2\pi/3$ and centered
along a different axis of the $\tilde N_{\alpha}$.

If we assume the weak energy condition for a perfect fluid 
(i.e., $\rho\geq 0$), then we 
restrict the state space ${\cal S}$ to the set  
$\tilde \Omega\geq 0$.  Since we are investigating
the behaviour of the Bianchi type IX brane-world models, we can restrict the state
space  
to $\tilde N_\alpha\geq 0$ without loss of 
generality.  
There are six matter 
invariant sets in addition to the
invariant sets associated with the geometry of the spacetime \cite{HCH}.
The evolution equation (\ref{Omega(U)_prime})  
for $\tilde\Omega_{\cal U}$ 
 implies that the
surface $\tilde\Omega_{\cal U}=0$ divides the 
state space into three distinct regions, ${\cal
U}^{+}=\{{\bf X}\in {\cal S}|\tilde\Omega_{\cal U} > 0\}$, 
${\cal U}^{0}=\{{\bf X}\in {\cal S}|\tilde\Omega_{\cal U} = 0\}$, 
and ${\cal U}^{-}=\{{\bf X}\in {\cal S}|\tilde\Omega_{\cal U} <
0\}$.

From (\ref{constraint1}) we have that $-1\leq \tilde H \leq 1$ and $\tilde
N_1 \tilde N_2 \tilde N_3\leq 8$.   Furthermore, in the invariant sets
${\cal U}^+$ and ${\cal U}^0$, using equation (\ref{constraint2}), it can
be shown that
$$0\leq \tilde \Sigma_+^2,\tilde
\Sigma_-^2,\tilde\Omega,\tilde\Omega_{\Lambda},\tilde\Psi^2, \tilde \Phi
,\tilde V\leq 1.$$  
However, knowing that $0\leq \tilde V\leq 1$ and
$0\leq \tilde N_1 \tilde N_2 \tilde N_3\leq 8$ is not sufficient to place
any bounds on the $\tilde N_\alpha$'s or $D$.  Furthermore, in the
invariant set ${\cal U}^-$ we cannot place upper bounds on any of the
variables without some redefinition of the dimensionless variables 
(\ref{new_vars}).
It is possible to show that the function
\begin{equation}
W=(\tilde H^2-1)^{[-2\alpha-3\gamma \beta]}\,
\tilde\Omega_\Lambda^{[\alpha+(3\gamma-2)\beta]}\,\tilde\Omega^{2\beta}\,\tilde\Omega_{\cal
U}^\alpha
\end{equation}
(where $\alpha,\beta$ are arbitrary parameters)
is a first integral of the dynamical system; that is, $W'=0$
\cite{HCH}.
In the invariant set ${\cal U}^+ \cup {\cal U}^0$, it 
is possible to show that $\tilde q\geq-1$, which 
implies that $\tilde \Omega_\Lambda \rightarrow 0$ 
as $\tau \rightarrow -\infty$ for those models that 
expand for all time.  Using 
 invariants, we find 
that $\tilde H^2\rightarrow 1$ as $\tau \rightarrow \infty$,
so that for ever-expanding 
models  $\tilde \Omega_{\cal U}\rightarrow 0$ when $\gamma>4/3$.
We also note that the function defined by
\begin{equation}
Z \equiv (N_1 N_2 N_3)^2,
\end{equation}
satisfies the evolution equation
$Z ' = 6qZ$,
and is consequently monotone close to the singularity~\cite{WE}.

As mentioned earlier,
generically the Bianchi type IX models have a cosmological initial
singularity
in which $\rho \rightarrow \infty$, and consequently $\Omega_b$ dominates
as $ t  \rightarrow 0^{+}$.
This can be proven by more rigorous
methods~\cite{ren2,ringstrom1,ringstrom2}. It can also be shown by
qualitative methods
that the  spatial 3-curvature is negligable at the initial singularity.

\subsection{Initial Singularity}

We include a general potential for the scalar
field, $V$, and study what happens as $D\rightarrow\infty$ at the initial 
singularity.
We also include normal matter with $1\leq\gamma < 2$.
Equations (\ref{Psi_prime}) and (\ref{Phi_prime}) become
\begin{eqnarray}
\tilde \Psi ' &=& (\tilde q-2)\tilde H\tilde \Psi - \bar \epsilon
\tilde\Phi \\
\tilde \Phi ' &=& 2(\tilde q +1)\tilde H\tilde \Phi + 2 \bar\epsilon
\tilde\Phi\tilde\Psi
\end{eqnarray}
where $\bar \epsilon$ (related to the usual inflationary slow roll
parameter $\epsilon$) is
defined  by $\bar \epsilon \equiv {3 V_{\phi}}/{2 \kappa V}.$

From the Friedmann equation we have that
\begin{equation}
\tilde{\Omega}_{\lambda} \equiv \frac{c^2}{4}D^2(\tilde{\Omega}+\tilde{\Phi}+\tilde{\Psi}^2)^2\leq 1
\end{equation}
and hence each term $D\tilde{\Omega}$, $D\tilde{\Phi}$, $D\tilde{\Psi}^2$
is bounded (since the left-hand side is the sum of positive definite terms).
Hence, as $D\rightarrow\infty$, $\tilde{\Omega}$, $\tilde{\Psi}^2$,
$\tilde{\Phi}\rightarrow 0$.
It is easy to show that $\tilde{\Omega}_\Lambda$, $\tilde{\Omega}_{\cal U}
\rightarrow 0$ as $D\rightarrow \infty$. Hence (\ref{qdef}) becomes
\begin{equation}
\tilde{q}=2(\tilde{\Sigma}_+^2+\tilde{\Sigma}_-^2)+ {\cal
A}D^2,\label{qdef2}\end{equation}
where
\begin{equation}
{\cal A}\equiv
\frac{c^2}{4}[\tilde{\Omega}+\tilde{\Psi}^2+\tilde{\Phi}]
[(3\gamma-1)\tilde{\Omega}+5\tilde{\Psi}^2-\tilde{\Phi}] \label{A}
\end{equation}
Assuming $\tilde{H}>0$, equations (\ref{D_prime}) 
and (\ref{H_prime}) imply that as $\tau\rightarrow -\infty$ 
for $D\rightarrow \infty$, either $\tilde{q}\rightarrow 0$ 
or $\tilde{q}$ is positive in a neighbourhood of the singularity 
($\tilde{q}$ can oscillate around zero; indeed, it is the 
possible oscillatory nature of the variables that causes potential
problems). 
However, if $\tilde{q}\rightarrow 0$, Eq. (\ref{Omega_prime}) implies
\begin{equation}
\tilde{\Omega}'=\tilde{H}\tilde{\Omega}(2-3\gamma)
\end{equation}
which implies a contradiction for $\gamma>\frac{2}{3}$ (i.e.,
$\tilde{\Omega} \not \rightarrow 0$ as $\tau\rightarrow-\infty$). Hence,
as $\tau\rightarrow-\infty$, $\tilde{q}>0$, where $D\rightarrow\infty$ and
\begin{equation}
\tilde{H}'=-\tilde{q}(1-\tilde{H}^2)
\end{equation}
and hence $\tilde{H}\rightarrow 1$ (assuming positive expansion)
monotonically. [Note that this implies the existence of a monotonic function,
and hence there are no periodic orbits  close to the singularity near the
set $\tilde{H}=1$ --- all orbits approach $\tilde{H}=1$.] In addition,
Eq. (\ref{constraint1}) gives $(\tilde{N}_1\tilde{N}_2\tilde{N}_3)\rightarrow
0$.

From Eqs. (\ref{Omega_prime}), (\ref{OmegaL_prime}), 
(\ref{Psi_prime}) and (\ref{Phi_prime}) we have that 
as $\tau \rightarrow -\infty$
\begin{equation}
\frac{\tilde{\Phi}}{\tilde{\Psi}^2}\rightarrow 0,
\end{equation}
that is, the scalar field becomes effectively massless, and
\begin{equation}
\frac{\tilde{\Omega}}{\tilde{\Psi}^2}\rightarrow 0 \ \ \ \ \text{if} \ \ \
\ \gamma<2.
\end{equation}
This follows directly from the evolution equation in the case of an
exponential scalar field potential, $\bar \epsilon =\sqrt{\frac{3}{2}}k$,
and follows for any physical potential for which $\bar \epsilon$ is
bounded as $\tau \rightarrow-\infty$. Hence we obtain
\begin{equation}
\tilde{q}=2(\tilde{\Sigma}_+^2+\tilde{\Sigma}_-^2)+(3\Gamma-1)C^2
\end{equation}
(where $\Gamma=2$, $C^2=\frac{c^2}{4}D^2\tilde{\Psi}^4$ 
if $\gamma<2$; $\Gamma=2$, $C^2=\frac{c^2}{4}D^2(\tilde{\Omega}+
\tilde{\Psi}^2)^2$ if $\gamma=2$;  $\Gamma=\gamma$ and $C^2=\frac{c^2}{4}
D^2\tilde{\Omega}^2$ if there is no scalar field), where 
$\tilde\Omega_{\lambda}\rightarrow C^2$ as $\tau\rightarrow -\infty$, and
the Friedmann equation becomes
\begin{equation}
1-(\tilde{\Sigma}_+^2+\tilde{\Sigma}_-^2)-C^2
=\frac{1}{12}((\tilde{N}_1^2+\tilde{N}_2^2+\tilde{N}_3^2)-2(\tilde{N}_1\tilde{N}_2+\tilde
N_1\tilde N_3+ \tilde N_2\tilde N_3 ))\geq 0
\end{equation}
From Eqs. (\ref{Omega_prime}) and (\ref{Psi_prime}) we then obtain
\begin{equation}\label{C}
(C^2)'=2C^2\tilde{H}[\tilde{q}-(3\Gamma-1)].
\end{equation}

We still have the possibility of $\tilde{\Sigma}_\pm$ or $\tilde{N}_i$
oscillating as the
singularity is approached (as in the Mixmaster models).  A rigorous proof
that oscillatory
behaviour does not occur can be presented using the techniques 
of Rendall and Ringstr\"{o}m
\cite{ren2,ringstrom1} (using analytic approximations to the
Brane-Einstein
equations for $\tilde{\Sigma}_\pm$, $\tilde{N}_\alpha$; i.e., estimates
for these quantities
that hold uniformly in an open neighbourhood of the initial singularity).
We can then prove
that $\tilde{\Sigma}_\pm \rightarrow 0$ as $\tau\rightarrow -\infty$,
$\tilde\Omega_{\lambda}\rightarrow 1$, and we obtain the BRW source.

To determine the dynamical behaviour close to the initial singularity, 
we need to examine what happens as $D\rightarrow \infty$.  
Let us present a heuristic analysis (and include ordinary matter).
We define a new bounded variable
\begin{equation}
d=\frac{D}{D+1}, \ \ \ \ \ \ \ \ \ \ 0\leq d\leq 1
\end{equation}
and we examine what happens as $d \rightarrow 1$ (assuming $\tilde{H}>0$). 
From above we have that $d \rightarrow 1$
and $ \tilde{H}\rightarrow 1$
monotonically, and hence we need to consider the equilibrium 
points in the set $d=1$.

The analysis depends on whether the quantity ${\cal A}$ defined by
(\ref{A}) that occurs in the expression for
 $\tilde q$ in equation (\ref{qdef2}) is zero or not
 in an open neighbourhood of the singularity.  
 Without loss of generality,
 assuming that ${\cal A}>0$, 
 we define a new time variable by
\[ \dot{f}=\frac{(1-d)^2}{{\cal A}\tilde{H}}f'\]
and the remaining evolution Eqs. (on $d=1$) become
\begin{eqnarray}
\dot{\tilde{\Sigma}}_+&=&\tilde{\Sigma}_+,\quad
\dot{\tilde{\Sigma}}_-=\tilde{\Sigma}_-,\quad
\dot{\tilde{N}}_1=\tilde{N}_1,\quad
\dot{\tilde{N}}_2=\tilde{N}_2,\quad
\dot{\tilde{N}}_3=\tilde{N}_3,\nonumber\\
\dot{\tilde{\Omega}}&=&2\tilde{\Omega},\quad
\dot{\tilde{\Omega}}_\Lambda=2\tilde{\Omega}_\Lambda,\quad
\dot{\tilde{\Psi}}=\tilde{\Psi},\quad
\dot{\tilde{\Phi}}=2\tilde{\Phi},\quad
\dot{\tilde{\Omega}}_{\cal U}=2\tilde{\Omega}_{\cal U}.
\end{eqnarray}
Therefore, the only equilibrium point is
\[\tilde{\Sigma}_+=\tilde{\Sigma}_-=\tilde{N}_1=\tilde{N}_2=\tilde{N}_3=\tilde{\Omega}=\tilde{\Omega}_\Lambda=\tilde{\Psi}=\tilde{\Phi}=\tilde{\Omega}_{\cal
U}=0\]
and this is a local source. This equilibrium point corresponds 
to the BRW solution \cite{Coley:2002a} with
\[\frac{c^2}{4}D^2\tilde{\Psi}^2=1, 
\qquad\qquad D^2\tilde{\Phi}=0 \qquad (D^2\tilde{\Omega}=0).\]
In the next two sections we will consider the two cases of 
a scalar field and a perfect fluid separately. 

Due to the quadratic nature of the brane corrections to the 
energy momentum tensor, a
rich variety of intermediate behaviour is possible in these 
two fluid models.  We note
that a dark radiation density ${\cal U}$ 
together with a scalar field is
similar to previous scaling models \cite{Billyard1,Billyard2}.  
Here the
equivalent equation of state is $p_{\cal U}=
\frac{1}{3}{\cal U}$.  It is known that
the bifurcation value for these scaling models is 
$k^2=3\gamma$, which for $\gamma=4/3$
corresponds to a value $k^2=4$, a bifurcation value 
found in the analysis here.
The intermediate behaviour of these multifluid 
models is extremely
complex \cite{HCH}.  A more complete analysis  
of the Bianchi type II models  (in
which some of the intermediate behaviour is outlined)
was studied in \cite{vandenHoogenAbolghasem:2003,HI}.

\section{Qualitative Analysis}
\subsection{Scalar Field Case}

In the invariant set ${\cal U}^+ \cup {\cal U}^0$, we can show that
$\tilde q\geq-1$.
This implies that the set $\tilde \Omega_{\Lambda}=0$ is the invariant set
containing all of the past asymptotic behaviour for
ever-expanding models in ${\cal U}^+ \cup {\cal U}^0$.  It can be argued
that a scalar
field becomes essentially massless as it evolves backwards in time, hence
it will
dominate the dynamics at early times.  In an
effort to understand the dynamical behaviour at early times we shall 
first assume that there is no perfect fluid (${\tilde \Omega=0}$) and the
4D cosmological constant is zero.

If $\tilde\Omega=\tilde\Omega_\Lambda=0$ (i.e., for a scalar
field only), then the dynamical system simplifies
(effectively the evolution equations for $\tilde\Omega$ and
$\tilde\Omega_\Lambda$
are omitted). The curvature terms
$\tilde S_+$ and $\tilde S_-$ are defined as before,
and the deceleration parameter $\tilde q$ becomes
  \begin{equation}
\tilde q \equiv   2\tilde \Sigma_+^2+2\tilde\Sigma_-^2+2\tilde\Psi^2-
\tilde\Phi
+{c^2}\Biggl[\frac{1}{4}D^2(\tilde\Psi^2+\tilde\Phi)(5\tilde\Psi^2-\tilde\Phi)
+\tilde\Omega_{\cal U}\Biggr].
\end{equation}
The two constraint equations  are given explicitly by
\begin{eqnarray}
     1    &=&   \tilde H^2 +\frac{1}{4}(\tilde N_1\tilde N_2 \tilde
N_3)^{2/3}
                           \label{SFconstraint1}\\
\tilde H^2&=&   \tilde \Sigma_+^2+\tilde \Sigma_-^2+ \tilde\Omega^{total}
+\frac{1}{12}\left((\tilde N_1^2+\tilde N_2^2+\tilde N_3^2)
        -2(\tilde N_1\tilde N_2+\tilde N_2\tilde N_3+\tilde N_1\tilde N_3)
                   \right) \label{SFconstraint2}
\end{eqnarray}
where
\begin{equation}
\tilde \Omega^{total}\equiv\frac{\kappa^2\rho^{total}}{3D^2}=
\tilde\Phi+\tilde\Psi^2+
c^2\Biggl[\frac{1}{4}D^2(\tilde\Phi+\tilde\Psi^2)^2+
\tilde\Omega_{\cal U}\Biggr].
\end{equation}

We first consider the equilibrium points at finite $D$.  
We define $\bar{\bf X} = \{D,\tilde
H,\tilde \Sigma_+,\tilde \Sigma_-,\tilde N_1,\tilde N_2,\tilde N_3,\tilde
\Psi, \tilde\Phi, \tilde
\Omega_{\cal U}\}$ and we restrict the state space accordingly to be $\bar
{\cal S} = \{ {\bf X}\in
{\cal S} | \tilde\Omega=0, \tilde\Omega_{\Lambda}=0\}$.   (Note,
$\epsilon$ is a discrete parameter where $\epsilon =1$ corresponds to
expanding
models, while $\epsilon = -1$ corresponds to contracting models.)
There are a number of saddle points \cite{HCH}.
The zero curvature, power law inflationary Robertson-Walker equilibrium
point $P_{\epsilon}$,
defined by\\
 $\bar {\bf
X}_0=[0,\epsilon,0,0,0,0,0,-\frac{k\epsilon\sqrt{6}}{6},1-\frac{k^2}{6},0]$
is a local sink in the eight dimensional phase space
$\bar {\cal S}$  if $k^2<2$ (if $\epsilon=-1$, this point is a local
source).  When $2<k^2$ this point becomes a saddle.
We note that there are models that recollapse. 
There are no sources for finite values of $D$. 

We considered the equilibrium points at  infinity in the  
previous section.
We now show that (for no perfect fluid and a scalar field with an 
exponential potential) the
BRW solution is always an equilibrium point of the system and a local
stability analysis establishes that it is a local source.
In order to analyze the dynamical system for 
large values of $D$, we define the following 
new variables
$$\tilde \Phi = r^2\sin^2\theta,\qquad \qquad \tilde\Psi=r\cos\theta,
\qquad \tilde\Omega_\lambda=\frac{1}{4}c^2D^2(\tilde\Phi+\tilde\Psi^2)^2=
\frac{1}{4}c^2D^2r^4,$$
so that the variable $D$ is essentially replaced by the bounded 
variable $\tilde \Omega_{\lambda}$ in the set ${\cal U}\cup{\cal U}^0$ 
(hence in this set the only variables that remain unbounded 
are the $\tilde N_\alpha$'s).
The dynamical system becomes
\begin{eqnarray}
  \tilde H '       &=& \tilde q (\tilde H^2-1) \\
  \tilde \Sigma_+' &=& (\tilde q-2)\tilde H\tilde\Sigma_+-\tilde S_+\\
  \tilde \Sigma_-' &=& (\tilde q-2)\tilde H\tilde\Sigma_--\tilde S_-\\
  \tilde N_1 '     &=&  \tilde N_1(\tilde H\tilde q -4\tilde\Sigma_+)\\
  \tilde N_2 '     &=&  \tilde N_2(\tilde H\tilde q
                        +2\tilde\Sigma_++2\sqrt{3}\tilde\Sigma_-)\\
  \tilde N_3 '     &=&  \tilde N_3(\tilde H\tilde q
                        +2\tilde\Sigma_+-2\sqrt{3}\tilde\Sigma_-)\\
  \tilde\Omega_{\cal U} ' &=& 2\tilde H\tilde\Omega_{\cal U}(\tilde q -1)
                    \\
  \tilde\Omega_{\lambda}' &=& 2\tilde H\tilde\Omega_{\lambda}(\tilde q +1
-6\cos^2\theta) \\
  r'    &=& r\tilde H (\tilde q-2+3\sin^2\theta)\\
  \theta' &=& \sin\theta(3\tilde H\cos\theta+\frac{\sqrt{6}}{2}k r)
\end{eqnarray}
where
\begin{equation}
\tilde q \equiv   2\tilde
\Sigma_+^2+2\tilde\Sigma_-^2+r^2(3\cos^2\theta-1)
 +\tilde\Omega_{\lambda}(6\cos^2\theta-1) +c^2\tilde\Omega_{\cal U}
\end{equation}
The two constraint equations become
\begin{eqnarray}
     1    &=&   \tilde H^2 +\frac{1}{4}(\tilde N_1\tilde N_2 
     \tilde N_3)^{2/3} \\
\tilde H^2&=&   \tilde \Sigma_+^2+\tilde \Sigma_-^2+ 
\tilde\Omega_{\lambda}+c^2\tilde\Omega_{\cal U}+r^2 \nonumber \\
&&\qquad +\frac{1}{12}\left((\tilde N_1^2+\tilde N_2^2+\tilde N_3^2)
        -2(\tilde N_1\tilde N_2+\tilde N_2\tilde N_3+\tilde N_1\tilde N_3)
                   \right)
\end{eqnarray}
From the constraint equations, we have 
$0\leq \{\tilde H^2,{\tilde \Sigma_+}^2,{\tilde\Sigma_-}^2,
\tilde\Omega_{\cal U},\tilde\Omega_{\lambda},r\}
\leq 1$, $0\leq \theta\leq \pi$, and 
$\tilde N_1\tilde N_2\tilde N_3 \leq 8$ (assuming 
$\tilde \Omega_{\cal U}\geq 0$).

The BRW solution   is
represented by an equilibrium point in the set $\tilde
\Omega_{\lambda}\not = 0$
($D\to \infty$).  If we let
$\tilde {\bf X}=\{\tilde H,\tilde\Sigma_+,\tilde\Sigma_-,
\tilde N_1,\tilde N_2,\tilde N_3,
\tilde\Omega_{\cal U},\tilde \Omega_{\lambda},r,\theta\}$, then 
the relevant equilibrium points are
$\tilde{\bf
X}_0=[\epsilon,0,0,0,0,0,0,1,0,\frac{\pi}{2}\pm\frac{\pi}{2}]$.
Using the constraint equation to eliminate $\tilde\Omega_{\cal U}$, 
the eigenvalues of the linearization at the points $\theta=0,\pi$ 
are $$\epsilon(10,10,5,5,5,3,3,3,3).$$
A value of $\theta$ that satisfies $\theta'=0$ in a 
neighbourhood of an equilibrium point corresponds to
a tangent plane to an invariant surface passing 
through that equilibrium point.  In the analysis above,
the directions $\theta=0$ and $\theta=\pi$ correspond 
to the $\tilde \Phi =0$ invariant surface (i.e.,
the massless scalar field models).   
We observe that this equilibrium point is a source
that strongly repels away from $\tilde \Psi=0$
(for $\epsilon > 0)$.  That is, 
when traversed in a time reverse direction,
typical orbits would asymptotically approach a 
massless scalar field BRW solution.

In summary, the isotropic BRW solution is a global source
and the past asymptotic behaviour of the Bianchi IX brane
world containing a scalar field
is qualitatively different from what is found in GR.

We note that
assuming $\tilde \Omega_{\cal U}\geq 0$, the future asymptotic 
behaviour of the Bianchi IX brane
world containing a scalar field with an exponential potential is 
not significantly different from
what is found in GR \cite{HCH}.  For $0<k<\sqrt{2}$, the
future asymptotic state of ever-expanding models is characterized 
by the power-law inflationary
solution, and if $k>\sqrt{2}$ there no longer exists any 
equilibrium point representing an expanding
model that is stable to the future.  We therefore conclude that 
if $k>\sqrt{2}$, then the Bianchi IX
models must recollapse.  In \cite{vandenHoogen:1999} it was 
shown that if $k>\sqrt{2}$, then a
collapsing massless scalar field solution is a stable equilibrium 
point.  In the brane-world scenario,
we have that this final end-point is the BRW solution.  
However, if $\tilde \Omega_{\cal U}< 0$, then a
variety of new behaviours are
possible, including possible oscillating cosmologies
\cite{CamposSopuerta:2001b}.

\subsection{Perfect Fluid Case: No Chaos in Brane-World Cosmology}

We shall now discuss the asymptotic dynamical evolution of spatially
homogeneous
brane-world cosmological models with no scalar field close to the initial
singularity in more detail.
Due to the existence of monotone functions, it is known
that there are no periodic or recurrent orbits in orthogonal spatially
homogeneous
Bianchi type IX models in GR.
In particular, there are no sources or sinks
and generically Bianchi
type  IX  models have an oscillatory behaviour with chaotic-like
characteristics, with the matter density becoming dynamically negligible
as one
follows the evolution into the past towards the initial singularity.
Using qualitative techniques,  Ma and Wainwright \cite{WE} have
shown that the orbits of the associated cosmological dynamical system are
negatively asymptotic to a lower two-dimensional attractor. This is the
union
of three ellipsoids in ${\bf R}^5$ consisting of the Kasner ring joined by
Taub
separatrices; the orbits spend most of the time near the  self-similar
Kasner vacuum
equilibrium points. More rigorous global results are possible.
Ringstr\"om has proven that a curvature invariant is unbounded in the
incomplete directions of inextendible null geodesics for generic vacuum
Bianchi models, and has rigorously shown that the Mixmaster attractor
is the past attractor of Bianchi type IX models with an orthogonal perfect
fluid \cite{ringstrom1,ringstrom2}.

All spatially homogeneous Bianchi models in GR expand 
indefinitely except for the type IX models.
Bianchi type IX models
obey the ``closed universe recollapse'' conjecture
 \cite{LW1,LW2}, whereby initially expanding models enter a
contracting phase and recollapse to a future ``Big Crunch''.  All orbits
in the Bianchi IX invariant
sets  are
positively departing; in order to analyse the future asymptotic states of
such models it is necessary to compactify phase-space.  The description of
these models in terms of conventional Hubble- or expansion-normalized
variables is only
valid up to the point of maximum expansion (where $H = 0)$.
 An appropriate set of alternative normalised variables
leading to the compactification of Bianchi IX state space, which 
were suggested in \cite{WE}
and utilized in 
\cite{CG,IbanezvandenHoogenColey,vandenHoogen:1999,Kitada,Uggla-zurMuhlen},
was discussed in the previous section.


Let us discuss the Bianchi  type IX brane-world models 
with a perfect fluid source.  Unlike earlier, and
following \cite{CamposSopuerta:2001a,CamposSopuerta:2001b,WE}, 
we define Hubble-normalized shear variables
$\Sigma_+,\Sigma_-$, curvature variables $N_1,N_2,N_3$  and matter
variables
(relative to a group-invariant orthonormal frame),
and a logarithmic (dimensionless) time variable, $\tau$, defined by $d\tau
= H dt$.
These variables do not lead to a global compact phase space, but they are
bounded close to the singularity \cite{WE}.
The governing evolution  equations for these quantities are then
\begin{eqnarray}
 \Sigma_+' = (q-2)\Sigma_+ -S_+\\
 \Sigma_-' = ( q-2)\Sigma_- - S_-\\
  N_1 '     = ( q -4\Sigma_+) N_1\\
  N_2 '     =  (q+2\Sigma_++2\sqrt{3}\Sigma_-)N_2\\
  N_3 '     =  (q +2\Sigma_+-2\sqrt{3}\Sigma_-)N_3,
\end{eqnarray}
where a prime denotes differentiation with respect to $\tau$, and $K$,
$S_+$ and
$S_-$ are curvature terms that are defined as follows:

\begin{mathletters}
\begin{eqnarray}
{K}&\equiv&  \frac{1}{12}\left((N_1^2+N_2^2+N_3^2)
        -2( N_1N_2+N_2N_3+ N_1N_3)\right)\\
{{S}_{+}}&\equiv&{\displaystyle \frac {1}{6}}\, \left( \! \,
        {{N}_{2}} - {{N}_{3}}\, \!  \right) ^{2}
        - {\displaystyle \frac {1}{6}}\,{{N}_{1}}\,
        \left( \! \,2\,{ {N}_{1}} - {{N}_{2}} -
        {{N}_{3}}\, \!  \right) \\
{{S}_{-}}&\equiv& \frac{1}{6}\,\sqrt {3}\, \left( \!
\,{{N}_{2}} -
         {{N}_{3}}\, \!  \right) \, \left( \! \,
         - {{N}_{1}} + {{N}_{2}} + {{N}_{3}}\, \!  \right)
\end{eqnarray}
\end{mathletters}
The quantity $q$ is the deceleration parameter given by
\begin{equation}
q \equiv   2\Sigma_+^2+2\Sigma_-^2  + \frac{1}{2}\sum \Omega_i +
\frac{3}{2}\sum P_i.\label{q}
\end{equation}
The decoupled Raychaudhuri equation becomes
\begin{equation}
H '     = -(1 + q)H.\label{raych}
\end{equation}
In addition, the generalized Friedmann equation reduces to the constraint
\begin{equation}
1 =  \Sigma_+^2+
\Sigma_-^2+ \sum \Omega_i + K. \label{FRIEM}
\end{equation}
Due to the symmetries in the dynamical system, we can restrict ourselves
to
the set  $ N_1\geq 0$, $N_2\geq 0$, and
$N_3\geq 0$, without loss of generality.  
We again note that these new normalized variables, for which the
evolution equation for
$H$ has decoupled from the remaining equations, are bounded close to the
initial
singularity.

In the above
\begin{equation}
 \sum \Omega_i =  \frac{\kappa^2 \rho^{tot}}{3H^2},~~
 \sum P_i =  \frac{\kappa^2 P^{tot}}{3H^2},  \label{def}
\end{equation}
where the $\Omega_i$ ($i= 1-4$) are given by
\begin{mathletters}
\begin{eqnarray}
 \Omega_1 =  \Omega = \frac{\kappa^2 \rho}{3H^2}; ~ P = (\gamma -1)\Omega
\label{d1}\\
 \Omega_2 =  \Omega_b =  \frac{\kappa^2 \rho^2}{6 \lambda H^2}; ~ P_b =
(2\gamma -1)\Omega_b\label{d2}\\
 \Omega_3 =  \Omega_{\cal U} =  \frac{2 {\cal U}}{\kappa^2 \lambda H^2}; ~
P_{\cal U} = \frac{1}{3}\Omega_{\cal U}\label{d3}\\
 \Omega_4 = \Omega_\Lambda =  \frac{\Lambda }{3 H^2}; ~ P_\Lambda =
-\Omega_\Lambda, \label{d4}
\end{eqnarray}
\end{mathletters}
which satisfy the equations
\begin{equation}
\Omega_i ' = [2(1+q) -3 \Gamma_i] \Omega_i,\label{CE}
\end{equation}
where
\begin{equation}
\Gamma_1=\gamma, \Gamma_2 = \Gamma_b = 2 \gamma, \Gamma_3 = \Gamma_{\cal
U} = \frac{4}{3},
\Gamma_4= \Gamma_\Lambda = 0.
\end{equation}
An extensive analysis of scalar field models has shown that close to the
initial singularity the scalar field must be massless 
\cite{CG,IbanezvandenHoogenColey,Foster,R2000}: it is therefore plausible that
scalar field models can be approximated by a stiff perfect fluid
close to the
initial singularity (particularly regarding questions of stability).

\subsubsection{The isotropic equilibrium point}

We recall that generically the Bianchi type IX models have a cosmological
initial singularity
in which $\rho \rightarrow \infty$, and consequently $\Omega_b$ dominates,
as $ t  \rightarrow 0^{+}$. It can also be shown by qualitative
methods
that the  spatial 3-curvature is negligable at the initial singularity.
Hence, close to the singularity
$\rho^{tot} = \rho_b$, so that  $\sum \Omega_i = \Omega_b$ and
$\sum P_i = P_b$, where $P_b = (2\gamma -1)\Omega_b$.
Consequently, Eq. (\ref{FRIEM})
can be written as
\begin{equation}
\Omega_b = 1-\Sigma_+^2 - \Sigma_-^2 - K,
\end{equation}
which can be used to eliminate $\Omega_b$ from the governing equations. In
particular,
Eq. (\ref{q}) becomes
\begin{equation}
q = 3(1-\gamma)(\Sigma_+^2+\Sigma_-^2) + (3\gamma - 1)(1 - K) , \label{qqq}
\end{equation}
and the governing equations are given by the dynamical system above, where
$q$ is now given by Eq. (\ref{qqq}).

There is an equilibrium point of the dynamical system,
denoted by ${\cal F}_b$, given by $\Sigma_+ = \Sigma_- = 0$, and $ N_1=
N_2 = N_3 = 0$,
which corresponds to the BRW model. 
For the equilibrium point ${\cal F}_b$, the 5 eigenvalues are:
\begin{equation}
3(\gamma - 1), 3(\gamma - 1), (3\gamma -1), (3\gamma -1), (3\gamma -1)
\end{equation}
Hence, for all physically relevant values of $\gamma$ ($\gamma \geq 1$),
${\cal F}_b$ is a source (or past-attractor) in the brane-world scenario
and
the singularity is isotropic.
This contrasts with the situation in GR in which
anisotropy dominates for $\gamma<2$.
This is also consistent with previous analyses of Bianchi type I and V
models where
${\cal F}_b$ is always a source for $\gamma \ge 1$
\cite{CamposSopuerta:2001a,CamposSopuerta:2001b}.

Therefore,  in general the Bianchi IX brane-world models do not have
space--like and
oscillatory  singularities in the past, and consequently brane-world
cosmological models
do not exhibit Mixmaster and chaotic-like behaviour
close to the initial singularity.
We expect this to be a generic feature of more general cosmological models
in the brane-world scenario.
Following arguments similar to those above, it follows that there exists a
singularity of a similar nature in all orthogonal Bianchi brane-world
models. In general,
$\Omega_b$ will again dominate as the initial singularity is approached
into the past
and the qualitative results will follow. In particular, ${\cal F}_b$
will be a local source  in all spatially homogeneous models 
and in general the initial singularity will be isotropic.

\newpage
\section{Inhomogeneous Brane-World Models}

We have found that an isotropic singularity is a
past-attractor in all  orthogonal Bianchi models. 
It is plausible that
the cosmological singularity is typically isotropic in spatially
inhomogeneous models \cite{Coley:2002a}.   We shall study this further
by considering the dynamics
of a class of $G_2$ spatially inhomogeneous 
cosmological models with one spatial
degree of freedom in the brane-world scenario \cite{CHL}. 
We follow the formalism of
\cite{elst}, which utilizes
area expansion normalized scale-invariant
dependent variables, and we use the
timelike area gauge
to discuss the asymptotic evolution of the class of
orthogonally transitive $G_{2}$ cosmologies near the cosmological
initial singularity.  The
initial singularity can be shown to be characterized by
$E_{1}{}^{1} \rightarrow  0$ as
$\tau \rightarrow -\infty$, where $E_{1}{}^{1}$ is a normalized frame
variable \cite{elst}.
In particular, we shall discuss the numerical behaviour of the
local dynamical behaviour of this class of spatially inhomogeneous models
close to the singularity.

All of the bulk corrections may be consolidated into an effective
total energy density, pressure, anisotropic stress and energy
flux as follows. The modified Einstein equations take the
standard Einstein form with a redefined energy-momentum tensor (possibly
tilting now):
\begin{equation}
G_{\mu\nu}=\rho^{\text{tot}} u_\mu
u_\nu+p^{\text{tot}}h_{\mu\nu}+\pi^{\text{tot}}_{\mu\nu}
+2q^{\text{tot}}_{(\mu}u_{\nu)}\, ,   
\label{8}
\end{equation}
where
\begin{eqnarray}
\rho^{\rm tot} &=& \kappa^2\,\rho+{6\kappa^2 \over
\lambda}\left[{1\over 24}\left(2\rho^2 -3
\pi_{\mu\nu}\pi^{\mu\nu}\right) + {1\over
\kappa^4}{\cal U}\right]\label{a}\\
p^{\rm tot} &=&\kappa^2\, p+ {6\kappa^2\over \lambda}\left[{1 \over
24}\left(2\rho^2+4\rho p+
\pi_{\mu\nu}\pi^{\mu\nu}-4q_\mu q^\mu\right) +{1 \over
3}{1\over\kappa^4}{\cal U}\right] \label{b}\\
 \pi^{\rm tot}_{\mu\nu} &=&
\kappa^2\,\pi_{\mu\nu}+ {6\,\kappa^2\over\lambda}\left[{1\over
12}\left(-(\rho+3p)\pi_{\mu\nu}-3
\pi_{\alpha\langle\mu}\pi_{\nu\rangle}{}^\alpha+3q_{\langle\mu}q_
{\nu\rangle}\right) +{1\over \kappa^4}{\cal P}_{\mu\nu}\right]\label{c}\\
 q^{\rm tot}_\mu &=&\kappa^2\,q_\mu+ {6\,\kappa^2\over\lambda}\left[{1
\over 12}\left(2\rho
q_\mu-3\pi_{\mu\nu}q^\nu\right)+ {1\over \,\kappa^4}{\cal Q}_\mu
\right]\label{d}
\end{eqnarray}
We shall assume that ${\cal P}_{\mu\nu}=0$, so that
the evolution of ${\cal E}_{\mu\nu}$ is fully determined \cite{sms2,sms1}.
In the inhomogeneous cosmological models of interest here a non-zero
$\D_\mu{\rho}$ acts as a source for ${\cal Q}_{\mu}$, and
hence ${\cal Q}_{\mu}=0$ is not consistent with an inhomogeneous
energy density and we need to include a dynamical analysis of the
evolution of ${\cal Q}_{\mu}$. We shall make no further assumptions on
the models and include all terms in the numerical analysis.

\subsubsection{Initial singularity}

From the numerical analysis we find that the area expansion rate
increases without bound ($\beta \rightarrow \infty$)  and the normalized
frame variable  vanishes ($E_1{}^1 \rightarrow 0$) as
logarithmic time $\tau \rightarrow -\infty$. Since $\beta \rightarrow
\infty$ (and hence the Hubble rate diverges), there always
exists an initial singularity as $\tau \rightarrow -\infty$. Thus the
singularity is characterized by $E_1{}^1=0$, which allows both dynamical
and numerical results to be obtained.
As noted earlier, since ${\rho} \rightarrow \infty$
as $ \tau \rightarrow -\infty$, it then follows directly from the
conservation laws  that ${\mu}_b \sim {{\rho}_b}^2$
dominates as $ \tau \rightarrow -\infty$ and that all of the other
contributions to the brane energy density are negligible
dynamically as the singularity is approached. Hence close to the singularity the matter
contribution is effectively given by
${p}^{\rm tot}  = (2\gamma -1)  {\rho}^{\rm tot}$.

Models with an  isotropic initial singularity \cite{GW85a,GW85b}
satisfy $\lim_{\tau \rightarrow -\infty}\Omega_b = 1$, $\lim_{\tau
\rightarrow -\infty}v = 0$, $\lim_{\tau \rightarrow
-\infty}\Sigma^{2} = 0$. Their evolution near the cosmological
initial singularity is approximated by the BRW model 
corresponding to the `equilibrium point'
${\cal F}_b$, characterized by
\begin{equation} \Omega_b = 1; \, 0 = E_{1}{}^{1} = \Sigma_+ =\Sigma_- = 
\Sigma_{\times} = N_- = N_{\times} = v = Q_u = \Om=\Omega_u \ . 
\end{equation}
 It was argued above that for all physically relevant values of
$\gamma$ ($\gamma \geq 1$) ${\cal F}_b$ is a source (or
past-attractor),  and hence the singularity is
isotropic,  in all non-tilting  spatially homogeneous  brane-world models
\cite{Coley:2002a}. It was also shown that  ${\cal F}_b$ is a
local source  in  the family of spatially
inhomogeneous `non-tilting' $G_2$ cosmological models 
for $\gamma >1$ in \cite{Coley:2002a}.
Linearizing the evolution equations about ${\cal F}_b$,
using the same spatial reparametrisation as in \cite{elst}
(so that $E_{1}{}^{1} = \exp((3\gamma-1)\tau)$), 
the following general solution of the
linearized equations
was obtained \cite{Coley:2002a}:
\begin{eqnarray}
&&\Sigma_-  =  a_{1}(x)\exp{(3(\gamma-1)\tau)}, ~~ 
\Sigma_{\times} =  a_{2}(x)\exp{(3(\gamma-1)\tau)},\\
&& N_-  =  a_{3}(x)\exp{((3\gamma-1)\tau)},~~ N_{\times}  
=  a_{4}(x)\exp{((3\gamma-1)\tau)}, \\
&& \Omega_b =  1 + a_{5}(x)\exp{(3(\gamma-1)\tau)},
\; 
v = a_{6}(x)\exp{((3\gamma-1)\tau)}
\end{eqnarray}
where the $a_{i}(x)$ are arbitrary functions of the space coordinate.
In addition, it follows from the conservation laws that
$\Omega \rightarrow 0$, $\Omega_{\cal U} \rightarrow 0$  
and $\Omega_\Lambda \rightarrow 0$
as the initial singularity is approached.
The above linearized solution represents a general
solution in the neighbourhood of the initial singularity. 
Hence ${\cal F}_b$ is a local source or
past-attractor
in  this family of spatially inhomogeneous cosmological models for $\gamma
>1$.
In particular, we see that the shear and
curvature asymptote to zero as $\tau \rightarrow -\infty$, and hence the
singularity is isotropic.
We also note that, unlike the analysis of the perfect fluid GR models in
\cite{elst},
the Kasner equilibrium set ${\cal K}$ are found to be saddles in the class
of $G_{2}$
brane-world cosmological models.

\subsection{$G_{2}$ Brane Cosmology}\lb{sec:BG2C}

We can study the nature of the initial singularity in the 
spatially inhomogeneous brane
cosmological models.  An analysis
of the behaviour of spatially inhomogeneous solutions 
to Einstein's equations near an initial
singularity has been made in classical GR 
\cite{inv1,inv2,Garf,IsenKich,KichRen,WIB}.
We consider the class of  $G_{2}$ cosmologies with
two commuting Killing vector fields, which consequenly  admit one degree
of  spatial 
freedom \cite{elst}. The
evolution system of  Einstein field equations are partial differential
equations (PDE) in two independent variables. The {\em
orthonormal frame formalism\/} is utilized \cite{hveugg97,mac73} with the
result that  the governing equations are a first-order autonomous
equation system, and the  dependent variables are scale-invariant.
In particular, we define scale-invariant dependent variables by
normalisation with the area expansion rate of the $G_{2}$--orbits in order
to obtain
the evolution equations as a well-posed system of PDE, ensuring the local existence,
uniqueness and stability of solutions to the Cauchy initial value
problem for ${G}_{2}$ cosmologies. Following \cite{elst} we
assume that the Abelian $G_{2}$ isometry group acts 
orthogonally transitively on spacelike 2-surfaces, and
introduce a group-invariant orthonormal frame $\{\,\p_{a}\,\}$,
with $\p_{2}$ and $\p_{3}$ tangent to the $G_{2}$--orbits. The frame
vector field $\p_{0}$,  which defines a 
future-directed timelike reference congruence,
is orthogonal to the $G_{2}$--orbits and it is hypersurface
orthogonal and hence is orthogonal to a locally defined family of
spacelike 3-surfaces. We then introduce
a set of symmetry-adapted local coordinates $\{\,\tau, \,x, \,y,
\,z\,\}$
\[
\p_{0} = N^{-1}\,\ptl_{\tau} \ , \hsp5
\p_{1} = e_{1}{}^{1}\,\ptl_{x} \ , \hsp5
\p_{A} = e_{A}{}^{B}\,\ptl_{x^{B}} \ , \hsp5
A, \,B = 2, \,3 \ ,
\]
where the coefficients are functions of the independent variables
$\tau$ and $x$ only. The only non-zero {\em frame variables\/} are
thus given by
$N \ , \hsp5 e_{1}{}^{1} \ , \hsp5 e_{A}{}^{B}$,
which yield the following non-zero {\em connection variables:} 
$\alpha, \,\beta, \,a_{1}, \,\np, \,\sigm, \,\nc, \,\sigc,
\,\nm, \,\udot_{1}, \,\Om_{1}$. The variables $\alpha$, $\beta$, $\sigm$
and
$\sigc$ are related to the Hubble volume expansion rate $H$ and
the shear rate $\sig_{\alpha\beta}$ of the timelike reference
congruence $\p_{0}$; in particular, $ \Theta := 3H = \alpha + 2\beta$.
The variables $a_{1}$, $\np$, $\nc$ and $\nm$ describe the
non-zero components of the purely spatial commutation functions
$a^{\alpha}$ and $n_{\alpha\beta}$ \cite{WE}. Finally, the
variable $\udot_{1}$ is the acceleration of the timelike reference
congruence $\p_{0}$, while $\Om_{1}$ represents the rotational
freedom of the spatial frame $\{\,\p_{\alpha}\,\}$ in the $({\bf
e}_{2},{\bf e}_{3})$--plane. Setting $\Om_{1}$ to zero corresponds
to the choice of a Fermi-propagated orthonormal frame
$\{\,\p_{a}\,\}$.  Within the present
framework the dependent variables
\be
\{\,N, \,\udot_{1}, \,\Om_{1}\,\}
\ee
enter the evolution system as freely prescribable {\em gauge
source functions\/}.

Since the $G_{2}$ isometry group acts orthogonally transitively,
the 4-velocity vector field ${\bf u}$ of the perfect fluid is
orthogonal to the $G_{2}$--orbits, and hence has the form
\be
\lb{fluid4vel}
{\bf u} = \Gamma\,(\p_{0}+v\,\p_{1}) \ ,
\ee
where $\Gamma \equiv (1-v^{2})^{-1/2}$.
We assume that the ordinary matter is a perfect fluid with equation of
state
\be
    p_{fl} = (\gamma-1) \rho_{fl}\ .
\ee
In a tilted frame, we have
\be
    T_{\mu\nu} = \rho u_\mu u_\nu + p h_{\mu\nu} + \pi_{\mu\nu}
            + 2 q_{(\mu} u_{\nu)}\ ,
\ee
where
\be
\label{matter}
    \rho = \frac{G_+}{1-v^2} \rho_{fl},\quad
    p = \frac{1}{3} \frac{3(\gamma-1)(1-v^2) + \gamma v^2}{G_+}
        \rho,\quad
    q_\alpha = \frac{\gamma \rho}{G_+} v_\alpha,\quad
    \pi_{\alpha\beta} = \frac{\gamma \rho}{G_+} v_{\la\alpha}
    v_{\beta\ra}
\ee
and $G_+ = 1+(\gamma-1)v^2$.
The basic variables  are $\rho$ and $v^\alpha$. The
quadratic correction matter tensor $S_{\alpha\beta}$ is explicitly 
given in
\cite{CHL};
comparing with (\ref{matter}), we see that it is similar to a
perfect fluid with equation of state
$p_{fl} = (2 \gamma -1) {\rho}_{fl}\ $.
In addition,
\be
{\cal
E}_{\mu\nu} = -{6\kappa^2\over\lambda}(\rho^u u_\mu u_\nu + p^u h_{\mu\nu}
            + \pi^u{}_{\mu\nu} + 2 q^u{}_{(\mu} u_{\nu)}),
\ee
and is similar to a
perfect fluid with equation of state
$p^u{}_{fl} = \frac{1}{3} {\rho^u}_{fl}\ $,
but with non-zero $(q^u{}_{fl})_\mu$ and
$(\pi^u{}_{fl})_{\mu\nu}$. We shall use $\rho^u$ and $q^u{}_\mu$ as the
basic variables, and we can set $(\pi^u{}_{fl})_{\mu\nu}=0$
self-consistently \cite{CHL}.
We define  $q_t$, $q_b$, $q_u$, $\pi_t$, $\pi_b$ and $\pi_u$ as in
\cite{CHL}, so that
the orthonormal frame version of the field equations, 
matter equations and
non-local equations, when specialised to the orthogonally
transitive Abelian $G_{2}$ case with the dependent variables
presented above, can be written down.

\subsubsection{Scale-invariant reduced system of equations}
\lb{subsec:dimlesseq}
We introduce $\beta$-normalised frame, connection and curvature
variables as follows \cite{CHL,elst}:
\bea
\lb{dlframe}
(\,\cn^{-1}, \,E_{1}{}^{1}\,)
& := & (\,N^{-1}, \,e_{1}{}^{1}\,)/\beta \\
(\,\Udot, \,A, \,(1-3\Sigp), \,\Sigm, \,\Nc, \,\Sigc, \,\Nm,
\,\Np, \,R\,)
& := & (\,\udot_{1}, \,a_{1}, \,\alpha, \,\sigm, \,\nc,
\,\sigc, \,\nm, \,\np, \,\Om_{1}\,)/\beta \\
\lb{dlcurv1} (\,\Om, \,\Om_u, \,Q_u \,) & := & (\,\kappa^2\,\rho, \,
{6\kappa^2 \over \lambda}\rho_u,
\,{6\kappa^2 \over \lambda}q_u\,)/(3\,\beta^2)=(\,\kappa^2\,\rho, \,
\frac{6 {\cal U}}{\lambda\kappa^2},
\, \frac{6{\cal Q}}{\lambda\kappa^2}\,)/(3\,\beta^2)\\
  B & := & \sqrt{\frac{6}{\lambda\kappa^2}}\sqrt{3}\beta
\eea
 where $\beta$ is the area
expansion rate of the $G_{2}$--orbits. The new {\em dimensionless
dependent variables\/}  are invariant under arbitrary scale
transformations, and are linked to the $H$-normalised variables
through the relation $H = (1-\Sigp)\,\beta$ 
\cite{WE}. Note that  
$v$ is dimensionless.
It is necessary to introduce the
time and space rates of change of the normalisation factor
$\beta$; we algebraically define  $q$ and $r$ in terms of the remaining
scale-invariant dependent
variables:
\be
\cn^{-1}\,\ptl_{\tau}B = -\,(q+1)\,B, ~~~ 0 = (E_{1}{}^{1}\,\ptl_{x}+r)\,B
\ .
\ee
The various physical quantities are defined by
\bea
\lb{Om_tot}
\Om_{\text{total}}&=& \Om + \Om_b +\Om_u = \Om
+\frac{(1-v^2)}{12\,G_+{}^2}\left(1+(2\,\gam-1)v^2\right)\Om^2\,B^2
+\Om_u \\
P_{\text{total}}&=& P + P_b +P_u
=\frac{3(\gam-1)(1-v^2)+\gam\,v^2}{3\,G_+}\Om +
 \frac{3(2\,\gam-1)(1-v^2)+2\,\gam\,v^2}{36\,G_+{}^2}(1-v^2)\Om^2\,B^2
+\frac{1}{3}\Om_u      \\
Q_{\text{total}}&=& Q + Q_b +Q_u =\frac{\gam\,v}{G_+}\Om +
\frac{\gam\,v\,(1-v^2)}{6\,G_+{}^2}\Om^2\,B^2  + Q_u   \\
\lb{Pi_tot}
\Pi_{\text{total}}&=& \Pi + \Pi_b +\Pi_u
=\frac{2}{3}\frac{\gam\,v^2}{G_+}\Om +
\frac{\gam\,v^2\,(1-v^2)}{9\,G_+{}^2}\Om^2\,B^2  +
\frac{8\,v^2}{3(3+v^2)}\Om_u
\eea

\noindent
$\mbox{The scale-invariant system contains evolution equations for the 
dependent variables }
\{\,E_{1}{}^{1}, \,\Sigp, \,A, \,\Np, \,\Sigm, \,\Nc,$ $\,\Sigc,
\,\Nm, \,\Om_b, \,v,\, \,Q\,\} $,
but not for the gauge source functions
$\{\,\cn, \,\Udot, \,R\,\} \ $,
which are arbitrarily prescribable real-valued functions of the
independent variables $t$ and $x$. 
We can fix the spatial gauge by requiring
\be \lb{sgfix} \Np =
\sqrt{3}\,\Nm \ , \hsp5 R = -\,\sqrt{3}\,\Sigc \ ,\ee
which is
preserved under evolution and under a boost.
We fix the temporal gauge by adapting the evolution of the gauge
source function $\Udot$. The {\em separable area gauge} is
determined by imposing the condition
\be \lb{areagc} 0 = (r-\Udot) \ , \ee
which determines $\Udot$ algebraically through the defining equation for
$r$.
It follows immediately from the gauge fixing condition
that $\cn = f(\tau)$, and we can use the
$\tau$-reparametrisation to set $f(\tau) = \cn_{0}$, a constant, which
we choose to be unity; i.e.,
$\lb{areagn} \cn = \cn_{0} := 1$.
Therefore, $\tau$ is effectively a logarithmic proper time, and the
initial
singularity occurs for $\tau\rightarrow -\infty$.
From the area density $\ca$ of the $G_{2}$--orbits, defined by
$\ca^{-1} = e_{2}{}^{2}\,e_{3}{}^{3} - e_{2}{}^{3}\,e_{3}{}^{2}$,
we obtain the magnitude of the spacetime gradient 
$\lb{agrad} (\nabla_{a}\ca)\,(\nabla^{a}\ca) =
-\,4\beta^{2}\,(1-A^{2})\,\ca^{2} $,
so $\nabla_{a}\ca$ is timelike for $A^{2} < 1$.
For the class of $G_{2}$ cosmologies in which the spacetime
gradient $\nabla_{a}\ca$ is timelike  we can choose the
 gauge condition $ \lb{tlareagc} A = 0$,
which would be achieved by choosing $\p_{0}$ to be parallel to
$\nabla_{a}\ca$. It follows that $\ptl_{x}\ca
= 0$, and we obtain $\ca = \ell_{0}^{2}\,\e^{2\cn_{0}\tau} $, which
is a function of $\tau$ only. This is
refered to as {\em  the timelike area gauge} \cite {elst}.

\subsection{Governing Equations in Timelike Area Gauge}
\lb{subsec:gst}

Let us explicitly give the evolution system in the timelike area
gauge. In the timelike area gauge we can use equation (\ref{sgfix}) to
eliminate the
evolution equation for $\Np$, and the evolution equation for $A$ becomes
trivial. The relevant equations are:

\noindent {\em Evolution system\/}:

\bea \lb{Edlbetadot} \ptl_{\tau}B
& = & -(q+1)B \\
\lb{Edle11dot} \ptl_{\tau}E_{1}{}^{1}
& = & (q+3\Sigp)\,E_{1}{}^{1} \\
\lb{Edlsigpdot}
\ptl_{\tau}\Sigp
& = & (q+3\Sigp - 2)\,\Sigp-2\,(\Nm^2 + \Nc
^2)-\sfrac{1}{3}\,E_1{}^1\,\ptl_{x}r - \sfrac{3}{2}\,\Pi_{\text{total}}\\
\lb{Edlsigmdot} \ptl_{\tau}\Sigm + E_{1}{}^{1}\,\ptl_{x}\Nc & = &
(q+3\Sigp-2)\,\Sigm + 2\sqrt{3}\,\Sigc^{2}
- 2\sqrt{3}\,\Nm^{2} \\
\lb{Edlncdot} \ptl_{\tau}\Nc + E_{1}{}^{1}\,\ptl_{x}\Sigm
& = & (q+3\Sigp)\,\Nc \\
\lb{Edlsigcdot} \ptl_{\tau}\Sigc - E_{1}{}^{1}\,\ptl_{x}\Nm
& = & (q+3\Sigp-2-2\sqrt{3}\Sigm)\,\Sigc - 2\sqrt{3}\,\Nc\,\Nm \\
\lb{Edlnmdot} \ptl_{\tau}\Nm - E_{1}{}^{1}\,\ptl_{x}\Sigc & = &
(q+3\Sigp+2\sqrt{3}\Sigm)\,\Nm + 2\sqrt{3}\,\Sigc\,\Nc \eea
\bea \lb{Edlmudot} \frac{f_{1}}{\Om}\,(\ptl_{\tau} +
\frac{\gam}{G_{+}}\,v\,E_{1}{}^{1}\,\ptl_{x})\,\Om +
f_{2}\,E_{1}{}^{1}\,\ptl_{x}v & = &
2\,\frac{\gam}{G_{+}}\,f_{1}\,[\ \frac{G_{+}}{\gam}\,(q+1)
- \,\frac{1}{2}\,(1-3\Sigp)\,(1+v^2) - \ 1 \,] \\
\lb{Edlvdot} \frac{f_{2}}{f_{1}}\,\Om\,(\ptl_{\tau} -
\frac{f_{3}}{G_{+}G_{-}}\,v\,E_{1}{}^{1}\,\ptl_{x})\,v +
\,f_{2}\,E_{1}{}^{1}\,\ptl_{x}\Om & = &
-\,\frac{f_{2}}{f_{1}G_{-}}\,\Om\,(1-v^{2})\,
[\ \frac{(2-\gam)}{\gam}\,G_+\,r \nonumber\\
& & \hspace{15mm} + \,(2-\gam)\,(1-3\Sigp)\,v - 2\,(\gam-1)\,v  \,]
\\
\lb{Edlqxdot}
        \ptl_\tau \Omega_u + \EEE \ptl_x Q_u
        &=& \left[2(q-1)+12\frac{1+v^2}{3+v^2}\Sigp\right]\Omega_u +vX
\\
\lb{Edlqdot}
        \ptl_\tau Q_u + \EEE \ptl_x (P_u + \Pi_u)
        &=& 2(q+3\Sigp-1)Q_u -\frac{2(1-v^2)}{3+v^2} r \Omega_u +X
\eea
where
\begin{equation}
        \ptl_x (P_u + \Pi_u) =
        \frac{1+3v^2}{3+v^2} \, \ptl_x \Omega_u
                + \frac{16 v}{(3+v^2)^2}\Omega_u \, \ptl_x v \, ,
\end{equation}
and
\bea
        X &\equiv& \frac{\gamma(1-v^2) B^2 \Omega^2}{6 {G_+}^2 G_-}
        \Big[ \gamma v (1-v^2)(1-3\Sigp) +2\gamma v +  2(1-v^2) r \Big]
\nonumber\\
        &&\quad - \frac{\gamma (1-v^2)^2}{6{G_+}^2G_-} B^2 \Omega
        \,\EEE \ptl_x \Omega
        + \frac{\gamma^2v(1-v^2)(3+(\gamma-1)v^2)}{6{G_+}^3G_-}
                B^2 \Omega^2 \,\EEE \ptl_x v
\eea
{\em Constraint equations\/}:
\bea
\lb{Efried} 0 & =&  2\Sigp -
1 + \Sigm^2 + \Sigc^2  +
\Nc^{2} + \Nm^{2}+\Om +
\frac{(1-v^2)}{12\,G_+{}^2}\,\left(1+(2\gam-1)\,v^2\right)\Om^2 B^2 +\,
\Omega_u \\
\lb{constraint2} 0 & = & r + 3\,(\Nc\,\Sigm-\Nm\,\Sigc) +\sfrac{3}{2}
\left(\frac{\gam\,v}{G_+}\Om +
\frac{\gam\,v\,(1-v^2)}{6\,G_+{}^2}\Om^2\,B^2 + Q_u  \right)
\eea
\noindent {\em Defining equations for $q$ and $r$\/}:
\bea \lb{Ehdecel} q & := & \sfrac{1}{2}
+ \sfrac{3}{2}\,(\Sigm^{2}+\Nc^{2}+\Sigc^{2}+\Nm^{2}) \nonumber\\
& & + \sfrac{3}{2}\,\left(
\frac{2\gam-1+v^2}{12\,G_+{}^2}\,(1-v^2)\,\Om^2\,B^2
+\,\frac{1+3\,v^2}{3+v^2}\Om_u + \frac{\gam-1+v^2}{G_+}\Om  \right)  \\
\lb{Eheq3sigp}
q+3\,\Sigp &=& 2 +\sfrac{3}{2}(-\Om_{\text{total}} + P_{\text{total}}
+\Pi_{\text{total}}) \nonumber\\
 &=& 2+
3\,(1-v^2)\,\left(\frac{(\gamma-1)(1-v^2)\Om^2B^2}{12\,G_+{}^2}-\frac{\Om_u}{3+v^2}
-\frac{(2-\gamma)\Om}{2\,G_+}\right)\\
\lb{Ecodac1} r & := & -\frac{E_1{}^1\ptl_{x} B}{B}\ ,
\eea
where
\begin{equation}
\lb{fdef}
G_\pm := 1\pm (\gam-1)\,v^2 \; , 
f_{1} := \frac{(\gam-1)}{\gam G_{-}}\,(1-v^{2})^{2} \; , 
f_{2} := \frac{(\gam-1)}{G_{+}^{2}}\,(1-v^{2})^{2} \; , 
f_{3} := (3\gam-4)-(\gam-1)\,(4-\gam)\,v^{2} \ .
\ee

\subsubsection{Numerical results}\label{sectionnumerical}

We have written the governing equations as a system of evolution
equations subject to the constraint equations (\ref{Efried}) and
(\ref{constraint2}). We can use (\ref{Efried}) to obtain $\Om_u$ and
(\ref{constraint2}) to solve for $Q_u$, and thus treat the governing system
as a system of evolution equations without constraints. 
From the numerical analysis  \cite{CHL} we find that the area expansion 
rate increases
without bound ($\beta \rightarrow \infty$)  and the normalized frame
variable vanishes ($E_1{}^1 \rightarrow 0$) as $\tau
\rightarrow
-\infty$. Since $\beta \rightarrow \infty$ (and hence the Hubble rate
diverges), there always exists an initial singularity. In addition, we
find that $\{\Om, \Sigm,\Nc, \Sigc, \Nm, r\}\rightarrow 0 $ as
$\tau\rightarrow -\infty$ for all $\gam>1$. In the case $\gam >4/3$, the
numerics indicate that $\{v, \Om_u, Q_u, \Sigp\} \rightarrow 0$ (and
$\Om_b
\rightarrow 1$) for {\em all} initial conditions. In the case of
radiation ($\gam=4/3$), the models still isotropize as $\tau\rightarrow
-\infty$, albeit slowly. This degenerate case was investigated in
more detail \cite{CHL}. For $\gam<4/3$, $\{v, \Om_u, Q_u, \Sigp\}$ tend to
constant
but non-zero values as $\tau\rightarrow -\infty$. It is interesting to
note
that $v^2 \not \rightarrow 1$; i.e., the tilt does not tend to an extreme
value.

The numerical results support the fact that all of the models
 have an initial singularity and that for the range of values of
the equation of state parameter $\gamma \geq 4/3$ the singularity is
isotropic. Indeed, the singularity is isotropic for {\em  all} initial
conditions (and not just for models close to ${\cal F}_b$), indicating
that for $\gam >4/3$ this is the global behaviour. The numerical
results also support the exponential decay (to the past) of the
anisotropies. Indeed, we found no evidence that models
do not isotropize to the past for  $\gamma > 4/3$.

In the case $\gam < 4/3$ we find numerically that $\beta^{-1} \rightarrow
0$, so that
there is always an initial singularity as $\tau \rightarrow -\infty$.  In
addition, we
find that $E_1{}^1$, $r$, $\Nm$, $\Nc$, $\Sigm$ $\Sigc$, $\Om$ all vanish
as the
initial singularity is approached.  However, the
initial singularity is not, in general, isotropic. 

$\mbox{We can derive the asymptotic dynamics of the models.
We have that}
\lim_{\tau\rightarrow -\infty} (\EEE, B^{-1}, r,
\Nm, \Nc, \Sigm, \Sigc, \Sigp,$ $\Om, v, \Om_u, Q_u, \Om_b-1)= \mathbf{0}$
from the numerics, and that
$\ptl_x (\EEE, B^{-1}, r,
\Nm, \Nc, \Sigm, \Sigc, \Sigp, \Om, v, \Om_u, Q_u, \Om_b-1)$
are bounded as $\tau \rightarrow -\infty$, and
$V= {\cal O} (f(\tau))$, which implies that $ \ptl_x V =
{\cal O} (f (\tau))$,  when $\gam > 4/3$.
In particular, since $E_1{}^1 \rightarrow 0$ as $\tau \rightarrow
-\infty$,
we can follow the analysis in \cite{lim0306118} and use equations
(\ref{Edlbetadot})-(\ref{Ecodac1}) to obtain the following asymptotic
decay rates \cite{CHL}:

\be
\lb{Brate}
\EEE = {\cal O}(\e^{[(3\gam-1)-\eps]\tau}), 
\quad \Nc = {\cal O}(\e^{[(3\gam-1)-\eps]\tau}), 
\quad    B^{-1} = {\cal O}(\e^{[3\gam-\eps]\tau}),
\ee
\be
    \Nm = {\cal O}(\e^{[(3\gam-1)-\eps]\tau}), \quad
    \Sigc = {\cal O}(\e^{[3(\gam-1)-\eps]\tau}), \quad
    \Sigm = {\cal O}(\e^{[3(\gam-1)-\eps]\tau}),
\ee

\be
    \Om = {\cal O}(\e^{[3\gam-\eps]\tau}), \quad
    r = {\cal O}(\e^{[(3\gam-1)-\eps]\tau})\ ,
\ee
\be
        v = {\cal O}(\e^{[(3\gam-4)-\eps]\tau}), \quad
        \Sigp = {\cal O}(\e^{[3(\gam-1)-\eps]\tau}
                                +\e^{[2(3\gam-4)-\eps]\tau}), \quad
        Q_u =  {\cal O}(\e^{[(3\gam-4)-\eps]\tau}), \quad
        \Om_u =  {\cal O}(\e^{[2(3\gam-4)-\eps]\tau})\ ,
\ee

for any $\epsilon >0$. The
asymptotic dynamics for $\gam=4/3$ are also given in \cite{CHL}.

\subsection{Discussion}

\subsubsection{The invariant set $E_1{}^1=0$}\label{sectionanisotropic}

From numerical experiment we find that as $\tau \rightarrow -\infty$,
$E_1{}^1 \rightarrow 0$ very rapidly,  $\Om,\,\Sigm,\, \Sigc,
\Nc,\Nm,\,r\, \rightarrow 0$ and $B \rightarrow \infty$.
Let us discuss the dynamics in the invariant set $E_1{}^1=0$ at
early times. In order to find the early time behaviour of $\Omega_b$ in
the invariant set $\EEE=0$,  we obtain the evolution equation for
$\Omega_b$:
\be
\partial_\tau \Omega_b = \partial_v \Om_b\,\partial_\tau v + 
 \partial_\Omega \Omega_b\,\partial_\tau \Omega +  
 \partial_B \Omega_b\,\partial_\tau B ,
\ee
where $\partial_\tau B$,  $\partial_\tau \Om$ and $\partial_\tau v$ are
given
separately by (\ref{Edlbetadot}), (\ref{Edlmudot}) and (\ref{Edlvdot})
with $E_1{}^1=0$. \\

Equations (\ref{Edlbetadot})-(\ref{Edlqdot}) imply that
$\Sigm=\Sigc=\Nc=\Nm=\Om=0$ is an invariant subset of the
invariant set $\EEE=0$. In this invariant subset
\be
 -\frac{3}{2}(Q_b+Q_u)=  r \equiv  -\frac{\EEE \ptl_x B}{B}=0
\ee

The dynamics in the invariant subset $\tilde{E} :
E_1{}^1=\Sigm=\Sigc=\Nc=\Nm=\Om=0$
(in which $r=Q_b+Q_u=0$) is given by
\bea
\lb{hedotOb}
\partial_\tau\Om_b &=&
2\left\{\frac{2+6\Om_b+6\Om_u-3\gam(1+\Om_b+\Om_u)}{(1+(2\,\gam-1)\,v^2)\,G_-\,G_+}\right\}\,\gam^2\,v^4\,\Om_b
\nonumber\\
 && +2\,\Om_b\,\left\{(q+1) -
3(1-\Sigp)\,(1+\frac{3(\gam-1)(1-v^2)+\gam\,v^2}{3\,G_+})
+\Sigp\frac{2\gam v^2}{G_+}\right\}  \\
\lb{hedotv}
\partial_\tau v &=&
-\frac{v(1-v^2)}{G_-}\,\left((2-\gam)(1-3\Sigp)-2(\gam-1)\right)  \\
\lb{hedotOu}
\partial_\tau\Om_u &=&
2(q+1)\,(\Om_b+\Om_u)-3(1-\Sigp)\left\{\Om_b+\Om_u+\frac{1}{3}\left(\frac{3(2\gam-1)(1-v^2)+2\gam\,v^2}{(1+(2\,\gam
-1)\,v^2)}\right)\Om_b + \frac{1}{3}\Om_u\right\} \nonumber\\
 & & +3\Sigp\,\left\{\frac{4}{3}\frac{\gam\,v^2}{(1+(2\gam-1)v^2)}\Om_b
+\frac{8\,v^2}{3\,(3+v^2)}\Om_u \right\}
-2\left\{\frac{2+6\Om_b+6\Om_u-3\gam(1+\Om_b+\Om_u)}{(1+(2\,\gam-1)\,v^2)\,G_-\,G_+}\right\}\,\gam^2\,v^4\,\Om_b
\nonumber\\
 && -2\,\Om_b\,\left\{(q+1) -
3(1-\Sigp)\,(1+\frac{3(\gam-1)(1-v^2)+\gam\,v^2}{3\,G_+})
+\Sigp\frac{2\gam v^2}{G_+}\right\}
\eea
where
\bea
\lb{hesigp}
\Sigp &=& \frac{1}{2}\,(1-\Om_b-\Om_u)\\
\lb{heq}
q&=& \frac{1}{2}
+\frac{1}{2}\,\left\{\frac{3(2\gam-1)(1-v^2)+2\gam\,v^2}{(1+(2\,\gam
-1)\,v^2)}\Om_b + \Om_u +\frac{4\gam\,v^2}{(1+(2\gam-1)v^2)}\Om_b
+\frac{8\,v^2}{(3+v^2)}\Om_u\right\}
\eea
and we note 
that $Q_u = -Q_b = -{2\,\gam\, v} \,\Om_b/{(1+(2\,\gam -1)\,v^2)}$.

Since $\Om_b \sim \Om^2\,B^2$, in this invariant set we have a
closed system of ODE (\ref{hedotOb})-(\ref{hedotOu}) in terms of the three
variables $\Om_b$, $v$ and $\Om_u$.
The  equilibrium points are summarized in Table II in \cite{CHL}.
The sources are ${\cal F}_b$
($\Om_b=1,\, v=0,\, \Om_u=0$) for
$\gam > 4/3$  and ${\cal P}_\pm$
($\Om_b=F_1(\gam),\, v=\pm F_2(\gam), \, \Om_u= F_3(\gam))$ for $\gam< 4/3$
(where the
constants $F_1,F_2,F_3$, which depend on $\gamma$, 
are given explicitly in \cite{CHL}).

Therefore, in the invariant subset $\tilde{E}$,
 ${\cal F}_b$ is the global source for $\gam>4/3$, and ${\cal P}_\pm$
($\pm$
 depending on sign of $v$) are anisotropic sources for $\gam<4/3$.  There
is a
 bifurcation at $\gam=4/3$, where ${\cal P}_\pm$ and ${\cal F}_b$
coincide;
 from the dynamics
 in the center manifold it was found \cite{CHL} that 
 the nonhyperbolic equilibrium point is a source
 in the 3D invariant subset.  This is
 consistent with the decay rates given above and with 
 the numerical analysis.
Hence models isotropize
 `slowly' in the radiation case.

In \cite{CHL} it was shown that  
when $\gamma \geq 4/3$, ${\cal F}_b$ 
is also a source for the full state space and that when $\gam<
4/3$, ${\cal P}_\pm$ 
are sources for the full state space.  We
observe that at ${\cal F}_b$
\[q+1=3\gam>0, \, \,\,\,\,\, q+3\Sigp-2=3(\gam-1)>0, \,\,\,\,\,\,
T_1=\frac{3}{2}>0,\,\,\,\,\,\, T_2=3\gam-1>0 ,\]
 and at ${\cal P}_\pm$
\[q+1=\frac{2\gam}{2-\gam}>0, \,\,\,\,\,\,
q+3\Sigp-2=\frac{2(\gam-1)}{(2-\gam)}>0,
 \,\,\,\,\,\,
{T}_1=-\frac{(3\gam-2)\gam}{(2\gam^2-7\gam+4)}>0, \,\,\,\,\,\, 
{T}_2=\frac{2}{2-\gam}>0, \]
where $T_1$ and $ T_2$ are given explicitly in \cite{CHL}.
So near the equilibrium points
${\cal F}_b$ and ${\cal P}_\pm$, $q+1>0$ and $q+3\Sigp-2>0$ and 
thus as $\tau\rightarrow -\infty$, we
have $B\rightarrow +\infty$ and $E_1{}^1\rightarrow 0$.  
Near ${\cal F}_b$ and ${\cal P}_\pm$, we
can neglect the terms with `$E_1{}^1\partial_x$' in the equations
(\ref{Edlbetadot})-(\ref{Edlqxdot}) and treat the PDE as a system of ODE.
Then from the linearization of equations
(\ref{Edlsigmdot})-(\ref{Edlmudot}) near ${\cal F}_b$ and ${\cal P}_\pm$, 
and using the fact that
$q+3\Sigp-2>0$, $T_1>0$ near ${\cal F}_b$ and ${\cal P}_\pm$, 
we find that as $\tau\rightarrow
-\infty$, $\Sigm$, $\Sigc$, $\Nc$, $\Nm$ and $\Om$ (and hence $Q$) will
decrease monotonically
towards zero.  We also observe that near ${\cal F}_b$ and ${\cal P}_\pm$
\bea
\partial_\tau
(Q_b+Q_u)&=&2(q+3\Sigp-1)(Q_b+Q_u)
\left(2(q+3\Sigp-1)-\frac{3}{2}(-\Om_b-\Om_u+ P_b+P_u+\Pi_b+\Pi_u)
\right)(Q_b+Q_u) \nonumber\\ &&
+\left(3(\Nm\Sigc-\Nc\Sigm)-\frac{3}{2}Q\right)(-\Om_b-\Om_u+
P_b+P_u+\Pi_b+\Pi_u)\,.
\eea
Linearizing this evolution equation for $Q_b+Q_u$ near ${\cal F}_b$ and
${\cal P}_\pm$ and using
$T_2 >0$, we conclude that $Q_b+Q_u$, and thus $r$,
 decrease towards zero as 
$\tau\rightarrow -\infty$. Therefore,
 as $\tau\rightarrow -\infty$, all orbits
in the full state space evolve back towards the invariant subset 
$\tilde{E}$.
Because ${\cal F}_b$ (when $\gam\geq
4/3$) and ${\cal
P}_\pm$ (when $\gam<4/3$) are sources in this invariant subset, orbits
 will shadow orbits in the invariant subset and evolve back
towards ${\cal F}_b$
(when $\gam\geq 4/3$) or ${\cal P}_\pm$ (when $\gam<4/3$).  Therefore,
${\cal F}_b$ is a source
for the full state space 
when $\gam\geq 4/3$ and ${\cal P}_\pm$ are sources for the full state
space when $\gam<4/3$.  
The numerical experiments have confirmed that ${\cal F}_b$ is a source in
the full state space
when $\gam \geq 4/3$.  The numerical simulation is consistent
with the decay rates \cite{CHL} and with the analysis given above
that ${\cal F}_b$ is a source for the full state space when $\gam=4/3$.

\subsubsection{Separable invariant sets}

We set $\Sigm=\Sigc=0$. We can then show that
$E_{1}{}^{1} = E(x) F(\tau)$ (i.e., $E_{1}{}^{1}$ is separable). It
follows that
 $\Nm^2 + \Nc ^2 =  F{^2}(\tau)$ and
 $(q+3\Sigp-2) = {\dot{F}}/F$. 
Therefore, setting $\Sigm =\Sigc =B = 0$ and taking 
$\Om(\tau)$, $\Om_u(\tau)$, $v(\tau)$, $F(\tau)$ 
as independent functions of $\tau$, we obtain 
\bea
\Nm   &=& \frac{F(\tau)}{\sqrt{1+e^{2g_2(x)}}}, \\
\Nc   &=& \frac{F(\tau)e^{g_2(x)}}{\sqrt{1+e^{2g_2(x)}}}, \\
\EEE  &=& \frac{F(\tau)e^{g_1(x)}}{\sqrt{1+e^{2g_2(x)}}},
\eea
and
\bea
Q_u   &=& -\frac{\gamma v(\tau) \Om(\tau)}{1+(\gamma-1) v^2(\tau)},\\
\Sigp(\tau) &=&
\frac{1}{2}\left(1-F^2(\tau)-\Om(\tau)-\Om_u(\tau\right),\\
q(\tau)  &=& \frac{1}{2}+\frac{3}{2}F^2(\tau)+
\frac{3}{2}\left(\frac{1+3 v^2(\tau)}{3+v^2(\tau)}\Om_u(\tau)
+\frac{(\gamma-1)+
v^2(\tau)}{1+(\gamma-1)v^2(\tau)}\Om(\tau)\right)
\eea
where $g_2(x)$ satisfies the following ODE:
\bea
\label{odeofg2}
\frac{d g_2(x)}{dx} &=& -2\sqrt{3}
(1+e^{2g_2(x)})e^{-g_1(x)}e^{-g_2(x)}.
\eea
Hence  $\Om(\tau)$, $\Om_u(\tau)$, $v(\tau)$ and $F(\tau)$  
satisfy the following ODE
\bea
\frac{dF(\tau)}{d\tau} &=&(q+3\Sigp)F(\tau)\\
\frac{d\Om(\tau)}{d\tau}&=&\frac{2\gamma\Om(\tau)}{G_+}
\left(\frac{G_+}{\gamma}(q+1)-\frac{1}{2}(1-3\Sigp)(1+v^2)-1 \right)\\
\frac{d v(\tau)}{d\tau} &=& -\frac{(1-v^2)}{G_-}
\left((2-\gamma)(1-3\Sigp)v-2(\gamma-1)v \right)\\
\frac{d\Om_u(\tau)}{d\tau} &=& \frac{2\Om_u}{3+v^2}  
\left((q+6\Sigp-1) v^2 +3(q+2\Sigp-1)  
\right)
\eea

Therefore, the PDE (211)-(225) have separable solutions (i.e.,  
there is a subset of solutions of  the PDE which are
separable).
Because there is no term containing $\partial_x\EEE$ in these 
equations, 
there is no 
differential equation for $g_1(x)$. 
The only equation for $g_2(x)$ and $g_1(x)$ 
is equation (\ref{odeofg2}). Hence we can take any $g_1(x)$, 
and then determine $g_2(x)$ from 
equation (\ref{odeofg2}), provided $g_2(x)$ is a real function for a given
real 
function $g_1(x)$.  We can choose coordinates
to set $ E(x)=1$,
which leads to 
 $e^{g_1(x)}=\sqrt{1+e^{2g_2(x)}}$. Thus  Eq. (\ref{odeofg2}) becomes
\[
\frac{d g_2(x)}{dx} =
-2\sqrt{3}\frac{(1+e^{2g_2(x)})}{\sqrt{1+e^{2g_2(x)}}}\,e^{-g_2(x)}= 
-2\sqrt{3}\sqrt{(1+e^{2g_2(x)})}\,e^{-g_2(x)}
\]
which has real solutions given by 
\[g_2(x)=-\sinh (2\sqrt{3} x+C).\]
Therefore, it is clear that the spatial 
dependence is gauge, and that these models are in fact spatially
homogeneous.  Consequently, we can obtain non-trivial spatially
homogeneous models (that are difficult to describe due to
gauge difficulties).  Other separable cases can be investigated.


\subsubsection{Remarks}
All models have an initial singularity as $\tau\rightarrow -\infty$. In
addition, we find that $\{\Om, \Sigm,\Nc, \Sigc, \Nm, r\}\rightarrow 0 $
as $\tau\rightarrow -\infty$ for all $\gam>1$. In the case $\gam >4/3$,
the
dynamical and numerical analysis indicates that $\{v, \Om_u, Q_u, \Sigp\}
\rightarrow 0$ (and $\Om_b \rightarrow 1$) for {\em all} initial
conditions. In the case of radiation ($\gam=4/3$), the models still
isotropize as $\tau\rightarrow -\infty$, albeit slowly. For $\gam<4/3$,
$\{v, \Om_u, Q_u, \Sigp\}$ tend to constant but non-zero values as
$\tau\rightarrow -\infty$.

In the invariant set $v=0$, all models isotropize to the past (for
$\gam>1$). 
Thus in the spatially homogeneous case with no tilt (with $\EEE=0$),  it
follows that there exists an isotropic
singularity in all orthogonal Bianchi brane-world models in which
$\Omega_b$ dominates as the initial singularity is approached into the
past 
(consistent with the results of \cite{Coley:2002b}).
In particular, ${\cal F}_b$ is a local source and in general the initial
singularity is isotropic.  The  linearized solution representing a general
solution in the neighbourhood of the initial singularity in a class of
$G_2$ models was given in \cite{Coley:2002a}; it was found that ${\cal
F}_b$ is a local source or
past-attractor
in  this family of spatially inhomogeneous cosmological models for $\gamma
>1$.
Let us now assume that $v\not=0$ (the general case). For the $G_2$ models
in the timelike area gauge we recover the orthogonally transitive
{\em tilting} Bianchi type VI$_0$  and VII$_0$ models in the spatially
homogeneous limit
with one tilt variable (note that from above other Bianchi models
can occur in certain limits). There are no sources in the Bianchi type
VI$_0$  
and VII$_0$ models
with tilted perfect fluid in GR; the past attractor is an infinite sequence
of orbits between Kasner points. 
The Bianchi
models have bifurcations at $\gamma = 2/3$ and $4/3$ (e.g., the
dimension of the unstable and stable manifolds of the
equilibrium points change) \cite {Hewitt}. 
We note that this is
consistent with our bifurcation $\gam=4/3$ (with $\gam \rightarrow 2\gam$
in brane-world models).

\newpage
\section{Covariant and Gauge\hs Invariant Linear Perturbations}

We have found that,
unlike standard GR where the generic cosmological 
singularity is anisotropic,
the past attractor in the brane for homogeneous anisotropic
models and a class of inhomogeous models  is a simple BRW model
${\cal F}_b$.
It has been conjectured \cite{Coley:2002a,Coley:2002b} that an isotropic
singularity could be a generic feature of brane cosmological models. 
In \cite {dunsbyetal} it was shown that this conjecture is true, within a
perturbative
approach and in the large\hs scale and high energy regime, 
through a detailed analysis of generic linear
inhomogeneous and anisotropic perturbations \cite{EB2,EB1,Goode}
of this past attractor ${\cal F}_b$ by using the full set
of linear 1+3 covariant propagation and constraint 
equations;
these equations are split into their {\it scalar}, {\it vector}
and {\it tensor} contributions  
which govern the complete perturbation dynamics.

From a dynamical systems point of view the past attractor ${\cal F}_b$ for
spatially homogeneous cosmological brane-world models is a
fixed point in the phase space of these models. This phase space may be
thought of as an invariant submanifold  within a higher dimensional phase
space for more general inhomogeneous  models. The 
linear perturbation analysis can be
thought of as an exploration of the neighbourhood of ${\cal F}_b$ out of
the invariant spatially homogeneous submanifold.
The analysis is restricted to large\hs scales, at a time when
physical scales of perturbations are much larger than the Hubble
radius, $\lambda\gg H^{-1}$. This is not very
restrictive, since this is the case for the non\hs inflationary perfect
fluid models (that are relevant here) 
in which any wavelength $\lambda<H^{-1}$ at a given time becomes much
larger
than $H^{-1}$ at early enough times. Hence the analysis is
 valid for any $\lambda$ as $t\rightarrow 0$.

Let us restrict the analysis to the case of vanishing background
dark
radiation (${\cal U}=0$) \cite{dunsbyetal2}.
We define dimensionless expansion normalised quantities for
the shear $\sigma_{ab}$, the vorticity $\omega_a$, the
electric $E_{ab}$ and magnetic $H_{ab}$ parts of the Weyl
tensor as follows \cite{Goode}:
\begin{equation}
\Sigma_{ab}=\frac{\sigma_{ab}}{H}\;,~~
W_a=\frac{\omega_a}{H}\;,~~
{\cal E}_{ab}=\frac{E_{ab}}{H^2}\;,~~
{\cal H}_{ab}=\frac{H_{ab}}{H^2}\; .
\end{equation}
It is convenient 
to
characterise the vorticity of the fluid flow
by using the
dimensionless variable
\begin{equation}
W^*_a = a\omega_a
\end{equation}

The appropriate dimensionless density and expansion gradients
which describe the {\it scalar} and {\it vector} parts of
density perturbations are given by (see \cite{Maartens1,Maartens2} for
details of definitions)
\bea {\Delta}_a=\frac{a}{\rho} \D_a \rho\;,~~
Z^*_a=\frac{3a}{H}\D_a H\;, \label{eq:grads}
\eea
and for the brane\hs world contributions we define the following
dimensionless gradients describing inhomogeneity in the non\hs
local quantities
\be
U^*_a=\frac{\kappa^2\rho}{H^2}U_a\;,~~
Q^*_a=\frac{\kappa^2 a\rho}{H}Q_a\;,
\ee
where $U_a$ and $Q_a$ are defined in equation (27) of \cite{gm}.
It is also
useful to define a set of auxiliary variables corresponding to the
curls of the standard quantities defined above:
\bea
\bar{W}^*_a&=&\frac{1}{H}\curl{W^*_a}\;,~~ \bar{\Sigma}_{ab}=\frac{1}{H}
\curl{\Sigma_{ab}}\;,   \\
\bar{{\cal E}}_{ab}&=&\frac{1}{H}\curl{\cal E}_{ab} \;,~~
\bar{\cal H}_{ab}=\frac{1}{H}{\cal H}_{ab}\;,\\
\bar{Q}^*_a&=&\frac{1}{H}\curl{Q}^*_a\;.
\eea
Finally, we  use the dimensionless time derivative
$d/d{\tau}=d/d~(ln~a)$ (denoted by a prime) to analyse the past
evolution of the perturbation dynamics.

We use the harmonics defined in~\cite{EB2,EB1} to expand the above
tensors $X_a\;,X_{ab}$ in terms of scalar (S), vector (V) and tensor (T)
harmonics $Q$. This yields a covariant and gauge invariant splitting
of the evolution and constraint equations for the above quantities
into three sets of evolution and constraint equations for scalar, vector
and tensor modes respectively:
\bea
X&=&X^S Q^S \\
X_a&=&k^{-1}X^S Q_{a}^S+X^V Q_{a}^V \\
X_{ab}&=&k^{-2}X^S Q_{ab}^S+k^{-1}X^VQ_{ab}^V+X^T Q_{ab}^T\;.
\eea
The  evolution equations
for {\it scalar perturbations} (suppressing the label S) are given by:
\bea
\Sigma'&=&(3\gamma-2)\Sigma-{\cal E}\;,\\
 {\cal E}'&=&(6\gamma-3){\cal E}-3\gamma\Sigma\;,\\
{\Delta}'&=&(3\gamma-3)\Delta-\gamma{Z^*}\;,\\
{Z^*}'&=&(3\gamma-2){Z^*}-3(3\gamma+1)\Delta -{U^*}\;,\\
{Q^*}'&=&(3\gamma -3){Q^*}-\case{1}/{3}U^*-6\gamma\Delta\;,\\
{U^*}'&=&(6\gamma-4){U^*}\; ,
\eea
and the {\it scalar} constraint equations are:
\bea
2{\Sigma}&=&2Z^*-3{Q^*}\;,\label{eq:Sig}\\
2{\cal E}&=&6\Delta-3{Q^*}+{U^*} \label{eq:divE} \;.
\eea
The vector and tensor perturbations on large\hs scales,
 together with their curls
and constraints, are given in \cite{dunsbyetal}.

Solutions to these three sets of equations are 
presented in Table I in \cite{dunsbyetal}. 
This work was extended in \cite{dunsbyetal2}
 by providing a complete analysis of
{\it scalar}, {\it vector} and {\it tensor} perturbations for
different stages of the brane-world evolution.
We conclude that ${\cal F}_b$ is
stable in the past (as $t\rightarrow 0$ or $\tau \to - \infty$) to generic
inhomogeneous
and anisotropic perturbations provided the matter is described by
a non\hs inflationary perfect fluid with $\gamma$\hs law equation of state
parameter satisfying $\gamma > 4/3$ (and we use the large\hs scale
approximation). In particular, unlike in GR, both the 
shear and density gradient tend to
zero at early times if $\gamma>4/3$. Thus the high\hs energy models
isotropize in the past for realistic equations of state when we
include generic linear inhomogeneous and anisotropic perturbations. 

We have discussed here only the case of vanishing background Weyl energy
density, ${\cal U}=0$. This assumption considerably simplifies the
analysis,
but the results remain true for ${\cal U}\not =0$.
Indeed, when ${\cal U}\not=0$ ${\cal F}_b$ still remains a past attractor. 
In other words, our analysis is restricted to
the invariant submanifold ${\cal U}=0$ of the larger phase space
with ${\cal U}\not =0$, but this submanifold is asymptotically stable
against ${\cal U}\not =0$ perturbations.
This is consistent with the dynamical analysis of the 
class of spatially inhomogeneous
$G_{2}$ cosmological models in the brane-world scenario presented 
earlier.

\newpage
\section{Conclusion}

The governing field equations induced on the brane, using
the Gauss-Codazzi equations, matching conditions and $Z_2$
symmetry were given \cite{Maartens1,Maartens2,sms2,sms1}.  
We are particularly interested in the physical case of a
perfect fluid with $p=(\gamma-1)\rho$ satisfying $\gamma \ge
4/3$, where $\gamma = 4/3$ corresponds to radiation and $\gamma
=2$ corresponds to a massless scalar field close to the initial
singularity. The dynamical equations on the 3-brane differ from
the GR equations  in that there are nonlocal
effects from the free gravitational field in the bulk, transmitted
via the projection ${\cal E}_{\mu\nu}$ of the bulk Weyl tensor,
and  the local quadratic energy-momentum corrections, which are
significant at very high energies and particularly close to the
initial singularity.
As a consequence of the form of the bulk energy-momentum tensor
and of $Z_2$ symmetry, it follows \cite{sms2,sms1} that the brane
energy-momentum tensor separately satisfies the conservation
equations; i.e., $\nabla^\nu T_{\mu\nu}=0$. Consequently, the
Bianchi identities on the brane imply that the projected Weyl
tensor obeys the non-local constraint $\nabla^\mu{\cal
E}_{\mu\nu}=\widetilde{\kappa}^4\nabla^\mu S_{\mu\nu}$. The
evolution of the anisotropic stress part is {\em not} determined
on the brane; the correction terms must be consistently derived
from the higher-dimensional equations. We
have  assumed here that ${\cal P}_{\mu\nu}=0$.

The asymptotic dynamical evolution of spatially homogeneous
brane-world cosmological models close to the initial singularity,
where the energy density of the matter is larger than the brane
tension and  the behaviour deviates significantly from the
classical GR  case, has been studied in detail. 
It has been found that for perfect fluid models
with a linear barotropic $\gamma$-law equation of state an
isotropic singularity is a past-attractor in all
orthogonal Bianchi models and is also a local past-attractor in a
class of {\em inhomogeneous} brane-world models (for $\gam \ge
\frac{4}{3}$).
The dynamics of
the class of inhomogeneous brane-world models considered above, together
with the
BKL conjectures 
\cite{bkl1,bkl2,bkl3}, indicate that a wide class of inhomogeneous 
brane-world models will
have an isotropic initial singularity.

More detailed information has been obtained from a study of
the dynamics of a class of {\em spatially
inhomogeneous $G_{2}$} cosmological models with one spatial degree
of freedom in the brane-world scenario 
near the cosmological initial singularity \cite{CHL}. 
 Since the normalized frame variables
were found to vanish asymptotically,
the singularity is characterized by the fact that spatial
derivatives are dynamically negligible.
The local dynamical behaviour of this class of spatially
inhomogeneous models close to the singularity was then studied
{\em numerically} \cite{CHL}. It was found that the area expansion
rate increases without bound (and hence
$H \rightarrow \infty$) as $\tau \rightarrow -\infty$,
so that there always exists an initial singularity. For $\gam
>4/3$, the numerics indicate that as
$\tau\rightarrow -\infty$ 
the singularity is isotropic for {\it all} initial
conditions (indicating that ${\cal F}_b$ is a
global past-attractor).
In the case of radiation ($\gam=4/3$), the
models were still found to isotropize as $\tau\rightarrow -\infty$,
albeit slowly. 
From the numerical analysis we conclude that there is an initial
isotropic singularity  in all of these $G_2$ spatially
inhomogeneous brane cosmologies for a range of parameter values
which include the physically important cases of radiation and a
scalar field source. The numerical results are supported by a
qualitative dynamical analysis and a detailed calculation of the
past asymptotic decay rates \cite{CHL}.

From numerical and dynamical considerations we concluded
 that the Hubble expansion rate increases without bound
as logarithmic time $\tau \rightarrow
-\infty$, and hence there always exists an initial
singularity in $G_2$ cosmologies. In addition, the normalized 
frame variable vanishes  as $\tau \rightarrow -\infty$. These conditions allow
us to calculate the exponential decay rates close to the initial
singularity.  The constrained evolution system of equations for
{\em general} inhomogeneous (denoted $G_0$) brane-world models was
given in \cite{lim0306118} in terms of the variables
$\Omega_b,{\bf X}$ (in the separable volume gauge using
Hubble-normalized quantities), where {\bf X} represents the
independent variables corresponding to the shear and curvature and
additional matter terms,  and $\Omega_b \equiv \mu_b/H^2$, where
essentially ${\mu}_b \sim {\rho}^2$. The exponential decay rates
in the case $\gam>4/3$ were calculated 
\cite{CHL}, and it  was found that as $\tau \rightarrow -\infty$ the
variables $(\Omega_b-1),{\bf X}$ have decay rates that depend
linearly on $\{\e^{(3\gam-1)\tau}, \e^{3\gam \tau}, \e^{3(\gam-1)\tau},
\e^{2(3\gam-4)\tau}, \e^{2\tau}, \e^{(3\gam-4)\tau},
\e^{2(3\gam-2)\tau}\}$.
There is also evidence of isotropization (albeit slowly) in the
degenerate case $\gam=4/3$. This supports the conjecture that, in
general, brane-world cosmologies have an isotropic singularity into
the past for $\gam \ge 4/3$.

The evolution of models with an isotropic cosmological
initial singularity is approximated by the flat
self-similar, spatially homogeneous and isotropic
brane-world BRW model 
\cite{BinDefLan:2000a,BinDefLan:2000b,BinDefLan:2000c},
corresponding to the `equilibrium state' ${\cal F}_b$,
characterized by $\Omega_b = 1$, $\bf{X}= \bf{0}$.
It follows immediately that the total energy density ${\rho}
\rightarrow \infty$ as $ \tau \rightarrow -\infty$ \cite{Coley:2002a}, so
that ${\mu}_b \sim {\rho}^2$ dominates as $ \tau \rightarrow
-\infty$, and that all of the other contributions to the brane
energy density are negligible dynamically as the singularity is
approached. Hence
close to the singularity the matter contribution is given by
${\rho}^{\rm tot} = {1\over 2\lambda}{\rho}^2 \equiv {\mu}_b ;
~~{p}^{\rm tot} = {1\over 2\lambda}({\rho}^2+2{\rho} {p}) =
(2\gamma -1){\rho}^{\rm tot}$,
so that the effective equation of state at high densities is
$(2\gamma -1)$, which is greater than unity in cases of physical
interest. For clarification, let us describe the dynamics
heuristically. As in GR, the anisotropies (for example) grow into
the past as the initial singularity is approached. The energy
density also diverges in GR, but at a slower rate than the
anisotropies and consequently the initial singularity is generically
anisotropic. However, in brane-world models, at higher energies close
to the initial singularity the dominant term in the total energy
density is effectively $\rho^2$, which now diverges much more
rapidly than the anisotropies, so that the ratio (i.e., the
normalised anisotropies) actually decay, with the result that the
anisotropies (and all other contributions to the total energy
density) are negligible dynamically as the initial singularity is
approached.

These results are further supported by an
analysis of {\em linear perturbations} of the BRW 
model  using the covariant and gauge invariant
approach \cite{dunsbyetal,dunsbyetal2}. In particular, a detailed analysis
of
generic linear inhomogeneous and anisotropic perturbations
\cite{EB2,EB1} of the past attractor ${\cal F}_b$ was carried out by
deriving a full set of linear 1+3 covariant  propagation and
constraint equations for this background. 
Solutions to
the set of perturbation equations were presented
\cite{dunsbyetal}, and it was concluded that ${\cal F}_b$ is stable in
the past to generic inhomogeneous and anisotropic perturbations
for physically relevant values of $\gamma$.
In addition, it follows
immediately that the expansion normalised shear vanishes as the
initial singularity is approached, and isotropization occurs.

We have argued that, unlike the situation in GR, it is plausible that
typically the initial singularity is isotropic in brane-world
cosmology. Such a `quiescent' cosmology~\cite{Barrow}, in which
the universe began in a highly regular state but subsequently
evolved towards irregularity, might offer an explanation of why
our Universe might have began its evolution in such a smooth
manner and may provide a realisation of Penrose's ideas on
gravitational entropy  and the second law of thermodynamics in
cosmology~\cite{Penrose79}. More importantly, it is therefore
possible that a quiescent cosmological period occuring in brane
cosmology provides a physical scenario in which the universe
starts off smooth and that naturally gives rise to the conditions
for inflation to subsequently take place. Cosmological
observations indicate that we live in a Universe which is
remarkably uniform on very large scales. However, the spatial
homogeneity and isotropy of the Universe is difficult to explain
within the standard GR framework since, in the presence of matter,
the class of solutions to the Einstein equations which isotropize
is essentially a set of measure zero. In the inflationary
scenario, we live in an isotropic region of a potentially highly
irregular universe as the result of an accelerated expansion phase in the
early universe thereby solving many of the problems of cosmology.
Thus this scenario can successfully generate a spatially
homogeneous and isotropic universe from initial conditions which,
in the absence of inflation, would have resulted in a universe far
removed from the one we live in today. However, still only a
restricted set of initial data will lead to smooth enough
conditions for the onset of inflation.

Let us discuss this in a little more detail. Although inflation
gives a natural solution to the horizon problem of the big-bang
universe, it requires homogeneous initial conditions over the
super-horizon scale; i.e., in general inflation itself requires
certain improbable initial conditions. When inflation begins to
act, the Universe must already be smooth on a scale of at least
$10^5$ times the Planck scale. Therefore, we cannot say that it is
a solution of the horizon problem, although it reduces the problem
by many orders of magnitude.  Many people have investigated how
initial inhomogeneity affects the onset of inflation, and it has
been found that including spatial inhomogeneities accentuates the
difference between models like new inflation and those like
chaotic inflation.  Goldwirth and Piran \cite{GP1,GP2,GP3}, who solved the
full Einstein equations for a spherically symmetric  spacetime,
found that {\em small-field} inflation models of the type of {\em
new inflation} are so sensitive to initial inhomogeneity that they
require homogeneity over a region of several horizon sizes. {\em
Large-field} inflation models such as {\em chaotic inflation} are
not as severely affected by initial inhomogeneity but require a
sufficiently high average value of the scalar field over a region
of several horizon sizes 
\cite{Brandenberger3,Brandenberger1,Brandenberger2}. Therefore,
inhomogeneities further reduce the measure of initial conditions
yielding new inflation, whereas the inhomogeneities have
sufficient time to redshift in chaotic inflation, letting the zero
mode of the field eventually drive successful inflation.
In conclusion, although inflation is a possible causal mechanism for
homogenization and isotropization, there is a fundemental problem
in that the initial conditions must be sufficiently smooth in
order for inflation to subsequently take place \cite{Coley:2002b}. We
have found that  an isotropic singularity in brane-world cosmology
might provide for  the
 necessary sufficiently smooth initial
conditions to remedy this problem.

The earliest investigations of the initial singularity used only
isotropic fluids as a source of matter. However, the
study of the behaviour of spatially homogeneous brane-worlds close
to the initial singularity in the presence of both local and
nonlocal stresses indicates that for physically relevant values of
the equation of state parameter there exist two local past
attractors for these brane-worlds, one isotropic and one
anisotropic (although the anisotropic models are likely unphysical). 
In particular, Barrow and Hervik \cite{BH}
studied a class of Bianchi type I brane-world
models with a pure magnetic field and a perfect fluid with a linear
barotropic $\gamma$-law equation of state.
They found that when $\gamma \ge \frac{4}{3}$,  the equilibrium point
${\cal F}_b$
is again a local source (past-attractor), but that there exists a second
equilibrium point,
denoted $PH_1$, which corresponds to a new brane-world solution with a
non-trivial
magnetic field, which is also a local source. When $\gamma < \frac{4}{3}$,
$PH_1$
is the only local source (the equilibrium point ${\cal F}_b$ is unstable
to magnetic field
perturbations and hence is a saddle). This
was generalized in \cite{hervik}, in which it was shown that for a
class of spatially homogeneous brane-worlds with anisotropic
stresses, both local and nonlocal, the brane-worlds could have
either an isotropic singularity or an anisotropic singularity for
$\gam>4/3$. Finally, we note that Bianchi type I and IX models have also
been studied in
Ho\v{r}ava-Witten cosmology in  which the fifth coordinate is a $S^1/Z_2$
orbifold while the remaining
six dimensions have already been compactified on a Calabi-Yau space, and
it was argued \cite{Dabr}
that there is no chaotic behaviour in such Bianchi IX  Ho\v{r}ava-Witten
cosmologies (also see \cite{BDDH2}).

It would be of interest to study general inhomogeneous ($G_0$) brane-world 
models further. The exponential decay rates in the case $\gam>4/3$
were
calculated in \cite{CHL}. The decay rates are essentially the same as
in the $G_2$ case presented earlier. This
supports the possibility that in general brane-world cosmologies have an
isotropic singularity. We hope to further study $G_0$ brane-world models
numerically
in the future.
The dynamics of
inhomogeneous brane-world models, together with the
BKL conjectures \cite{bkl1,bkl2,bkl3}, perhaps indicate that a wide class of 
inhomogeneous brane-world models will
have an isotropic initial singularity.

The results might also be applicable in a number of 
more general situations. For example, in theories with field 
equations with higher-order curvature
 corrections (e.g., the 4D brane-world in the 
 case of a Gauss-Bonnet term in the  bulk spacetime 
 \cite{MT,Paza}), 
 the results concerning stability are expected to be 
 essentially valid in regimes in which 
 the curvature is negligible (e.g., close to the initial 
 singularity).
The single-brane cosmological model can be generalized to include
stresses other than $\Lambda$ in the bulk.
The simplest example arises from considering a charged bulk
black hole, leading to the Reissner-Nordstr\"om AdS$_5$ bulk
metric~\cite{bv}. The effect of the black hole charge on the brane
arises via the junction conditions and leads to a modified
Friedmann equation, which can cause a bounce
at high energies.
Another example is provided by the Vaidya-AdS$_5$ metric,
in which a RW brane moves in a 4-isotropic bulk (which is
not static), with either a radiating bulk black hole or a
radiating brane ~\cite{ckn1,ckn3,ckn2,ckn4}. The metric satisfies the 5D
field
equations with a null-radiation energy-momentum tensor. The
modified Friedmann equation has the standard form,
but with a dark radiation term that no longer behaves strictly
like radiation.
A more complicated bulk metric arises when there is a
self-interacting scalar field $\Phi$ in the bulk. The simplest
case, when there is no coupling between the bulk field and brane
matter, was studied in \cite{mw3,mw1,mw2}.
When there is coupling between brane matter and the bulk scalar
field, then the Friedmann and conservation equations are more
complicated.
The scalar field could represent a bulk dilaton of the
gravitational sector, or a modulus field encoding the dynamical
influence on the effective 5D theory of an extra dimension other
than the large fifth dimension 
\cite{rbbd2,rbbd3,rbbd5,rbbd7,rbbd4,rbbd6,rbbd1}. For two-brane models,
the brane separation introduces a new scalar degree of freedom,
the radion. For general potentials of the scalar field which
provide radion stabilization, 4D Einstein gravity is recovered at
low energies on either brane~\cite{2b3,2b4,2b2,2b1}. 
Inflation in brane-world cosmology
and the issue of stabilization in models with scalar fields in the bulk
has
been studied \cite{hs10}.

\[ \]
{\bf Acknowledgements:}\\

I would like to thank Maria Fe Elder for help
in preparing this manuscript, Sigbjorn Hervik for useful comments
and NSERC for financial support.
This review includes work done in collaboration with 
Marco Bruni, 
Chris Clarkson, Peter Dunsby, Noreen Goheer,
Yanjing He, Sigbjorn Hervik,
Robert van den Hoogen and Woei Chet Lim.


\newpage

 \begin{references}
 
 
 \bibitem[*]{byline} Electronic address: aac@mathstat.dal.ca


\bibitem{Ruth}
J. M. Aguirregabiria, L. P. Chimento and R. Lazkoz,
 Class. Quantum Grav. {\bf 21}, 823 (2004).


\bibitem{Brandenberger3}
A. Albrecht, R.H. Brandenberger and
R. Matzner, Phys.\ Rev.\ D{\bf 35}, 429 (1987).

\bibitem{cline15}
L. Anchordoqui, C. Nunez and K. Olsen,
J. High Energy Phys. {\bf 10}, 050 (2000).



\bibitem{rubakov5}
I. Antoniadis, Phys.
Lett. {\bf B246}, 377 (1990).


\bibitem{rubakov4}
I. Antoniadis, N. Arkani-Hamed, S. Dimopoulos and G.
Dvali, Phys. Lett. B{\bf 436}, 257 (1998). 



\bibitem{rubakov3}
N. Arkani-Hamed, S. Dimopoulos and G. Dvali, Phys. Lett. B{\bf 429},
263 (1998).


\bibitem{HDDK1}
N. Arkani-Hamed, S. Dimopoulos, G. Dvali and N. Kaloper, Phys.
Rev. Lett. {\bf 84}, 586 (2000).


\bibitem{hs11}
P.R. Ashcroft, C. van de Bruck and A.-C. Davis, Phys. Rev. D{\bf 69},
063519 (2004).


\bibitem{bernandis2}
A. Balbi
{\it et al.}, Astrophys. J. {\bf 545}, L1 (2000).


\bibitem{mw3} 
C. Barcelo and M.
Visser, J. High Energy Phys. {\bf 10}, 019 (2000).


\bibitem{bv}
C. Barcelo and M. Visser, Phys. Lett. {\bf B482}, 183 (2000).

\bibitem{barreiro}
T. Barreiro, E.J. Copeland and N.J. Nunes, Phys. Rev. 
{\bf 61}, 127301 (2000).


\bibitem{cla-bar99b}
R.K. Barrett and C.A. 
Clarkson,
Class. Quantum Grav. {\bf 17}, 5047 (2000).


\bibitem{Barrow} 
J.D. Barrow, Nature {\bf 272}, 211 (1978).


\bibitem{BH} 
J.D. Barrow and S. Hervik, Class. Quantum Grav. 
{\bf 19}, 155 (2002).


\bibitem{Barrow:2002b} 
J.D.
Barrow and R. Maartens,  Phys. Lett. B{\bf 532} 153 (2002).

\bibitem{ek5a}
M. Bastero-Gil, E.J. Copeland, J. Gray, A. Lukas and M. Plumacher,
Phys. Rev. D{\bf 66}, 066005 (2002). 


\bibitem{bkl1}
V.A. Belinskii, I.M.  Khalatnikov and E.M. Lifshitz,
Adv. Phys. {\bf 19}, 525 (1970).


\bibitem{bkl2} 
V.A. Belinskii, I.M. Khalatnikov and E.M. Lifshitz,
Sov. Phys. Usp. {\bf 13},  745  (1971).
 

\bibitem{bkl3} 
V.A. Belinskii, I.M. Khalatnikov 
and E.M. Lifshitz,
Adv. Phys. {\bf31}, 639 (1982).

 


\bibitem{inf3}
M.C. Bento and O. Bertolami, Phys. Rev. D{\bf
65}, 063513 (2002).

\bibitem{inf4}
M.C. Bento, O. Bertolami
and A.A. Sen, Phys. Rev. D{\bf 67}, 023504 (2003).


\bibitem{inv1}
B.K. Berger and V. Moncrief, 
Phys. Rev. D{\bf 57},  7235 (1998).

\bibitem{inv2}
B.K. Berger and D. Garfinkle, Phys. Rev. D{\bf57}, 4767 (1998).







\bibitem{bernandis1}
P. de Bernandis {\it et al.}, Nature {\bf 404}, 955 (2000).


\bibitem{Billyard1}
A.P. Billyard, A.A. Coley and R.J. van den Hoogen, 
Phys. Rev. D{\bf 58}, 123501 (1998).
 


\bibitem{BinDefLan:2000a}
P. Bin$\acute{\mbox{e}}$truy, C. Deffayet and D. Langlois, 
Nucl. Phys. {\bf B565},  269  (2000).


\bibitem{BinDefLan:2000b}
P. Bin$\acute{\mbox{e}}$truy, C. Deffayet, U. Ellwanger 
and D. Langlois, Phys.
  Lett. B{\bf 477},  285  (2000).
  
  \bibitem{bub1}
J.J. Blanco-Pillado and M. Bucher, Phys. Rev. D{\bf 65}, 083517
(2002). 
  
  \bibitem{msm2}
 P. Bowcock, C. Charmousis and R. Gregory, Class.
Quantum Grav. {\bf 17}, 4745 (2000).



 

\bibitem{rbbd2}
P. Brax, C. van de
Bruck, A.-C. Davis and C.S. Rhodes, in Proc. Marseille 
Fund. Phys. Conf. (2003) [hep-ph/0309181]. 


\bibitem{Brandenberger1}
R. Brandenberger, G. Geshnizjani and S. Watson. Phys. 
Rev. D{\bf 67}, 123510 (2003).


 



\bibitem{new3}
C. van de Bruck,
M. Dorca, R.H. Brandenberger and A. Lukas,  Phys. Rev. D{\bf 62},
123515 (2000). 

\bibitem{new6}
C. van de Bruck, M. Dorca, C.J. Martins and 
M. Parry, Phys. Lett. {\bf B495}, 183
(2000). 


\bibitem{rbbd3}
R. Brustein, S.P.
de Alwis and E.G. Novak, Phys. Rev. D{\bf 68}, 043507 (2003). 





\bibitem{Cadeau:2000t}
C. Cadeau and E. Woolgar, Class. Quantum Grav. {\bf 527} (2001).




\bibitem{CamposSopuerta:2001a}
A. Campos and C.F. Sopuerta, Phys. Rev. D{\bf 63}, 104012 (2001).

\bibitem{CamposSopuerta:2001b}
A. Campos and C.F. Sopuerta, Phys. Rev. D{\bf 64}, 104011 (2001).


\bibitem{CMMS} 
A. Campos, R. Maartens, D. Matravers and 
C.F. Sopuerta,
Phys.\ Rev.\ {\bf D68}, 103520 (2003).


\bibitem{CarrColey:1999} 
B.J. Carr and  A.A. Coley ,  Class. Quantum Grav. {\bf 16}, R31
(1999) [gr-qc/9806048]. 






\bibitem{HDDK2}
A. Chamblin and G.W. Gibbons, 
Phys. Rev. Lett. {\bf 84}, 1090 (2000).


\bibitem{chr}
A. Chamblin, S.W. Hawking and H.S. Reall, Phys. Rev. D{\bf 61},
065007 (2000).


 




\bibitem{ckn1}
A. Chamblin, A. Karch and A. Nayeri, Phys. Lett. {\bf B509}, 163
(2001). 


\bibitem{CHSS} 
A. Chamblin, H.S. Reall, H. Shinkai and T. Shiromizu, Phys. Rev. 
D{\bf 63}, 064015 (2001).


\bibitem{hs3}
C. Charmousis, Class. Quantum Grav. {\bf 19}, 83 (2002).


\bibitem{cline9}
D.J.H. Chung and K. Freese, Phys.
Rev. D{\bf 61}, 023511 (2000). 


\bibitem{cla-bar99a}
C.A. Clarkson and R.K. Barrett,
Class. Quantum Grav. {\bf 16}, 3781 (1999).





\bibitem{cline1}
J.M. Cline, C. Grojean and G. Servant, Phys. Rev. Lett. 
{\bf 83}, 4245 (1999).



\bibitem{coley} 
A.A. Coley, Phys. Rev. Letts.
{\bf 89}, 281601 (2002).


\bibitem{Coley:2002a}
A.A. Coley, Class. Quantum Grav., {\bf 19}, L45, (2002).

\bibitem{Coley:2002b}
A.A. Coley, Phys. Rev. D{\bf 66}, 023512, (2002).




 

\bibitem{COLEY2} 
A.A. Coley, {\em Dynamical Systems
and Cosmology} (Kluwer Academic Publishers, Dordrecht, 2003).


\bibitem{CG} 
A.A. Coley and M. Goliath, Phys. Rev. D{\bf 62}, 043526 (2000).




\bibitem{hervik} 
A.A. Coley and S. Hervik, Class. Quantum Grav. 
{\bf20}, 3061 (2003).


\bibitem{ColHer1} 
A.A. Coley and S. Hervik, 
Class. Quantum Grav. {\bf 21}, 5759 (2004).

\bibitem{ColHer2} 
A.A. Coley and S. Hervik,
Class. Quantum Grav. {\bf 21}, 4193 (2004).



\bibitem{CHL} 
A.A. Coley, Y. He and W.C. Lim, Class. Quantum Grav. 
{\bf 21}, 1311 (2004).


\bibitem{classification2} 
A.A. Coley, R. Milson, N. Pelavas,
A. Pravda, A. Pravdova and R. Zhalaletdinov, 
Phys. Rev. D{\bf 67}, 104020 (2002).
 


\bibitem{classification1}
A.A. Coley, R. Milson, V. Pravda and A. Pravdova, 
 Class. Quantum Grav. (2004).
 Class. Quantum Grav. {\bf 21}, 5519 (2004).
 




\bibitem{coleymcmanus}
 A.A. Coley and D.J. McManus,  Class. Quantum
Grav. {\bf 11}, 1261 (1994).





\bibitem{steep7}
A.A. Coley
and R.J. van den Hoogen, Phys. Rev. D{\bf 62}, 023517 (2000).


\bibitem{IbanezvandenHoogenColey}
A.A. Coley, J. Ib{\'a}{\~n}ez, and R.J. van den Hoogen, J. Math. Phys.
{\bf 38},  5256  (1997).




\bibitem{ch73}
C.B. Collins and S.W. Hawking, Astrophys. J. {\bf 180}, 317
(1973).

\bibitem{ek5b}
E.J. Copeland, J. Gray, A. Lukas and D. Skinner, Phys. Rev. D{\bf 66},
124007 (2002).


 


\bibitem{steep1}
E.J. Copeland, A.R. Liddle and J.E. Lidsey, Phys. Rev. D{\bf 64},
023509 (2001).  



\bibitem{cline3}
C. Csaki, M. Graesser, C. Kolda, and J. Terning,
Phys. Lett. {\bf B462}, 34 (1999).


\bibitem{Dabr} 
M.P. Dabrowski, Phys. Lett. 
B{\bf 474}, 52 (2000).



\bibitem{BDDH2}
T. Damour
and M. Henneaux,  Phys. Rev. Letts.  {\bf 85}, 920 (2000).

\bibitem{DaRo}  
F. Dahlia and C. Romero, 
Class. Quantum Grav.  {\bf 21}, 927  (2004).


\bibitem{new8}
N. Deruelle, T. Dolezel
and J. Katz, Phys. Rev. D{\bf 63}, 083513 (2001).


\bibitem{kss1}
G. Dvali and S.-H.H. Tye, Phys. Lett. {\bf B450}, 72 (1999). 


\bibitem{EB2}
P.K.S. Dunsby, M. Bruni and G.F.R. Ellis, { Astrophys. J.}
{\bf 395}, 54 (1992).

\bibitem{dunsbyetal}
P.K.S. Dunsby, N. Goheer, M. Bruni and A.A. Coley, 
Phys. Rev. D{\bf 69}, 101303 (R)
 (2004).





\bibitem{EGS} 
J. Ehlers, P. Geren and R.K. Sachs,  J. Math. Phys. 
{\bf 9}, 1344 (1968).



\bibitem{ee3}
 G.F.R. Ellis, in {\em General Relativity and
Cosmology}, ed. R.K. Sachs (Academic: New York, 1971).

\bibitem{EB1}
G.F.R. Ellis and M. Bruni, Phys. Rev. D{\bf 40}, 1804
(1989). 

\bibitem{cov2}
 G.F.R. Ellis and H. van Elst, in {\em
Theoretical and Observational Cosmology}, ed. M. Lachieze-Rey
(Kluwer, Dordrecht, 1999).


\bibitem{egsother1}
G.F.R. Ellis, D.R. Matravers and R. Treciokas,
Ann. Phys. {\bf 150}, 455 (1983).


\bibitem{egsother2}
G.F.R. Ellis, D.R. Matravers and R. Treciokas,
Ann. Phys. {\bf 150}, 487 (1983).



\bibitem{hveugg97}
H. van Elst and C. Uggla,  Class. Quantum Grav.  
{\bf 14}, 2673 (1997).


\bibitem{elst} 
H. van Elst, C. Uggla and J. Wainwright, 
Class. Quantum Grav. {\bf 19}, 51 (2002).


\bibitem{FLSZ} 
A. Fabbri, D. Langlois, D.A. Steer and R. Zegers, JHEP
{\bf 0409}, 025 (2004).



\bibitem{BinDefLan:2000c}
\'E.\'E. Flanagan, 
S.-H.H. Tye, and
I. Wasserman, Phys. Rev. D{\bf 62},  044039  (2000).


 

\bibitem{Foster} 
S. Foster,  Class. Quantum Grav. 
{\bf 15}, 3485 (1998).



\bibitem{frolov} 
A.V. Frolov,
Phys. Lett. B{\bf 514},  213 (2001).

 

\bibitem{hs10}
 A.V. Frolov and L.
Kofman, Phys. Rev. D{\bf 69}, 044021 (2004). 


\bibitem{Garf}
D. Garfinkle, Int. J. Mod. Phys. D{\bf 13} 2261  (200).


\bibitem{perlmutter3}
P. Garnavich {\it et al.}, 
Astrophys. J. {\bf 493},
L53 (1998). 


\bibitem{bub2}
J. Garriga and T. Tanaka, Phys. Rev. D{\bf 65}, 103506 (2002). 

\bibitem{bub3}
U. Gen, A. Ishibashi and T.
Tanaka, Phys. Rev. D{\bf 66}, 023519 (2002).



\bibitem{Gerg}
L.A. Gergely and R. Maartens, Class. Quantum Grav. 
{\bf 19},  213 (2002).


\bibitem{Giddings} S.B. Giddings, S. Kachru and J. Polchinski,
Phys. Rev. D{\bf 66}, 106006 (2002).

\bibitem{Dunsby:2002}
N. Goheer and P.K.S. Dunsby, Phys. Rev. D{\bf 66}, 
043527 (2002).

\bibitem{naureen2}
N. Goheer and P.K.S. Dunsby,
Phys. Rev. D{\bf 67}, 103513 (2003).

 
\bibitem{dunsbyetal2}
N. Goheer, P.K.S. Dunsby, A.A. Coley and M. Bruni, Phys. Rev. 
D{\bf 70}, 123517 (2004).


\bibitem{GP1}
D.S. Goldwirth, Phys.\ Lett.\ B{\bf 243}, 41 (1990).


\bibitem{GP2} 
D. S. Goldwirth  and T. Piran, Phys. Rev. D{\bf 40}, 3263 (1989).

 

\bibitem{GP3}
D.S. Goldwirth and T. Piran, Phys. Rep.  {\bf 214}, 223 (1993).


\bibitem{Goode}
S.W. Goode, Phys. Rev. D{\bf 39}, 2882 (1989).


\bibitem{GW85a} 
S.W. Goode and J. Wainwright, Class. Quantum Grav. 
{\bf 2}, 99 (1985). 


\bibitem{GW85b}
S.W. Goode, A.A. Coley and J. Wainwright, Class. Quantum
Grav. {\bf 9}, 445 (1992).



\bibitem{gm}
C. Gordon and R. Maartens, Phys. Rev. D{\bf 63}, 044022 (2001).






\bibitem{La89d}
J.J. Halliwell, Phys. Letts. B{\bf 185}, 341 (1987).

  






\bibitem{cline6}
R.M. Hawkins and J.E. Lidsey, Phys. Rev. D{\bf 63}, 
041301 (2001).


\bibitem{inf6}
R.M. Hawkins and J.E. Lidsey, Phys. Rev. D{\bf 68}, 083505 (2003). 


\bibitem{Hewitt} 
C.G. Hewitt, R. Bridson and J. Wainwright,
Gen. Rel. Grav. {\bf 33}, 65 (2001).


\bibitem{cline16}
Y. Himemoto and 
M. Sasaki, Phys. Rev.
D{\bf 63}, 044015 (2001).

 

\bibitem{hs4}
Y. Himemoto, T. Tanaka and M. Sasaki, Phys. Rev.
D{\bf 65}, 104020 (2002) 


\bibitem{vandenHoogenAbolghasem:2003}
R.J. van den Hoogen and H. Abolghasem,  preprint, 2003.

\bibitem{HH}  
R.J. van den Hoogen and  A. Horne, preprint, gr-qc/0408014.


\bibitem{HI}
R.J. van den Hoogen and J. Ibanez,
Phys. Rev. D{\bf 67}, 083510 (2003).



\bibitem {vandenHoogen:1999}
R.J. van den Hoogen and I. Olasagasti, Phys. Rev. D{\bf 59}, 
107302 (1999).


\bibitem{HCH}
R.J. van den Hoogen, A.A. Coley and Y. He,
Phys. Rev. D{\bf 68}, 023502 (2003).


\bibitem{Billyard2}
R.J. van den Hoogen, A.A. Coley, and D. Wands, Class. 
Quantum Grav. {\bf 16}, 1843 (1999).



\bibitem{Horava} 
P. Horava and E.
Witten, Nucl. Phys. B{\bf 460}, 506 (1996).




 

\bibitem{steep6}
G. Huey and J.E. Lidsey,
Phys. Lett. {\bf B514}, 217 (2001). 



\bibitem{cline4}
D. Ida, J. High Energy Phys.
{\bf 09}, 014 (2000). 

\bibitem{IsenKich}
J. Isenberg and S. Kichenassamy,  Phys. Rev. D{\bf58}, 064023 (1998).




\bibitem{Ishihara} H. Ishihara,
Phys. Rev. D{\bf66}, 023513 (2002).


\bibitem{js86}
L.G. Jensen and J. Stein-Schabes, Phys. Rev. D{\bf 34}, 931 (1986).


\bibitem{rbbd5}
S. Kachru, R.
Kallosh, A. Linde, J. Maldacena, L. McAllister and S.P. Trivedi, JCAP
{\bf 10}, 013 (2003). 


\bibitem{rbbd7}
S. Kachru, R.
Kallosh, A. Linde and S.P. Trivedi, 
Phys. Rev. D{\bf 68}, 046005 (2003).

\bibitem{mtheory1}
R. Kallosh, Stephen Hawking's 60th Birthday Conference 
(2002) [hep-th/0205315].


\bibitem{ek2} 
R. Kallosh, L. Kofman and A. Linde, Phys. Rev. D{\bf 64}, 
123523 (2001).  





\bibitem{cline7}
N. Kaloper, Phys. Rev. D{\bf 60}, 123506 (1999).

\bibitem{kss2}
S. Kanno, M. Sasaki and J. Soda, Prog. Theor. Phys.
{\bf 109}, 357 (2003).



\bibitem{cline8}
P. Kanti, I.I. Kogan, K.A. Olive, and 
M. Pospelov, Phys.
Lett. {\bf B468}, 31 (1999).

\bibitem{Katz} N. I. Katzourakis, preprint, math.DG/0411630.


\bibitem{ek1}
J. Khoury, B.A. Ovrut, P.J. Steinhardt and N. Turok, Phys. Rev. 
D{\bf 64}, 123522 (2001).

\bibitem{KichRen}
S. Kichenassamy and A.D. Rendall, Class. Quantum Grav.
{\bf15}, 1339 (1998).


\bibitem{Kitada}
Y. Kitada and K. Maeda, Class. Quantum Grav. {\bf 10},  703  (1993).


\bibitem{hs1}
S. Kobayashi, K. Koyama and J. Soda, Phys. Lett. {\bf B501}, 157
(2001). 

\bibitem{new2}
H. Kodama, A. Ishibashi and O. Seto, Phys. Rev. D{\bf 62}, 
064022 (2000).


\bibitem{rbbd4}
L. Kofman, preprint astro-ph/0303614.


\bibitem{new4}
K. Koyama and J. Soda, Phys. Rev. D{\bf 62},
123502 (2000). 


\bibitem{hs5}
K. Koyama and K.
Takahashi, Phys. Rev. D{\bf 67}, 103503 (2003).

\bibitem{cline10}
P. Kraus, J. High Energy Phys. 
{\bf 9912}, 011 (1999).


\bibitem{Brandenberger2}
J.H. Kung  and H. Brandenberger,  Phys. Rev. D{\bf 42}, 
1008 (1990).

\bibitem{La89a}  
D. La and P.J. Steinhardt, Phys. Rev. Letts. 
{\bf 62}, 376 (1989).



\bibitem{bernandis3}
A.E. Lange
{\it et al.}, Phys. Rev. D{\bf 63}, 042001 (2001).

\bibitem{Lang}
D. Langlois, 6th RESCEU Int. Symp. (2003) [astro-ph/0403579].

\bibitem{new7}
D. Langlois, Phys. Rev. Lett. {\bf 86}, 2212 (2001).

\bibitem{hs6}
D. Langlois and M. Sasaki, Phys. Rev. D{\bf 68}, 064012 (2003).

\bibitem{ckn3}
D. Langlois and L. Sorbo, Phys. Rev. D{\bf 68},
084006 (2003).

 
\bibitem{lmw1}
D. Langlois, R. Maartens, and D. Wands, Phys. Lett. {\bf B489},
259 (2000).

\bibitem{lmw2}
D. Langlois, R. Maartens, M. Sasaki and D. Wands, 
Phys. Rev. D{\bf 63},
084009 (2001).

\bibitem{ckn2}
D. Langlois, L. Sorbo and M.
Rodriguez-Martinez, Phys. Rev. Lett. {\bf 89}, 171301 (2002).  






\bibitem{ckn4}
E. Leeper, R. Maartens, C.
Sopuerta, Class. Quantum Grav. {\bf 21}, 1125 (2004).



\bibitem{2b3}
J. Lesgourgues and  L. Sorbo, Phys.
Rev. D{\bf 69}, 084010 (2004). 





\bibitem{lidsey3}
A.R. Liddle and D.H.
Lyth, {\it Cosmological Inflation and Large Scale Structure} 
(Cambridge
University Press, 2000).

\bibitem{steep5}
A.R. Liddle and L.A. Urena-Lopez, Phys. Rev.
D{\bf 68}, 043517 (2003). 


\bibitem{lidrev}
J.E. Lidsey, Lect. Notes Phys. {\bf 646}, 357 (2004).



\bibitem{lidsey1}
J.E. Lidsey, A.R. Liddle, 
E.W. Kolb, E.J.
Copeland, T. Barreiro and M. Abney, Rev. Mod. Phys. 
{\bf 69}, 373 (1997). 


\bibitem{lim0306118} 
W.C. Lim, H. van. Elst, 
C. Uggla and J. Wainwright, Phys. Rev. D{\bf 69}, 103507 
(2004).


\bibitem{LW1} 
X.F. Lin and R.M. Wald, Phys. Rev. D{\bf 40}, 
3280 (1989).


\bibitem{LW2}
X.F. Lin and R.M. Wald, Phys. Rev. D{\bf 41}, 2444 (1989).




\bibitem{low2}
A. Lukas, B.A. Ovrut and D.
Waldram, Phys. Rev. D{\bf 60}, 086001 (1999).

\bibitem{low3}
A. Lukas, B.A. Ovrut and D.
Waldram, Phys. Rev. D{\bf 61}, 023506 (2000).

\bibitem{low1}
A. Lukas, B.A. Ovrut, K.S. Stelle and D. Waldram, 
Phys. Rev. D{\bf 59}, 
086001 (1999). 





\bibitem{lidsey2}
D.H. Lyth and A. Riotto, Phys. Rep. {\bf 314}, 1 (1999).







\bibitem{Maartens1}
R. Maartens, Phys. Rev. D{\bf 62}, 084023 (2000).


\bibitem{Maartens2}
R. Maartens, Living Reviews in Relativity 
{\bf 7}, 1 (2004).



\bibitem{cov1} 
R. Maartens, T. Gebbie and G.F.R. Ellis, Phys. Rev.
D{\bf 59}, 083506 (1999).

\bibitem{mss} 
R. Maartens, V. Sahni and T.D. Saini, Phys. Rev. D{\bf 63}, 
063509 (2001).


\bibitem{mwbh} 
R. Maartens, D. Wands, B.A.
Bassett and I.P.C. Heard, Phys. Rev. D{\bf 62}, 041301R (2000).




\bibitem{mac73} M.A.H. MacCallum, 
in {\em Carg\`{e}se Lectures in Physics Vol. 6\/}, ed. E. Schatzman
(Gordon and Breach, New York, 1973).

\bibitem{MT} 
K. Maeda and T. Torii, Phys. Rev. D{\bf 69}, 024002 (2004).


\bibitem{mw1}
K. Maeda and D. Wands, Phys. Rev. D{\bf 62}, 124009 (2000).


\bibitem{steep2}
A.S. Majumdar, Phys. Rev.
D{\bf 64}, 083503 (2001). 


\bibitem{hs9}
 J. Martin, G.N. Felder, A.V.
Frolov, M. Peloso and L. Kofman, 
Phys. Rev. D{\bf 69}, 084017 (2004).


\bibitem{cline5}
A. Mazumdar, Nucl. Phys. {\bf B597}, 561 (2001).


\bibitem{inf2}
A. Mazumdar, Phys. Rev. D{\bf 64}, 027304 (2001).


\bibitem{Mazumbar} 
A. Mazumbar, Phys. Rev. D{\bf 64}, 083503 (2001).



\bibitem{cline17}
L. Mendes and A.R. Liddle, Phys. Rev. D{\bf 62},
103511 (2000).


 

\bibitem{mw2}
A. Mennim and R.A. Battye, Class. Quantum Grav. {\bf 18}, 2171 (2001).



\bibitem{hs8}
M. Minamitsuji, Y. Himemoto and M. Sasaki, Phys. Rev. D{\bf 68}, 024016
(2003).

\bibitem{inf5}
S. Mizuno, K.I. Maeda and K.
Yamamoto, Phys. Rev. D{\bf 67}, 024033 (2003). 



\bibitem{cline2}
R.N. Mohapatra, A. Perez-Lorenzana and  C.A. de S. Pires, 
Phys. Rev. D{\bf 62}, 105030 (2000).


\bibitem{cline11}
S. Mukohyama, Phys. Lett. {\bf B473}, 241 (2000). 

 
\bibitem{new1}
S. Mukohyama, Phys. Rev. D{\bf 62}, 084015 (2000).

\bibitem{new5} 
S. Mukohyama, Class.
Quantum Grav. {\bf 17}, 4777 (2000).

 
\bibitem{2b4}
S. Mukohyama and
A.A. Coley, Phys. Rev. D{\bf 69}, 064029 (2004).

\bibitem{2b2}
S. Mukohyama and L. Kofman, Phys. Rev. D{\bf 65},
124025 (2002). 


\bibitem{cline14}
S. Mukohyama, T. Shiromizu and K. Maeda, Phys. Rev. D{\bf 61},
024028 (2000). 



\bibitem{steep4}
N.J. Nunes and E.J. Copeland, Phys. Rev. D{\bf 66}, 043524 (2002).

\bibitem{paul} 
B.C. Paul, Phys. Rev. D{\bf 64}, 124001
(2001).



\bibitem{Paza} E. Papantonopoulos and V. Zamaria, 
JCAP {\bf 0410},  001 (2004).

\bibitem{rbbd6}
M. Peloso and E. Poppitz,
Phys. Rev. D{\bf 68}, 125009 (2003).



\bibitem{Penrose79}
R. Penrose, in {\it General Relativity: 
An Einstein centenary survey},
eds. S.W. Hawking and W. Israel (Cambridge University Press, 1979).

\bibitem{perlmutter1}
S. Perlmutter {\it et al.}, Nature {\bf 391}, 51 (1998).

\bibitem{mtheory2}
J. Polchinski, Heisenberg Centennial Symposium (2001)
 [hep-th/0209105].



\bibitem{rubakov2}
J. Polchinsky, Phys. Rev. Lett. {\bf 75}, 4724 (1995).




\bibitem{randall1}
L. Randall and R. Sundrum, Phys. Rev. Lett. {\bf 83}, 3370 (1999); 


\bibitem{randall2}
L. Randall and R. Sundrum, Phys. Rev. Lett.
{\bf 83}, 4690 (1999).


\bibitem{ratra1}
B. Ratra and P.J.E. Peebles, Phys. Rev. D{\bf 37}, 3406 (1988).


\bibitem{ren2} 
A.D. Rendall, Class. Quantum Grav. {\bf 14},
2341 (1997).

\bibitem{R2000} 
A.D. Rendall, in {\it Proceedings of Journees 
Equations aux Derivees Partielles},
eds. N. Depauw, D. Robert and X. Saint-Raymond (Groupement de 
Recherche 1151 du CNRS, XIV, 2000).



\bibitem{rbbd1}
C.S. Rhodes, C. van de Bruck, Ph. Brax and A.C. Davis,  Phys. Rev.
D{\bf 68}, 083511 (2003). 


\bibitem{perlmutter2}
A.G. Riess {\it et al.},
Astron. J. {\bf 116}, 1009 (1998). 


\bibitem{inf7}
M. Rinaldi, Phys. Lett. {\bf B582}, 249 (2004).


\bibitem{ringstrom1}
H. Ringstr{\"o}m, Classical Quantum Gravity {\bf 17}
713 (2000).
 

\bibitem{ringstrom2}
H. Ringstr{\"o}m,
Ann. Inst. Henri Poincare, Ann. Non-Lin. {\bf 2}, 405
(2001).


\bibitem{rellis86}
T. Rothman and G.F.R. Ellis, Phys. Lett. {\bf B180}, 19 (1986).




\bibitem{rubakov1}
V. Rubakov and M.E. Shaposhnikov, Phys. Lett. B{\bf 159}, 22
(1985).

\bibitem{steep3}
V. Sahni, M. Sami and T.
Souradeep, Phys. Rev. D{\bf 65}, 023518 (2002).


\bibitem{La89c}
A. Salam and E. Sezgin, Phys. Letts. B{\bf 147}, 47 (1984).



\bibitem{Santos:2001}
M.G. Santos, F. Vernizzi and P.G. Ferreira, Phys. Rev. D{\bf 64}, 
063506 (2001).


 

\bibitem{sms2}
M. Sasaki, T. Shiromizu and K. Maeda, Phys. Rev. D{\bf 62},
024008 (2000).



\bibitem{Savchenko:2002} 
N. Yu
Savchenko, S.V. Savchenko  and A.V. Toporensky, 
Class. Quantum Grav. {\bf 19}, 4923 (2002).


\bibitem{perlmutter4}
B.P. Schmidt {\it et al.}, Astrophys. J. 
{\bf 507}, 46 (1998).

\bibitem{mtheory3}
J.H. Schwarz, Carnegie Inst. Centennial 
Symposium (2002) [astro-ph/0304507].




\bibitem{sms1} 
T. Shiromizu, K. Maeda and M. Sasaki,
Phys. Rev. D{\bf 62}, 024012 (2000).

 

\bibitem{La89b}
P.J. Steinhardt and F.S. Accetta, Phys. Rev. Letts. 
{\bf 64}, 2740 (1990).


\bibitem{ek3}
P.J. Steinhardt and N. Turok, Phys. Rev. D{\bf 65}, 126003 (2002).




\bibitem{cline13}
H. Stoica, S.-H. Henry Tye and I. Wasserman, Phys. Lett. 
{\bf B482}, 205 (2000).


\bibitem{2b1}
T. Tanaka and X. Montes, Nucl. Phys. {\bf B582}, 259 (2000). 



\bibitem{Toporensky:2001} 
A.V. Toporensky, Class. Quantum Grav. {\bf 18}, 2311 (2001).



\bibitem
{Uggla-zurMuhlen} C. Uggla and H. von zur M\"uhlen,
  Class. Quantum Grav. {\bf 7}, 1365 (1990).






\bibitem{cline12}
D.N. Vollick,
Class. Quantum Grav. {\bf 18}, 1 (2001).

 

\bibitem{ratra3}
L. Wang, R.R. Caldwell, J.P.
Ostriker and P.J. Steinhardt, Astrophys. J. {\bf 530}, 17 (2000).

\bibitem{WE} 
J. Wainwright and G.F.R. Ellis, 
{\it Dynamical Systems in Cosmology} (Cambridge University 
Press, 1997).

 
\bibitem{WIB}
M. Weaver, J. Isenberg and B.K. Berger, Phys. Rev. Letts. 
{\bf 80}, 2984 (1998).


\bibitem{ratra2}
C. Wetterich, Nucl.
Phys. B{\bf 302}, 668 (1998).


 


\end{references}
\end{document}